\documentclass{aa}  
\usepackage{twoopt}
\bibpunct{(}{)}{;}{a}{}{,}             %% natbib 
  \newcommandtwoopt{\citeads}[3][][]{\href{http://adsabs.harvard.edu/abs/#3}%
    {\def\hyper@linkstart##1##2{}%
     \let\hyper@linkend\@empty\citealp[#1][#2]{#3}}}
  \newcommandtwoopt{\citepads}[3][][]{\href{http://adsabs.harvard.edu/abs/#3}%
    {\def\hyper@linkstart##1##2{}%
     \let\hyper@linkend\@empty\citep[#1][#2]{#3}}}
  \newcommandtwoopt{\citetads}[3][][]{\href{http://adsabs.harvard.edu/abs/#3}%
    {\def\hyper@linkstart##1##2{}%
     \let\hyper@linkend\@empty\citet[#1][#2]{#3}}}
  \newcommandtwoopt{\citeyearads}[3][][]%
    {\href{http://adsabs.harvard.edu/abs/#3}
    {\def\hyper@linkstart##1##2{}%
     \let\hyper@linkend\@empty\citeyear[#1][#2]{#3}}}
\makeatother

\usepackage{graphicx}
\usepackage{footnote}
\usepackage{amsmath}
\usepackage{amssymb}
\usepackage{xcolor}
\usepackage{pdflscape}
\usepackage{rotating}
\usepackage{pdflscape}
\usepackage{url}
\usepackage[varg]{txfonts}
\usepackage{bm}		% Bold maths symbols, including upright Greek
\usepackage{siunitx} % For \degree symbol
\usepackage{booktabs}
\usepackage{txfonts}
\usepackage[pagewise]{lineno}
\DeclareGraphicsExtensions{.pdf,.png,.jpg}

\begin{document} 

   \title{MUSE-ALMA Haloes XIII}

   \subtitle{Molecular gas in $z\sim 0.5$ H\,\textsc{i}\,–selected galaxies}

   \author{Victoria~Bollo\inst{1\thanks{Email: victoria.bollo@eso.org}},
          C\'eline~P\'eroux\inst{1, 2},
          Martin~Zwaan\inst{1},
          Jianhang~Chen\inst{3},
          Varsha~P.~Kulkarni\inst{4},
          Capucine~Barfety\inst{3},\\
          Simon Weng\inst{2},
          Natascha~M.~Förster~Schreiber\inst{3}, 
          Linda~Tacconi\inst{3},
          Benedetta~Casavecchia\inst{5},
          Tamsyn~O'Beirne\inst{6, 1, 7},\\
          Laurent~Chemin\inst{8},
          Ramona Augustin\inst{9},
          \and
          Mitchell Halley\inst{4} 
          }

   \institute{European Southern Observatory, Karl-Schwarzschild-Str. 2, 85748 Garching near Munich, Germany \label{1}
              \and
             Aix Marseille Univ., CNRS, LAM, (Laboratoire d’Astrophysique de Marseille), UMR 7326, F-13388 Marseille, France \label{2}
            \and
            Max-Planck-Institut für Extraterrestrische Physik (MPE), Giessenbachstrasse 1, 85748 Garching near Munich, Germany \label{3}
            \and
            Department of Physics and Astronomy, University of South Carolina, Columbia, SC 29208, USA\label{4}
            \and
            Max-Planck-Institut für Astrophysik, Karl-Schwarzschild-Strasse 1, 85748 Garching near Munich, Germany \label{5}
            \and
            Centre for Astrophysics and Supercomputing, Swinburne University of Technology, Hawthorn, Victoria 3122, Australia \label{6}
            \and
             CSIRO Space \& Astronomy, PO Box 1130, Bentley Western Australia 6102, Australia \label{7}
            \and
            Observatoire Astronomique de Strasbourg, Université de Strasbourg/CNRS, 11 rue de l'Université, 67000 Strasbourg, France \label{8}
            \and
            Leibniz Institut fur Astrophysik Potsdam, An der Sternwarte 16, 14482 Potsdam, Germany\label{9}
            % \and
            % Space Telescope Science Institute, 3700 San Martin Drive, Baltimore, MD 21218, USA \label{10}
            % \and 
            % Institute for Astronomy, University of Edinburgh, Royal Observatory, Blackford Hill, Edinburgh EH9 3HJ, UK \label{11}
            % \and
            % School of Cosmic Physics, Dublin Institute for Advanced Studies, 31 Fitzwilliam Place, Dublin D02 XF86, Ireland \label{12}
            }

   \date{Received XXX; accepted XXX}

\titlerunning{Molecular gas in H\,\textsc{i}–absorbers at $z\sim 0.5$}
\authorrunning{V.~Bollo et al.}
% \abstract{}{}{}{}{} 
% 5 {} token are mandatory
 
\abstract{
We present further results from the MUSE-ALMA Haloes survey, which includes 79 galaxies associated with strong H\,\textsc{i} absorption at $z\sim 0.5$.
As part of this effort, our ALMA Cycle 10 Large Program contributed new observations of 39 systems. 
This expands on the initial set of {21 systems} in the MUSE-ALMA Haloes survey, bringing the total to {60} galaxies.
Among the newly observed systems, we detect CO line emission in 9 galaxies, corresponding to a 23\% detection rate in a sample not selected by metallicity.
When combined with prior MUSE-ALMA Haloes data, our total CO detection count rises to 12 out of 60 galaxies ({20}\%), effectively doubling the number of detected CO-emitting H\,\textsc{i}–selected galaxies at $z\sim0.5${ and probing a factor of $\sim1.2$ dex deeper in $M_{\mathrm{H}_2}$ than earlier absorber studies}.
These sources, selected based on known circumgalactic H\,\textsc{i} gas, span a wide range of stellar masses and metallicities, providing a unique view of gas-rich environments. 
By comparing the molecular gas properties, traced through CO(2$-$1) and CO(3$-$2) transitions with existing information of their physical properties, such as star formation rates (SFRs) and gas-phase metallicities from VLT/MUSE and HST spectroscopy, we investigate how these systems relate to the population of normal star-forming galaxies at similar redshift.
{Our deep, unbiased CO observations of H\,\textsc{i}-selected galaxies reveal a dual behaviour in star formation efficiency. Low-$M_{\rm H_2}$ systems form stars efficiently and follow the scaling relations of main-sequence galaxies, while high-$M_{\rm H_2}$ systems exhibit suppressed star formation and lower-than-expected stellar masses, likely reflecting ongoing gas accretion or environmental regulation. This diversity indicates that H\,\textsc{i} absorbers trace both evolved, actively star-forming galaxies and younger or dynamically influenced systems that are still building their gas reservoirs. By reaching molecular gas masses more than 1 dex below previous studies, our survey provides a key step toward completing the baryon census at $z\sim0.5$ and characterising the molecular phase of the broader H\,\textsc{i}-selected population.}
}
  % context heading (optional)
  % {} leave it empty if necessary  
  %  {}
  % % aims heading (mandatory)
  %  {}
  % % methods heading (mandatory)
  %  {}
  % % results heading (mandatory)
  %  {}
  % % conclusions heading (optional), leave it empty if necessary 
  %  {}


   \maketitle
%
%-------------------------------------------------------------------
\section{Introduction} \label{sec:introduction}

The evolution of galaxies is fundamentally regulated by the availability and cycling of gas, which serves as raw material for star formation. 
Cold gas plays a crucial role in this process, as it forms stars that enrich the interstellar medium (ISM) with metals and dust over time \citep{kennicuttStarFormationMilky2012, carilliCoolGasHighRedshift2013, tacconiEvolutionStarFormingInterstellar2020}. 
Feedback from supernovae (SNe) and active galactic nuclei (AGN) injects mechanical and radiative energy into the surrounding environment, influencing the efficiency of star formation and driving large–scale outflows \citep{oppenheimerMassMetalEnergy2008, lillyGasRegulationGalaxies2013, faucher-giguereKeyPhysicalProcesses2023}. 
These processes collectively shape the baryon cycle, which describes how gas is accreted, converted into stars, and ejected into the circumgalactic medium (CGM), with the potential for subsequent re-accretion \citep{tumlinsonCircumgalacticMedium2017, perouxCosmicBaryonMetal2020, perouxMultiScaleMultiPhaseCircumgalactic2024}.

A complete understanding of the baryon cycle requires studying the CGM, which acts as a reservoir of gas regulating galaxy growth \citep{werkCOSHALOSSURVEYPHYSICAL2014, nelsonResolvingSmallscaleCold2020, naabTheoreticalChallengesGalaxy2017, oppenheimerSimulatingGroupsIntraGroup2021}.
The CGM has been extensively probed using a multi-wavelength approach. 
Observations have revealed that the CGM is multiphase, comprising gas at a range of temperatures, densities, and ionisation states.
This includes cold neutral gas, warm ionised gas, and hot, X-ray emitting plasma.
Optical and ultraviolet absorption-line spectroscopy, particularly with the Cosmic Origins Spectrograph on the Hubble Space Telescope (HST) \citep[e.g.,][]{muzahidHSTCOSSurvey2015, richterHSTCOSLegacy2017} has been crucial in studying ionised gas phases. 
X-ray observations with Chandra, XMM-Newton, and eROSITA trace the hot component of the CGM  \citep[e.g.,][]{stricklandHighSpatialResolution2004, liEnvironmentalEffectsAtomic2013, bogdanHOTXRAYCORONAE2013, merloniEROSITAScienceBook2012, zhangHotCircumgalacticMedium2025, linInadequateTurbulentSupport2025}. 
Atomic hydrogen (H\,\textsc{i}) has been investigated both in emission and absorption across cosmic time. 
Emission observations using the 21 cm line have been key to studying the star formation efficiency, gas kinematics, and disk structure in nearby galaxies \citep[e.g.][]{haynesNeutralHydrogenIsolated1984, zwaanHIPASSCatalogueOHI2005, bigielStarFormationLaw2008, leroyStarFormationEfficiency2008, walterTHINGSNEARBYGALAXY2008, catinellaXGASSTotalCold2018, yuCentrallyConcentratedDistribution2022, sharmaGMRTHighResolution2023, wangFEASTSIGMCooling2023, wangFEASTSRadialDistribution2025}. However, direct detection of 21 cm emission becomes increasingly difficult beyond $z\sim 0.1$, and even harder for individual galaxies at $z>0.4$ \citep{fernandezHIGHESTREDSHIFTIMAGE2016, xiMostDistantGalaxies2024}.
To probe higher redshifts, statistical techniques such as spectral stacking and intensity mapping have been adopted \citep{masuiMEASUREMENT21Cm2013, rheeNeutralHydrogenGas2018, beraAtomicHydrogenStarforming2019, chowdhury21centimetreEmissionEnsemble2020}, and efforts are ongoing supported by SKA precursors and pathfinders \citep[e.g.,][]{blythLADUMA2016MeerKAT2016, adamsRadioSurveysNow2019, koribalskiASKAPRevealsRadio2024, maddoxMIGHTEEHIEmissionProject2021}.
Given the emission limitations, 21 cm absorption studies provide a powerful alternative, as sensitivity is independent of the galaxy's luminosity distance. FAST has recently enabled discoveries of new low-redshift absorbers \citep{suFASTDiscoveryFast2023, yuFASTObservationsNeutral2024}, while the Australian Square Kilometre Array Pathfinder (ASKAP) and MeerKAT have begun probing intermediate redshifts through surveys like The First Large Absorption Survey in H\,\textsc{i} (FLASH) and the MeerKAT Absorption Line Survey (MALS) \citep[e.g.,][]{allisonFLASHEarlyScience2020, hotanAustralianSquareKilometre2021, glowackiLookingDistantUniverse2022, dekaMeerKATAbsorptionLine2023, yoonFirstLargeAbsorption2024}.

The H\,\textsc{i} absorption selection method, in particular, has proved to be a powerful tool for detecting neutral gas in the CGM \citep{wolfireNeutralAtomicPhases1995, wolfeDAMPEDLYaSYSTEMS2005}. 
By observing the Ly$\alpha$ absorption lines against bright background sources, this method enables the identification of diffuse, cold gas that may be difficult to detect in emission. 
It is highly sensitive to low column densities, independent of redshift, luminosity, or star formation, and provides detailed kinematic information about the gas distribution \citep[e.g.,][]{perouxMetalrichDampedSubdamped2006, zwaanHIPASSCatalogueOHI2005, boucheImpactColdGas2010, bouchePOSSIBLESIGNATURESCOLDFLOW2016, frankObservableSignaturesLowz2012, kacprzakMorphologicalProperties052011, sternUniversalDensityStructure2016}.

The use of 3D integral field spectroscopy (IFS) has revolutionised our ability to study the connection between galaxies and the CGM, especially in systems selected via H\,\textsc{i} absorption, such as damped Lyman-$\alpha$ (DLA, $\log N$(H\,\textsc{i})/\text{cm}$^2 \gtrsim 20.3$), sub-damped Lyman-$\alpha$ systems (sub-DLA; $19.0 \lesssim \log N$(H\,\textsc{i})/\text{cm}$^2 < 20.3$){, and Lyman Limit systems  (LLS; $18.0 \lesssim \log N$(H\,\textsc{i})/\text{cm}$^2 < 19$,} \citealt{perouxEvolutionOHIEpoch2003}). 
These absorption-selected systems probe some of the most gas-rich environments in the Universe and have long been used to study gas metallicities \citep[e.g.,][]{rafelskiMETALLICITYEVOLUTIONDAMPED2012, kulkarniHubbleSpaceTelescope2005, kulkarniKeckVLTObservations2015}, kinematics \citep[e.g.,][]{prochaskaProtogalacticDiskModels1998}, gas temperatures \citep[e.g.,][]{kanekarSpinTemperatureHighredshift2014}, and chemical enrichment histories \citep[e.g.,][]{dessauges-zavadskyNewComprehensiveSet2007, quiretESOUVESAdvanced2016}.

Multiple optical surveys have used the IFS VLT/MUSE \citep{baconMUSESecondgenerationVLT2010}, such as MUSE-QuBES \citep[e.g.,][]{muzahidMUSEQuBESCalibratingRedshifts2020, duttaMUSEQuBESMappingDistribution2024, duttaMUSEQuBESKinematicsVibearing2025}, and MEGAFLOW \citep[e.g.,][]{bouchePOSSIBLESIGNATURESCOLDFLOW2016, schroetterMUSEGASFLOW2016, langanMusEGAsFLOw2023} that established statistical links between CGM absorbers and their associated galaxies. 
\cite{klimenkoBaryonicContentGalaxies2023} report results from an HST-COS survey with the opposite approach of determining the CGM properties of galaxies observed with IFS at $z < 0.1$ in the MaNGA survey. 
Parallel programs including MAGG \citep[e.g.,][]{duttaMUSEAnalysisGas2020, lofthouseMUSEAnalysisGas2020, fossatiMUSEAnalysisGas2021}, the Cosmic Ultraviolet Baryon Survey (CUBS) \citep[e.g.,][]{ boettcherCosmicUltravioletBaryon2021, cooperCosmicUltravioletBaryon2021, zahedyCosmicUltravioletBaryon2021}, and studies with the Keck/KCWI optical IFS \citep[e.g.,][]{ martinMultifilamentGasInflows2019, nielsenCGMCosmicNoon2020} have extended these insights to higher redshifts, broader dynamic ranges, and higher spectral resolution.

While significant progress has been made in studying the ionised and neutral atomic components of the CGM, the molecular gas content remains less explored. 
Molecular gas is a critical ingredient for star formation, and is closely connected with gas accretion and star formation \citep{tacconiEvolutionStarFormingInterstellar2020, walterEvolutionBaryonsAssociated2020, saintongeColdInterstellarMedium2022}. 
Observations with the Atacama Large Millimetre/submillimeter Array (ALMA), Northern Extended Millimeter Array (NOEMA), and Jansky Very Large Array (JVLA) have opened new avenues to detect molecular gas in and around galaxies.
Previous works have hinted at long gas depletion timescales and suppressed star formation efficiencies in some H\,\textsc{i}–selected systems, suggesting a distinct phase of galaxy evolution not well represented in emission–selected surveys \citep[e.g.,][]{kanekarMassiveAbsorptionselectedGalaxies2018, kanekarHighMolecularGas2020, neelemanII158mmEmission2017, neelemanMolecularEmissionGalaxy2018, neelemanIi158Mm2019, klitschALMACALAbsorptionselectedGalaxies2019, perouxMultiphaseCircumgalacticMedium2019, szakacsMUSEALMAHaloesVI2021, kaurNatureHiabsorptionselectedGalaxies2021, kaurHIabsorptionselectedColdRotating2024}. However, a systematic study of the molecular gas content of H\,\textsc{i}–selected galaxies with multiwavelength data is still lacking.

The MUSE-ALMA Haloes survey was designed to probe the multiphase nature of the circumgalactic medium by targeting 32 H\,\textsc{i} Ly$\alpha$ absorbers with column densities in the range $\log N(\text{H\,\textsc{i}})/\text{cm}^2 = 18.1 - 21.7$ at redshifts $0.2 < z < 1.4$. It combines new observations from the Multi Unit Spectroscopic Explorer (MUSE) on the Very Large Telescope (VLT), ALMA, and HST.
The scientific development of the survey builds on previous work and first results based on smaller subsamples of H\,\textsc{i} absorbers. These early studies, using precursor observations, investigated the environments of absorbers, revealing intragroup associations and multi-galaxy systems \citep{perouxNatureAbsorbingGas2017, perouxMultiphaseCircumgalacticMedium2019}, evidence for cold accretion \citep{rahmaniObservationalSignaturesWarped2018}, outflows \citep{rahmaniLymanLimitSystem2018}, and extended molecular gas beyond the ionised disks \citep{klitschALMACALIIICombined2018}.
Kinematic alignment between molecular and ionised gas phases was explored by \citet{szakacsMUSEALMAHaloesVI2021}, while \citet{hamanowiczMUSEALMAHaloesPhysical2020} reported absorber–galaxy correlations based on 14 systems.
The full statistical analysis of the MUSE data was presented by \citet{wengMUSEALMAHaloesVIII2022}, who identified 79 galaxies within $\pm$500 km/s of the absorbers in 19 MUSE fields, each covering a 1$\times$1 arcmin$^2$ field of view, providing a first comprehensive view of CGM galaxy environments in this sample. These 79 galaxies form the basis of the MUSE-ALMA Haloes survey sample \citep{perouxMUSEALMAHaloes2022}, on which subsequent studies have built.
\citet{karkiMUSEALMAHaloesIX2023} found that these absorption-selected galaxies lie on the star-forming main sequence with a $2\sigma$ scatter, and that higher H\,\textsc{i} column densities tend to be associated with more compact galaxies. They also showed that both emission and absorption metallicities correlate with stellar mass and specific SFR, suggesting that metal-poor absorbers trace galaxies with lower past star formation and more rapid current gas consumption.
\citet{wengMUSEALMAHaloesXI2023} applied 3D forward modelling of the ionised gas kinematics to explore the physical origins of the absorbers. In parallel, \citet{wengPhysicalOriginsGas2024} used the TNG50 simulation to track H\,\textsc{i} absorbers around galaxies at $z = 0.5$, finding that many absorbers at large impact parameters or low column densities arise from satellites, neighbouring haloes, or the IGM.
\citet{augustinMUSEALMAHaloesStellar2024} examined the stellar mass distribution of the sample, reporting an anti-correlation between stellar mass and H\,\textsc{i} column density, suggesting that more massive galaxies are surrounded by less neutral hydrogen.
Finally, Karki et al. (submitted) investigated 20 interacting galaxies associated with quasar absorbers, showing that tidal features trace galaxy-CGM interactions, revealing enhanced star formation, distinct kinematics, and extended CGM properties compared to non-interacting systems.

This paper presents the first results of an ALMA Large Program for MUSE-ALMA Haloes survey (Cycle 10, PI: C. Péroux), targeting galaxies selected as H\,\textsc{i} absorbers. 
The science goals of this survey include identifying the role of the molecular gas in H\,\textsc{i}-rich galaxies, comparing with the canonical scaling relations, and exploring their connection to other galaxy properties. By comparing their depletion timescales, mass-metallicity relations, and physical diversity with other galaxy populations, we aim to better understand the physical conditions that govern the efficiency of gas conversion into stars.

This paper is organised as follows. 
In Section~\ref{sec:data}, we describe the MUSE-ALMA Haloes ALMA Large program and summarise the science goals achieved with this multiwavelength dataset.
Section~\ref{sec:analysis} details our analysis and results.
Section~\ref{sec:results} presents the properties of H\,\textsc{i}–selected systems and compares our findings with existing literature.
Section~\ref{sec:discussion} discusses the nature of H\,\textsc{i}–selected galaxies.
In Section~\ref{sec:conclusions}, we summarise our key conclusions. 
Throughout this paper, we use $H_0 = 70$ kms$^{-1}$Mpc$^{-1}$, $\Omega_{\text{M}} =0.3$ and $\Omega_{\Lambda} = 0.7$. The present-day cosmological critical density is $\rho_{0, \text{crit}} \simeq 277.4$ $h^2$ M$_{\odot}$ kpc$^{-3}$. We adopt  $\log Z = 12 + \log{(\text{O/H})}_{\text{PP}04}$ following \cite{pettiniOIIINIIAbundance2004} and refer to it simply as $12 + \log{(\text{O/H})}$.

\section{Multi-wavelength data} \label{sec:data}
The analysis in this paper builds upon the extensive dataset assembled by the MUSE-ALMA Haloes survey \citep{perouxMUSEALMAHaloes2022}. In this section, we provide a brief overview of the ancillary datasets from VLT/UVES, VLT/MUSE, HST, and ALMA, and introduce the
new ALMA observations obtained as part of a Cycle 10 Large Program (PI: C. Péroux, ID: 2023.1.00127.L).
This rich dataset enables a comprehensive characterisation of the stellar, ionised, atomic, and molecular gas properties of absorption-selected galaxies, offering new insights into the baryon cycle and the physical processes governing galaxy evolution.

The MUSE-ALMA Haloes survey has measured absorber redshifts, H\,\textsc{i} column densities from Voigt profile fitting of high-resolution quasar spectra, and metal abundances of the absorbing gas. 
Within the MUSE field of view, a total of 3658 extra-galactic sources were found across all the fields, of which 79 galaxies were identified within $\pm500$ km/s of the absorbers and  {ranging from 8 to} $>100$ kpc in projected separation \cite{wengMUSEALMAHaloesVIII2023}. 
From galaxy morphologies, position angles and inclinations were used to calculate azimuthal angles between galaxies and background quasars \citep{hamanowiczMUSEALMAHaloesPhysical2020, perouxMUSEALMAHaloes2022, wengMUSEALMAHaloesXI2023}.

\subsection{Ancillary optical observations}

\subsubsection{VLT observations} \label{subsec:muse_data}

Observations from the Very Large Telescope (VLT) Multi-Unit Spectroscopic Explorer (MUSE) in 19 quasar fields 
allowed to identify 79 galaxies associated with strong H\,\textsc{i} absorbers at intermediate redshifts ($z \sim 0.5$) \citep{wengMUSEALMAHaloesVIII2022} (PIs: Péroux; ESO programs 96.A-0303, 100.A-0753, 101.A-0660, 102.A-0370, and Klitsch 298.A-0517).

Spatially resolved MUSE spectroscopy provided rest-frame optical emission lines ([O\,\textsc{ii}], H$\beta$, [O\,\textsc{iii}], H$\alpha$, [N\,\textsc{ii}]), yielding spectroscopic redshifts, star formation rates, and gas-phase metallicities via strong-line diagnostics calibrated by \citet{curtiNewFullyEmpirical2017} and \citet{maiolinoAMAZEEvolutionMassmetallicity2008}.

Complementary new high-resolution quasar spectroscopy from VLT/UVES (PI: Péroux, ID: 113.A-0369), along with archival Keck/HIRES and VLT/X-Shooter data, provides detailed measurements of metal absorption lines (e.g., Fe\,\textsc{ii}, Si\,\textsc{ii}, Zn\,\textsc{ii}, C\,\textsc{iv}). Column density estimates offer dust and ionisation corrections, and CGM metallicity measurements in neutral gas absorption from Voigt profile fitting (\citealp{wengMUSEALMAHaloesXI2023, karkiMUSEALMAHaloesIX2023, augustinMUSEALMAHaloesStellar2024}; Halley et al., in prep).

\subsubsection{HST observations}\label{subsec:hst_data}

The HST ancillary data comes primarily from broad-band imaging observed with the Wide Field Camera 3 (WFC3) during Cycle 27 under GO Program ID 15939 (PI: Péroux) using both optical (UVIS) and infrared (IR) detectors, as well as archival data from the Wide Field and Planetary Camera-2 (WFPC2) and WFC3 from programs 5098, 5143, 5351, 6557, 7329, 7451, 9173, and 14 594 (PIs: Burbidge, Macchetto, Bergeron, Steidel, Malkan, Smette, Bechtold, Bielby).
For a comprehensive description of the observational setup, data reduction, PSF subtraction, photometric measurements, and the master table of all targets, see \cite{perouxMUSEALMAHaloes2022}.

The high spatial resolution of HST imaging (resolution $\sim$0.04'') provided a crucial complement to MUSE observations, enabling the identification of structural features such as tidal tails and galaxy morphologies \citep{karkiMUSEALMAHaloesIX2023}.

\cite{augustinMUSEALMAHaloesStellar2024} performed Spectral Energy Distribution (SED) fitting on the multi-band HST photometry using the \textsc{Le Phare} code \citep{arnoutsMeasuringModellingRedshift1999, ilbertAccuratePhotometricRedshifts2006}. 
The fitting assumed \cite{bruzualStellarPopulationSynthesis2003} SED templates, a \cite{calzettiDustExtinctionStellar1994} dust extinction law, and a \cite{chabrierGalacticStellarSubstellar2003} initial mass function. 
The resulting stellar mass measurements span a wide range from $\log(M_{\star} / M_{\odot}) = 7.8 - 12.4$, highlighting the diversity of galaxies associated with H\,\textsc{i} absorbers. 
The most robust mass determinations (34 galaxies) had at least two detections in different HST filters, with higher-mass galaxies ($\log(M_{\star} / M_{\odot})>10$) typically having more reliable measurements due to their stronger detections across multiple bands.

\subsection{ALMA observations} \label{sec:alma_data}

\subsubsection{Previous ALMA data} \label{subsec:ancillary_alma_data}
In addition to the rich optical and UV dataset, we compile sub-millimetre observations from archival data.
\citet{kanekarMassiveAbsorptionselectedGalaxies2018} observed one absorber associated with one galaxy included in the MUSE-ALMA Haloes survey in the quasar field J0138-0005. 
The ALMA observations in Band 4 (program IDs: 2013.1.01178.S, 2015.1.00561.S, PIs: J. Prochaska, N. Kanekar) targeted the CO(2--1) line. 
They performed data reduction in the Common Astronomy Software Applications (CASA, \citealt{mcmullinCASAArchitectureApplications2007}), using \texttt{tclean} with a natural weighting, producing cubes with a spectral resolution of 100 km s$^{-1}$ and a rms noise of $\sim 0.20$ mJy beam$^{-1}$ in the data cubes.
They placed an upper limit on the CO luminosity for the galaxy associated with the absorber.

\citet{klitschALMACALIIICombined2018} used ALMA calibrator data from the ALMACAL survey \citep{zwaanALMACALSurveyingUniverse2022, bolloALMACALXIIData2024} for quasar J0423-0130, linked to four galaxies from the MUSE-ALMA Haloes survey.
They analysed CO(2-1) and CO(3-2) lines from Band 4 and Band 6 data (1333 s and 605 s total), reduced with CASA (Briggs weighting), achieving velocity resolutions of 33 and 22 km s$^{-1}$, and the rms noise levels in the data cubes are $\sim$0.28 and $\sim$0.37 mJy beam$^{-1}$, respectively.
One detection was reported, with two upper limits and one source outside the ALMA field of view.

Four quasar fields (Q1232-0224, Q0152-2001, Q1211+1030 and Q1130-1449) were observed with ALMA to cover CO(2--1) or CO(3--2) under the programs 2016.1.01250.S and 2017.1.00571.S (PI: C. Péroux) and 2018.1.01575.S (PI: A. Klitsch), presented in \cite{perouxMultiphaseCircumgalacticMedium2019} and \cite{szakacsMUSEALMAHaloesVI2021}.
The programs targeted 17 galaxies included in the MUSE-ALMA Haloes survey, which had been identified by MUSE and associated with absorbers at $z\sim 0.4$ with impact parameters ranging from 8 to 82 kpc. 
The primary spectral window was centred on the redshifted CO(3$-$2) frequency (345.796 GHz) in high spectral resolution mode, complemented by three additional spectral windows in low resolution mode (31.250 MHz). 
They performed data reduction in CASA, using \texttt{tclean} with a Briggs weighting scheme, reaching a spectral resolution of 50 km s$^{-1}$ per channel and an RMS sensitivity of $0.28$ mJy beam$^{-1}$. They reported two detections out of the 17 systems that are part of the MUSE-ALMA Haloes survey.

In total, we compiled ALMA archival data for 21 galaxies from the MUSE–ALMA Haloes Survey, all of which were observed in the CO(2$-$1) or CO(3$-$2) transitions.  
By design of the MUSE-ALMA Haloes survey, they have MUSE and HST data, as mentioned earlier in the text.

\subsubsection{MUSE-ALMA Haloes ALMA Large Program} \label{subsec:new_alma_data}

The ALMA data presented in this paper were obtained from the ALMA Large Program (2023.1.00127.L, Cycle 10, PI: C. Péroux). 
The observations were conducted between December 2023 and March 2024 using ALMA configurations C43-2, C43-3, and C43-4, which provided angular resolutions ranging from $0.6^{\prime\prime} $ to $ 1.1^{\prime\prime}$ at the observed frequencies.

This program targeted CO emission lines (specifically CO(2$-$1), CO(3$-$2), CO(4$-$3) transitions) using Band 4 ($125-163$ GHz) and Band 6 ($211-275$ GHz), reaching a rms of 0.16 mJy beam$^{-1}$ over 50 km\,s$^{-1}$. Our sample consists of 39 H\,\textsc{i}-rich galaxies associated with 17 quasar absorbers with known H\,\textsc{i} column densities ($\log$ $N$(H\,\textsc{i}) $> 10^{18}$ cm$^{-2}$) at redshifts $0.3 \lesssim z \lesssim 1.2 $ \citep{perouxMUSEALMAHaloes2022}. {This sample is drawn from the full parent sample of 79 galaxies identified in the MUSE data \citep{wengMUSEALMAHaloesVIII2022}. Of these 79 galaxies, 60 have been observed with ALMA to date: 39 in the present Large Program and 21 from pilot studies and archival data. The remaining 19 galaxies that have not yet been observed were not excluded for scientific reasons; rather, the Large Program strategy was designed to optimise telescope time by prioritising galaxies based on redshift and by targeting multiple absorbers within the same pointing.}

The data were calibrated and imaged using the standard ALMA pipeline in CASA (version 6.5.6), with phase and/or amplitude self-calibration applied to five fields where the background quasar was bright enough ($> 5$ mJy) to achieve a signal-to-noise ratio of five.
{The time interval used in \texttt{gaincal} varied from target to target, depending on the quasar flux density: in the most favourable cases, we used 30 s for phase-only self-calibration and 60 s for combined amplitude-and-phase calibration, while for the faintest sources, the solution interval corresponds to the full scan length.}
The images were built using \texttt{tclean} with pixel sizes adjusted according to the synthesised beam and a natural weighting scheme. 
Following these calibrations, we performed continuum subtraction only in one field (Q2131-1207), to mitigate low-level artefacts associated with the presence of a bright quasar in the field, using \texttt{uvcontsub} with a polynomial order of 2. Although higher than the commonly adopted zeroth- or first-order fits, the resulting continuum model is effectively linear across the ALMA spectral window. The difference between the first- and second-order continuum models is at the $\sim 10^{-7}$ Jy level in the image domain, far below the noise level of the data, indicating that the quadratic term does not introduce any artificial curvature or spurious structure. Tests on all fields confirm that the effect of the quadratic term is negligible, so no bias is introduced in the extracted spectra.
The final data cubes were produced and corrected for primary beam effects using \texttt{impbcor}.
Final data cubes were created with $\sim60$ km/s spectral binning for optimal line sensitivity.
Full details of the observations, calibration procedures, and data quality assessments are presented in a separate paper (Péroux et al., in prep.).
All subsequent analyses are based on data from the complete MUSE-ALMA Halos survey.

\section{Analysis and Results} \label{sec:analysis}

\begin{figure*}
    \centering
    \includegraphics[width=0.33\linewidth]{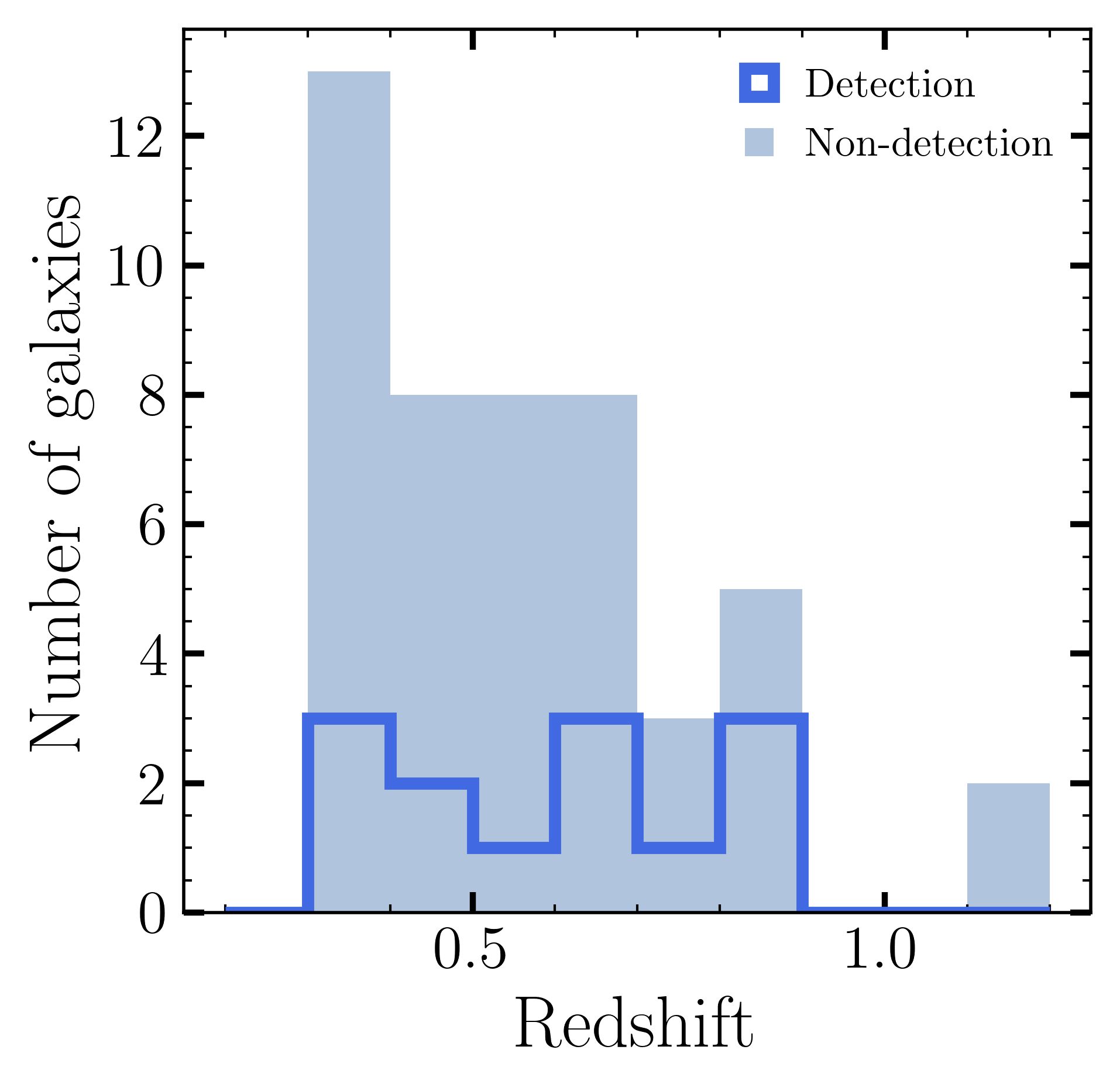}
    \includegraphics[width=0.32\linewidth]{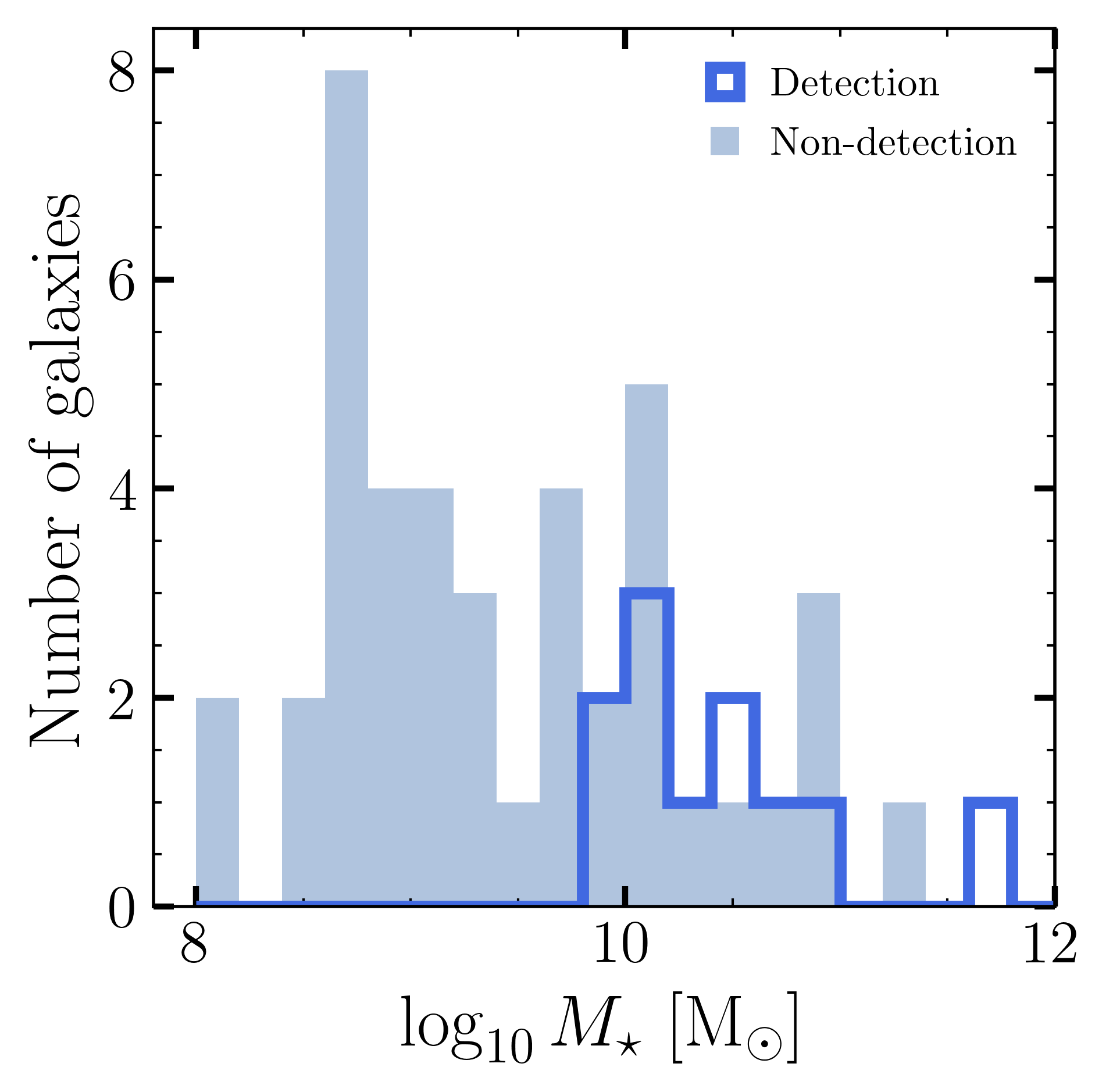}
    \includegraphics[width=0.32\linewidth]{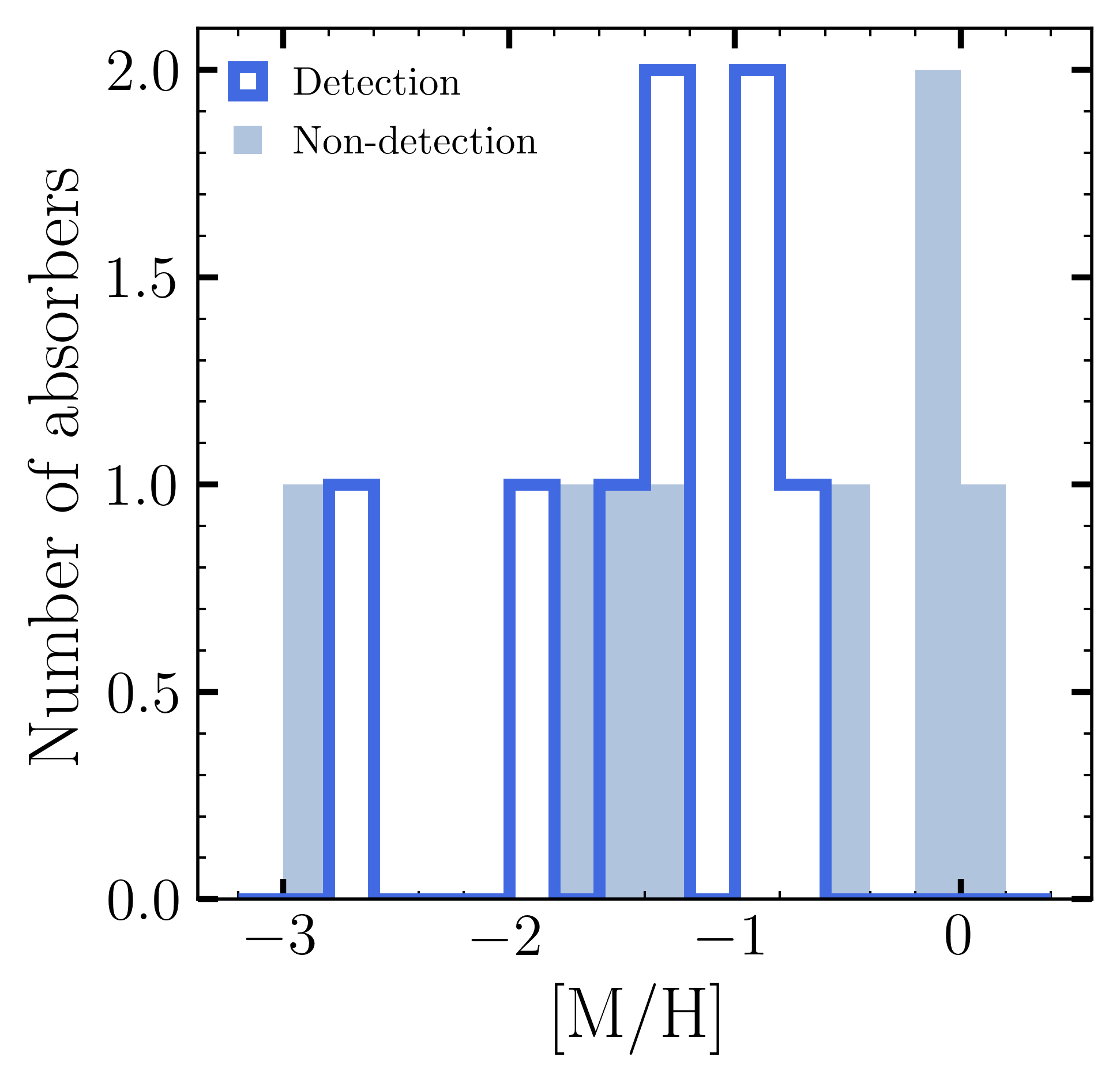}
    \caption{{Distribution of CO-detections (blue) and non-CO-detections (light blue) on galaxies as a function of redshift (left) and stellar mass (middle). The right panel is plotted as a function of metallicity corresponding to multiple galaxies associated with the absorbers (Halley et al., in prep)}. This figure represents the MUSE-ALMA Haloes survey, including all the sources from the new ALMA Large Program and previous ALMA data (see Section \ref{sec:alma_data}). 
    The middle and right panels have fewer numbers because only those with reliable $M_{\star}$ and [M/H] determinations are included.
    The median values for the detections are 0.6, 9.66 and $-1.23$ for the redshift, stellar mass and absorption metallicity, while the non-detections have median values of {0.5, 9.16, and {$-1.84$}}.
    % We report a CO detection rate of 24\% for the MUSE-ALMA Haloes survey.
    Overall, we report no clear correlation between the detection rate and any of these properties, suggesting that the detection of CO emission may be governed by a combination of global galaxy properties, rather than just one property alone. 
    }
    \label{fig:co_detection_properties}
\end{figure*}

\subsection{Emission line search identification} \label{subsec:emission_line}

We systematically searched for CO emission lines within the calibrated and cleaned ALMA data cubes. The search used two complementary approaches to maximise completeness while maintaining reliability.

First, we conducted a targeted search at the expected positions of 39 H\,\textsc{i}–selected galaxies previously identified in \cite{wengMUSEALMAHaloesVIII2022}. 
For each galaxy, we applied a systematic aperture optimisation procedure to maximise the likelihood of detecting CO emission.
We used elliptical apertures with semi-major and semi-minor axes ranging from 1 to 4 times the synthesised beam size. For each aperture size, we rotated the ellipse from $0^\circ$ to $90^\circ$ in steps of $5^\circ$, systematically exploring different orientations. 
At each combination of size and angle, we extracted spectra and computed the signal-to-noise ratio (SNR), estimated from the integrated line flux over the beam size, to identify the configuration that maximised the signal significance. 
This method allowed us to capture the full extent of potentially resolved CO emissions.
We defined emission line candidates where the signal exceeded $3\sigma$ {in at least one channel ($\sim 60$ km s$^{-1}$)}, where $\sigma$ represents the RMS noise level measured in {regions of the cube without emission}, which is the minimum expected line width based on detection for the previous ALMA data (\S\ref{sec:alma_data}).
{The aperture optimisation procedure increases the effective parameter space explored and can therefore detect spurious $\geq 3\sigma$ features (known as the look-elsewhere effect). To quantify this, we implemented a test in which the same aperture optimisation procedure was applied to 500 random positions within emission-free regions of the cubes. From this analysis, we determined the frequency of spurious $\geq 3\sigma$ detections and used it to assess the fidelity of our detection threshold. Based on this calculation, we report a detection threshold corresponding to a fidelity of $>90\%$.}
The aperture optimisation technique allowed us to successfully detect CO emission lines in nine sources from our targeted sample.

In parallel, we performed a blind search across all the fields using the Source Finding Application \textsc{sofia-2} \citep{serraSOFIAFlexibleSource2015, westmeierSOFIA2Automated2021a} that uses a number of different source detection algorithms to find emission lines in radio data cubes. 
In this process, we used the data cubes, not corrected from the primary beam, to model the noise level homogeneously across the field of view.
We use the \texttt{Smooth + Clip Finder}, which smooths the data in both spatial and spectral directions using several 3D smoothing kernels. 
Our setup consisted of a detection threshold of $3\sigma$, a minimum requirement of emission spanning one channel, and a reliability estimated by \textsc{sofia-2} of 0.9, which compares the distribution of positive and negative sources in parameter space, defined by signal-to-noise ratio and line width.
{This blind search recovered six sources coincident with our primary targets, thereby providing an independent confirmation of the majority of our targeted detections. The three targeted sources not identified by \textsc{sofia-2} correspond to lower SNR ($<5$) lines located near the edges of the primary beam, where the sensitivity drops and the reliability of blind searches decreases. We stress that the aim here is not to derive a blind detection rate, which requires a careful assessment of the false-positive rate and completeness, but rather to provide an independent check on our targeted search results.
Finally, this exercise demonstrates the potential of the dataset for a fully blind search. In addition to confirming targeted detections, the algorithm also identified several candidate CO emitters without optical counterparts. A full analysis of these serendipitous detections, including a detailed fidelity assessment of the blind search, will be presented in future work.}

In total, we detected CO emission from nine galaxies at a significance of at least $3\sigma$ out of our targeted sample of 39.
For the remaining 30 sources, we calculated $3\sigma$ upper limits based on the sensitivity reached in the data cube and assuming a line width of 200 km/s, typical for galaxies in our sample. These are stringent limits due to the depth of our survey.

\begin{figure*}
    \centering
    \includegraphics[width=0.49\linewidth]{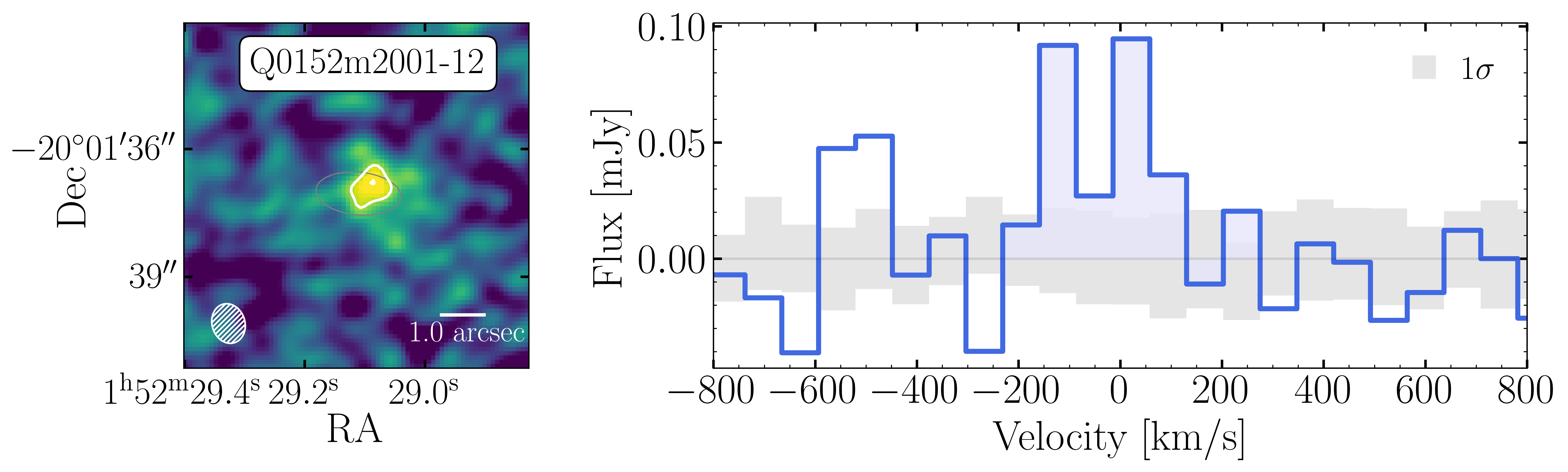}
    \includegraphics[width=0.49\linewidth]{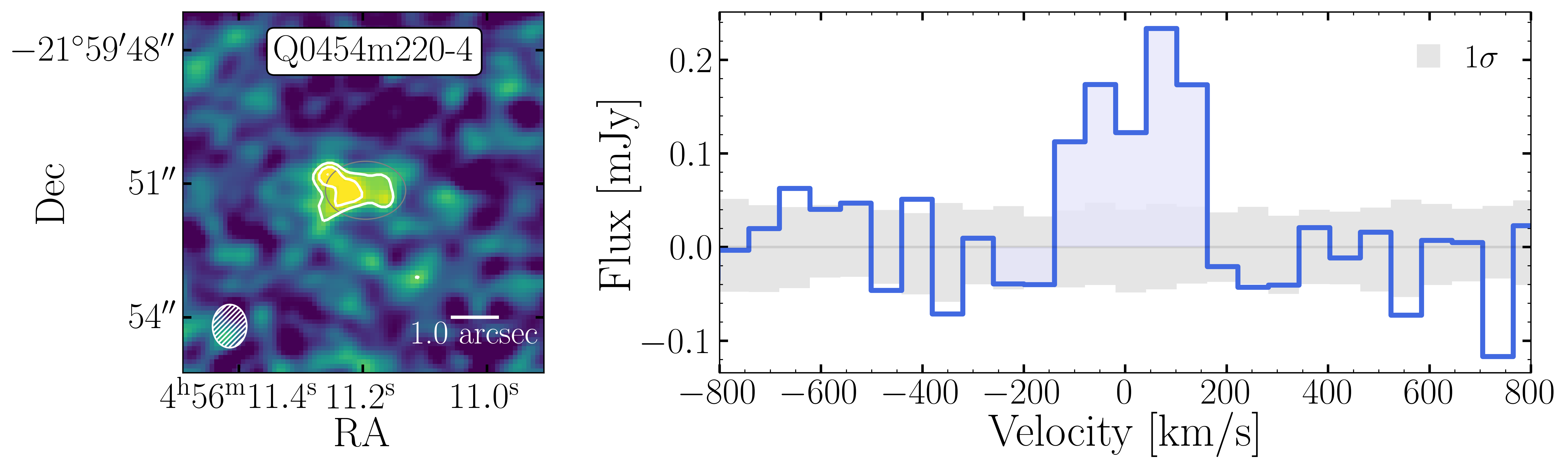}
    \includegraphics[width=0.49\linewidth]{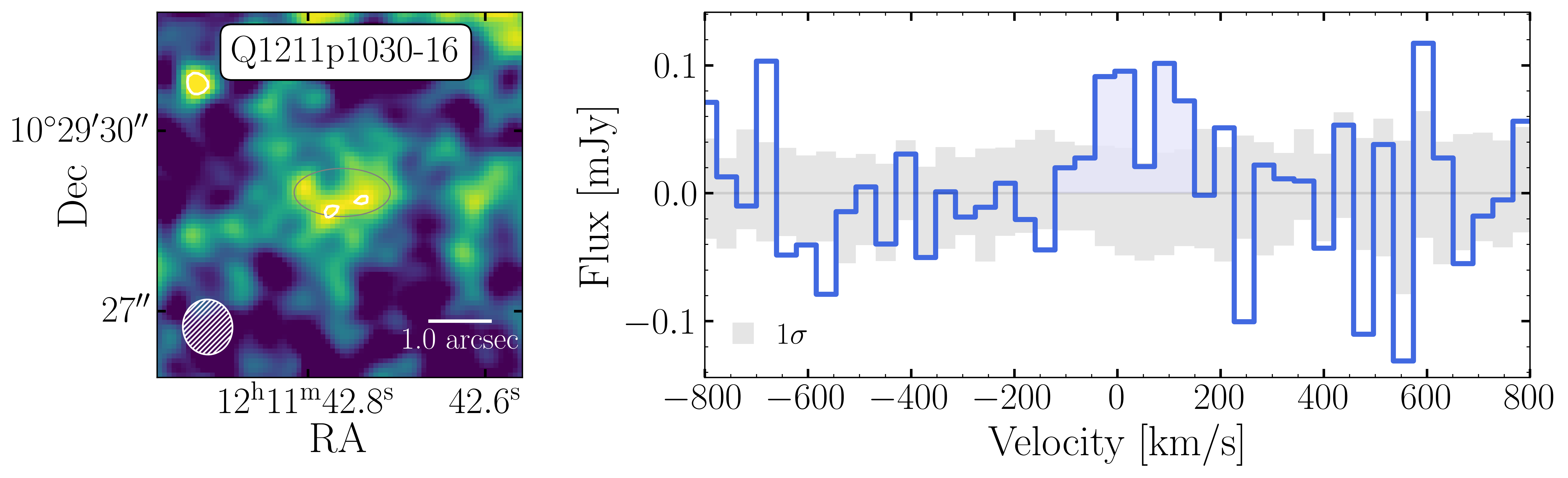}
    \includegraphics[width=0.49\linewidth]{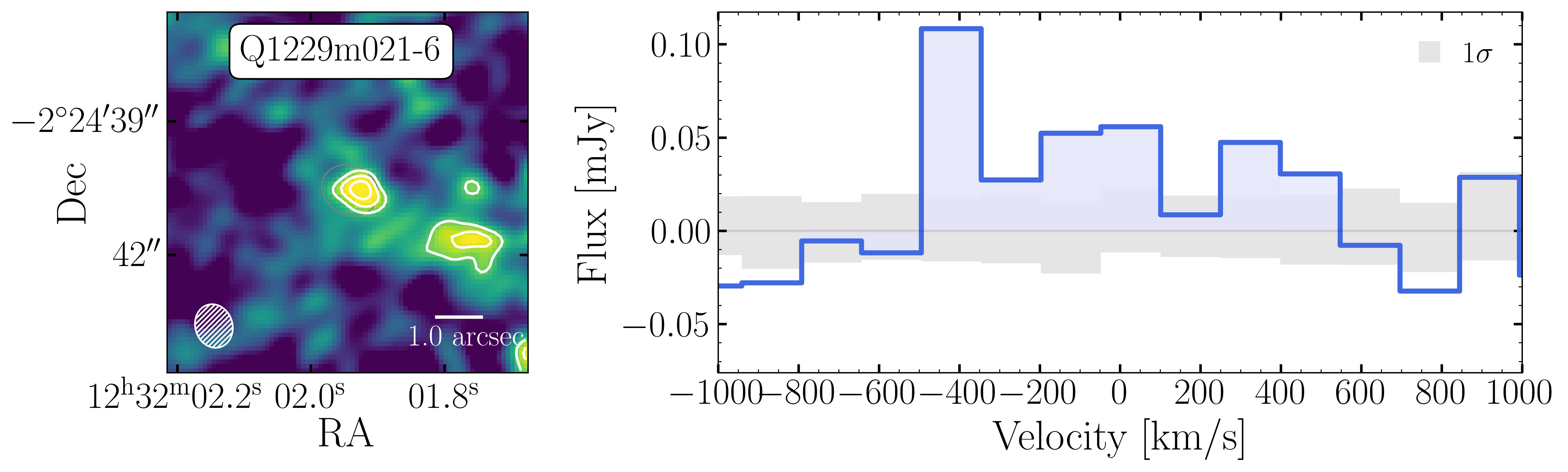}
    \includegraphics[width=0.49\linewidth]{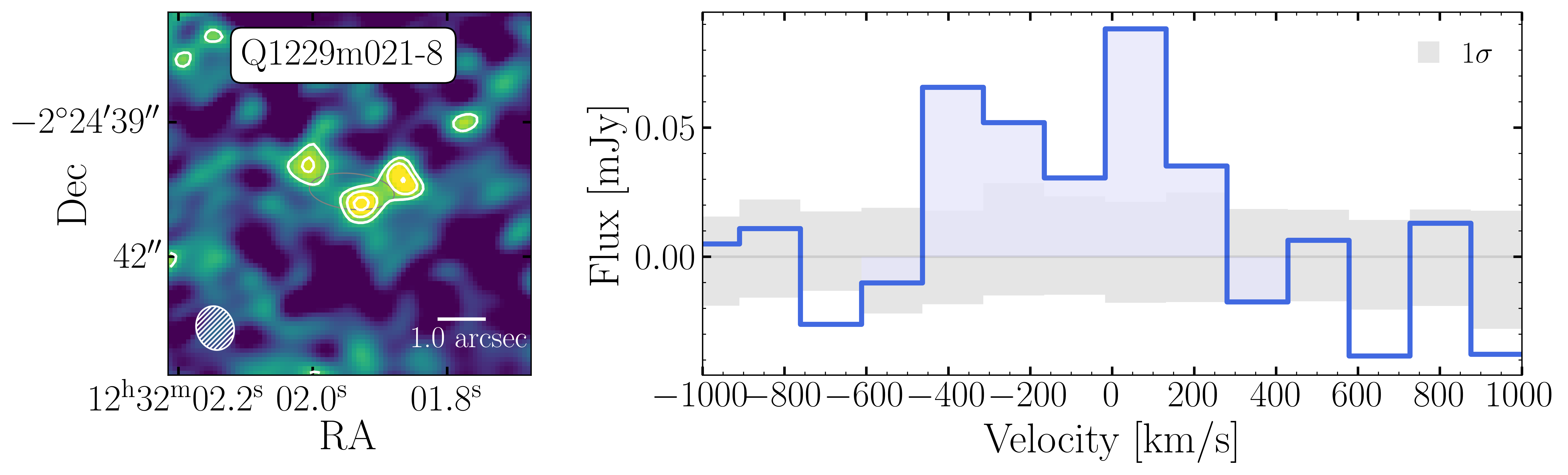}
    \includegraphics[width=0.49\linewidth]{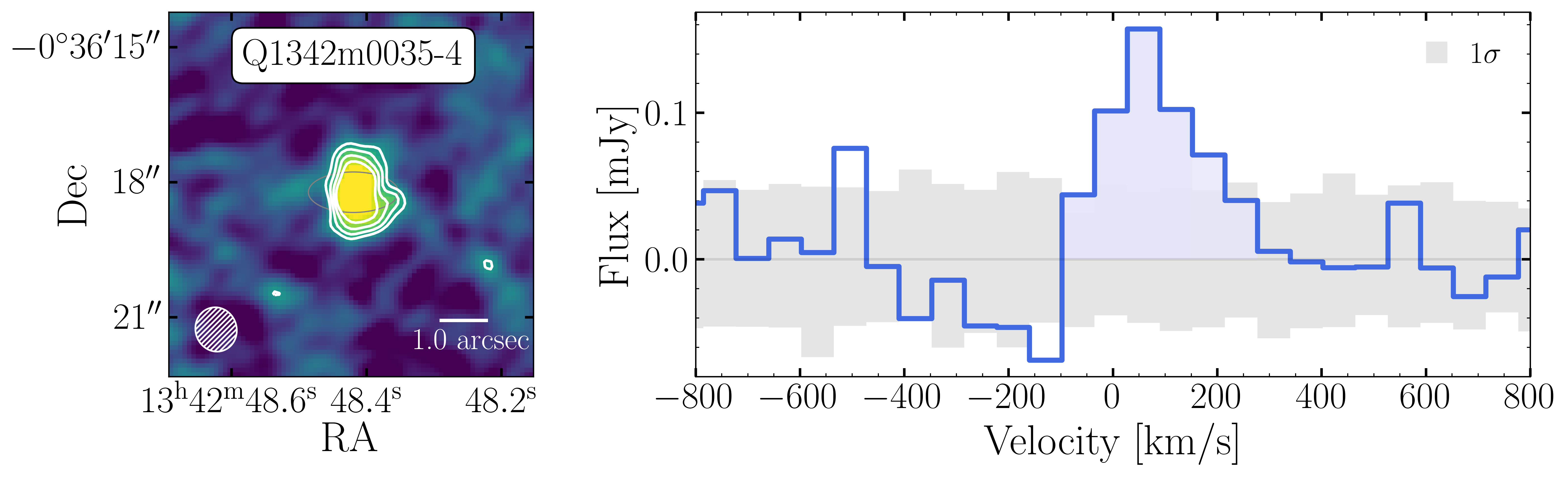}
    \includegraphics[width=0.49\linewidth]{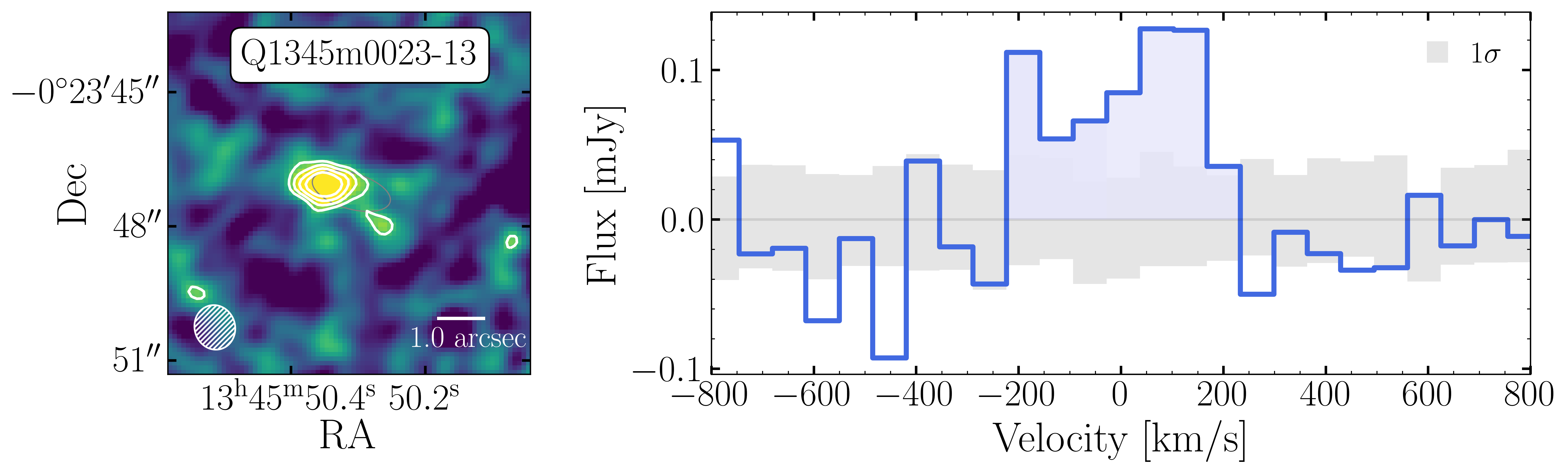}
    \includegraphics[width=0.49\linewidth]{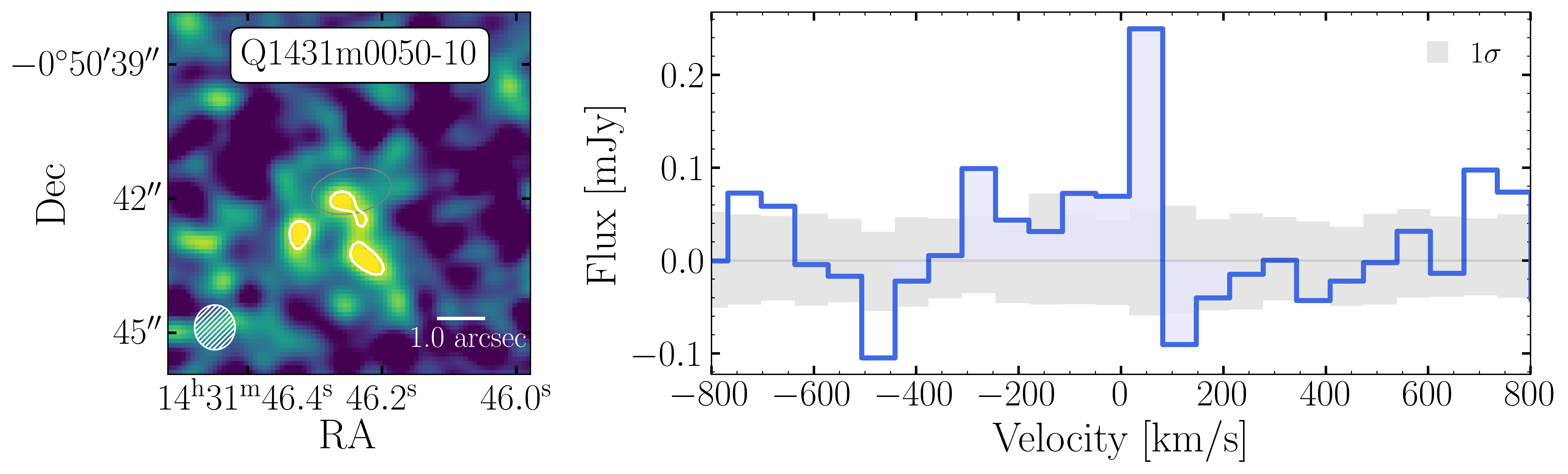}
    \includegraphics[width=0.49\linewidth]{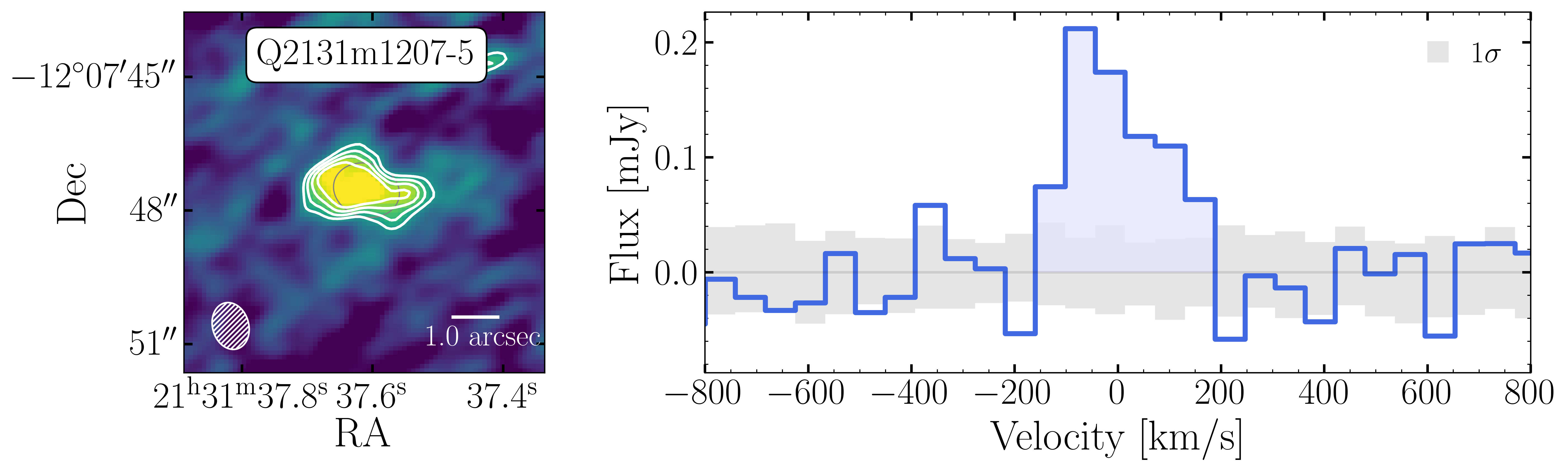}
    \includegraphics[width=0.49\linewidth]{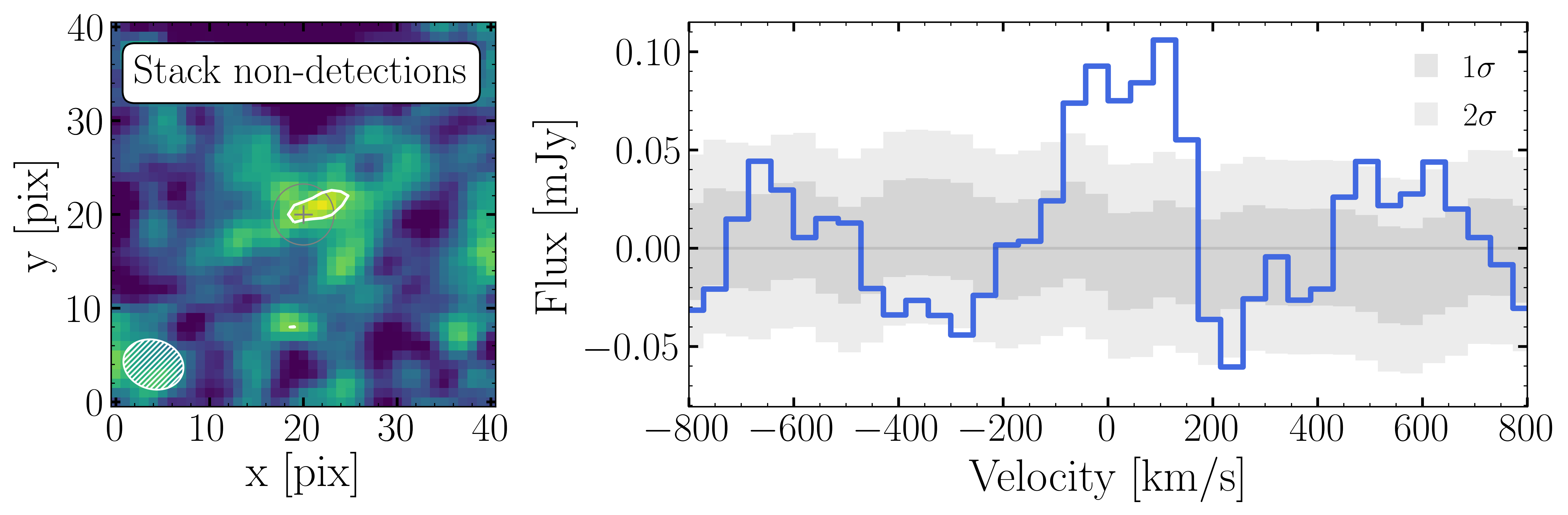}
    \caption{CO detections and spectral profiles of the new nine detected sources. Left panels show the moment zero maps with the $\pm$3, $\pm$4, and $\pm$5$\sigma$ level contours, while the right panels display the extracted CO spectra. The grey shaded regions around each spectrum represent the $\pm1\sigma$ RMS noise level estimate from the individual data cubes, and the blue shaded region indicates the velocity range used to create the moment maps. The final row, right figure, presents the rest-frame stacked spectrum of non-detections with $\pm1\sigma$ and $\pm2\sigma$ levels. A summary of the integrated CO fluxes and associated uncertainties for all detected sources is provided in Table \ref{table:co_prop}.}
    \label{fig:emission_lines}
\end{figure*}

When combined with previous CO observations from the MUSE-ALMA Haloes survey \citep{kanekarMassiveAbsorptionselectedGalaxies2018, klitschALMACALIIICombined2018, szakacsMUSEALMAHaloesVI2021}, which added three detections and eighteen upper limits, our comprehensive analysis provides a total of 12 CO detections out of 60 galaxies at $0.3 < z < 1.2$, as shown in the left panel of Figure \ref{fig:co_detection_properties}. 
Before the MUSE-ALMA Haloes survey, only a handful of H\,\textsc{i} at $z\sim0.5$ had been targeted in CO \citep[e.g.,][]{kanekarMassiveAbsorptionselectedGalaxies2018}. Our observations increased the number of targeted galaxies by a factor of 5 and nearly doubled the number of CO detections ever previously known.
Figure \ref{fig:emission_lines} shows the resulting moment map and extracted spectrum for each detected source.

\begin{figure}
    \centering
    \includegraphics[width=0.9\linewidth]{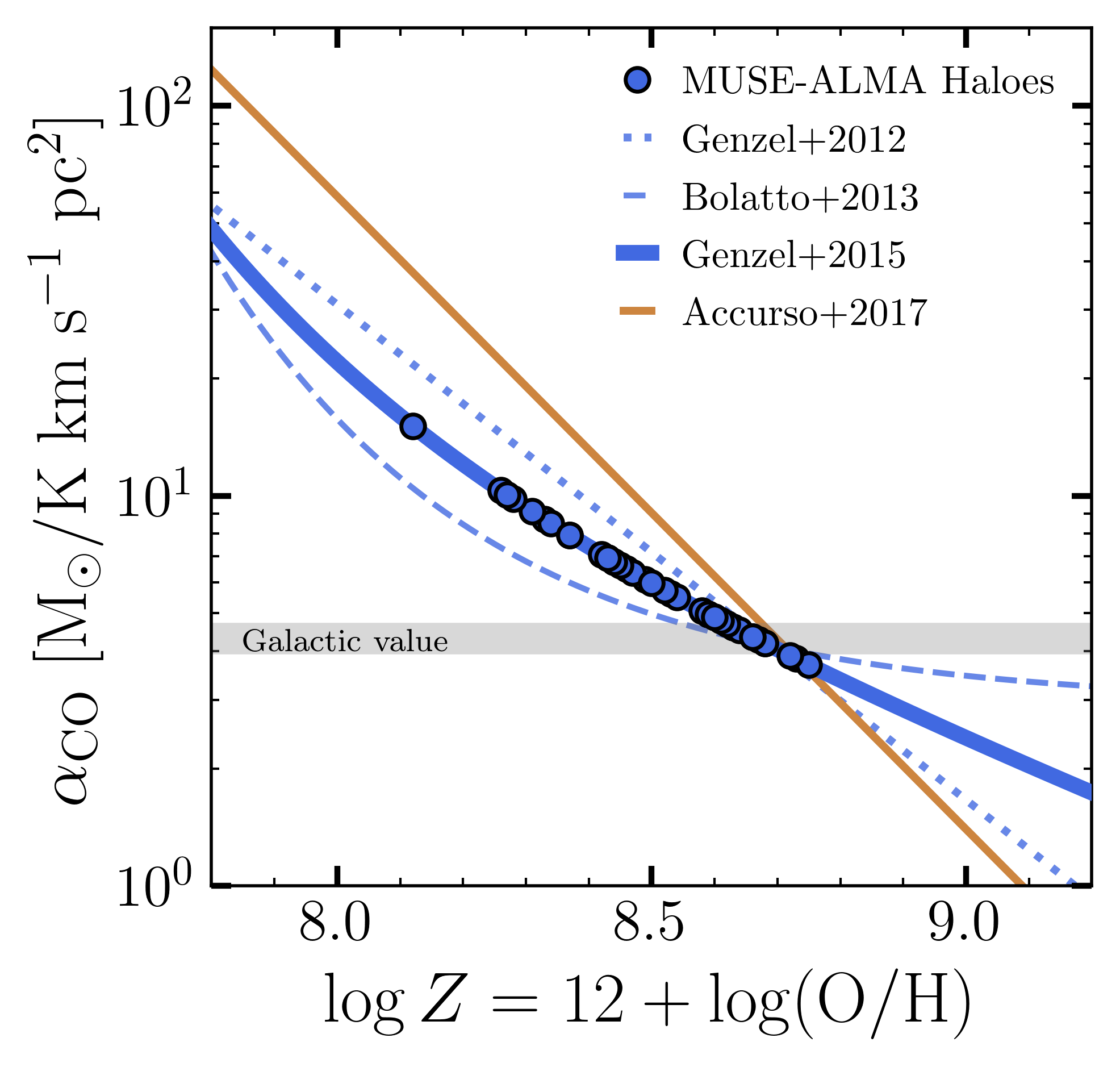}
    \caption{Different $\alpha_{\text{CO}}$ models across varying metallicites. In this work, we adopt the model proposed by \cite{genzelCombinedCODust2015}, shown in the blue points, which corresponds to the geometric mean of the prescriptions of \cite{bolattoCOtoH2ConversionFactor2013} and \cite{genzelMETALLICITYDEPENDENCECO2012}, as indicated in Eq. \ref{eq:alpha_co}. 
    As a reference, we also include the model proposed by \cite{accursoDerivingMultivariateACO2017}, which would lead to higher values for $\alpha_{\text{CO}}$ for the galaxies in our sample.
    To prevent extrapolation into poorly constrained regions of parameter space, we impose an upper limit on $\alpha_{\text{CO}}$, set to the lowest emission metallicity measured in our sample, which leads to an $\alpha_{\text{CO}} \sim 15$.}
    \label{fig:alpha_co}
\end{figure}
\subsection{Non-detections stacking}\label{subsec:stack}

{We performed a spectral stacking analysis for the sources without individual CO detections. To ensure consistency in the transition being stacked, we restricted the analysis to the CO(2–1) line. As a result, we stacked 27 sub-cubes corresponding to the non-detected sources.}
For each galaxy, we extracted a subcube centred on the expected position of the galaxy, using a spatial size of 50 kpc at its redshift.
The spectral axis spanned 1000 km s$^{-1}$, centred on the expected frequency of the CO line. 
All subcubes were then rebinned {and smoothed to the coarsest} spatial and spectral resolution.

The cubes were stacked, averaging each subcube weighted by the square of the luminosity distance to account for flux dimming with redshift. 
This weighting helps recover a quantity closer to the intrinsic luminosity by correcting for the cosmological decrease in observed flux with distance. 
We also applied a weight given by the sensitivity of each data cube.
We examined the resulting stacked cube for potential CO emission at the central position.

To extract the spectrum, we used {a circular aperture}, and the same aperture was used to estimate the significance of the emission line in the stacked cube. 
To quantify the noise, we extracted spectra from 100 random apertures placed across the stacked cube, excluding the central region.
These apertures matched the size determined during optimisation. 
The per-channel error was then computed as the standard deviation across all 100 spectra. The resulting moment map and spectrum of the stacked cube are presented in the bottom right panel of Figure \ref{fig:emission_lines}. The spectrum exceeds $3\sigma$ in two consecutive channels. We report a stacked CO emission line detected at a significance level of $\sim4.2\,\sigma$.

\subsection{CO Luminosity}

We calculated the CO luminosity from CO fluxes using the following equation, from \citet{solomonMolecularInterstellarMedium1997},

\begin{equation}
    L'_{\text{CO}} = 3.25 \times 10^7 \frac{F_{\text{CO}}}{(1+z)^3} \left (\frac{D_{\text{L}}}{\nu_{\text{obs}}}\right )^2 \hspace{3px} \text{[K km s}^{-1} \text{pc}^2\text{]} \hspace{3px},
\end{equation}
\noindent
where $L'_{\text{CO}}$ is in units of K km s$^{-1}$ pc$^2$, $\nu_{\text{obs}}$ is the observed frequency of the CO line in GHz, $D_{\text{L}}$ is the luminosity distance of the galaxy in Mpc, $z$ is the redshift, and $F_{\text{CO}}$ is the integrated flux in Jy km s$^{-1}$.
We converted the luminosity measured for mid- and high$-J$ CO into CO(1$-$0), scaling by the empirical conversion factors from \cite{fixsenCOBEFarInfrared1999, weissSpectralEnergyDistribution2005}: $r_{J\rightarrow 1} =$\{$2.5, 3.5, 2.5$\}, for $J = 2, 3, 4$. These values were also assumed in previous ALMA observations of the MUSE-ALMA Haloes survey published by \citet{kanekarMassiveAbsorptionselectedGalaxies2018, szakacsMUSEALMAHaloesVI2021}, and we scaled the values from \citet{klitschALMACALIIICombined2018} accordingly.
Table \ref{table:co_prop} shows the measured CO fluxes and full width at half maximum (FWHM) for all the detected sources. The FWHM was calculated from a Gaussian fit to the line profile, using the relation FWHM = $2.355\times \sigma$, where $\sigma$ is the standard deviation obtained from the fit.

\subsection{Molecular gas mass}

We derive the molecular gas mass of our sources using the CO(1--0) luminosity and the empirical conversion factor, $\alpha_{\text{CO}}$, following \citet{tacconiHighMolecularGas2010}:
\begin{equation} \label{eq:omega}
    M_{\text{mol\,gas}}/M_{\odot} = \alpha_{\text{CO}} \cdot L^{\prime}_{\text{CO}(1-0)} \hspace{6px} \hspace{3px},
\end{equation}
\noindent

The conversion factor $\alpha_{\text{CO}}$ increases with decreasing metallicity $Z$, since lower metallicity environments allow ultraviolet radiation to penetrate deeper into molecular clouds, leading to more extensive photodissociation of CO \citep{wolfireDarkMolecularGas2010, bolattoCOtoH2ConversionFactor2013}. From the different metallicity corrections proposed in the literature, we adopt the geometric mean of the recipes by \citet{bolattoCOtoH2ConversionFactor2013}
and \citet{genzelMETALLICITYDEPENDENCECO2012} as adopted by \citet{genzelCombinedCODust2015} and \citet{tacconiPHIBSSUnifiedScaling2018}:

\begin{equation} \label{eq:alpha_co}
\begin{aligned}
\alpha_{\mathrm{CO}} =\; & 4.36 \times 
\left[
0.67 \times \exp\left(0.36 \times 10^{-(12 + \log(\mathrm{O}/\mathrm{H}) - 8.67)}\right) \right. \\
& \left. \times\; 10^{-1.27 \times (12 + \log(\mathrm{O}/\mathrm{H}) - 8.67)}
\right]^{1/2} \; [M_\odot/(\mathrm{K\,km\,s^{-1}\,pc}^2)].
\end{aligned}
\end{equation}
where $12 + \log(\mathrm{O}/\mathrm{H})$ is the metallicity on the \citet{pettiniOIIINIIAbundance2004} scale. Figure \ref{fig:alpha_co} shows the comparison of the different prescriptions proposed by \cite{genzelMETALLICITYDEPENDENCECO2012, bolattoCOtoH2ConversionFactor2013, accursoDerivingMultivariateACO2017} and the adopted model for our sources from \cite{genzelCombinedCODust2015}.

The metallicities for our targets were derived by \citet{wengMUSEALMAHaloesVIII2023} using the R$_3$ strong-line calibration \citep{curtiNewFullyEmpirical2017, curtiMassMetallicityFundamental2020}, defined as \( \mathrm{R_3} = \log(\mathrm{[O\,\textsc{iii}]}\,\lambda5007 / \mathrm{H}\beta) \). This method is applicable up to redshift \( z \sim 0.85 \) and is largely unaffected by dust obscuration. For galaxies at \( z < 0.4 \), the {O3N2} diagnostic \citep{pettiniOIIINIIAbundance2004}, given by \( \log(\mathrm{[O\,\textsc{iii}]}\,\lambda5007 / \mathrm{H}\beta) - \log(\mathrm{[N\,\textsc{ii}]}\,\lambda6584 / \mathrm{H}\alpha) \), was also measured to cross-check R$_3$-based metallicities and, in some cases, resolve the degeneracy in the R$_3$ indicator. Errors were estimated by propagating flux uncertainties and using Monte Carlo sampling to solve the metallicity calibration polynomial. 

Out of the 39 galaxies included in the new ALMA observations, we have direct metallicity measurements for 24 from the MUSE data based on the optical emission lines to get the $R_3$ auroral strong line calibration. 
For the remaining 15, stellar mass estimates are available for 13 galaxies, allowing us to infer their metallicities using the mass-metallicity relation \citep{pettiniOIIINIIAbundance2004}.
To maintain consistency with previous studies \citep[e.g.,][]{saintongeXCOLDGASSComplete2017, tacconiPHIBSSUnifiedScaling2018}, we use the following mass–metallicity relation:
\begin{equation}
    \log Z = 8.74 - 0.087 \times (\log(M_*) - b)^2,
\end{equation}
with $b = 10.4 \,+ \,4.46 \,\times \,\log(1 + z) \,- \,1.78 \,\times \,(\log(1 + z))^2$ \citep{genzelCombinedCODust2015, erbHaObservationsLarge2006, maiolinoAMAZEEvolutionMassmetallicity2008, zahidFMOSCOSMOSSURVEYSTARFORMING2014, wuytsCONSISTENTSTUDYMETALLICITY2014}.
For the two galaxies without reliable stellar mass estimates (IDs: Q0454p039\_15, Q0454p039\_57), we investigated the metallicity estimated in absorption along the line of sight of a quasar background. The metal content of the absorbing gas is very poor ([M/H] $\lesssim -1.0$), so we take a lower limit in the metallicity set to the lowest observed value in our sample. This approach is more conservative than using an extrapolation of the prescriptions mentioned before, as the $\alpha_{\text{CO}}$ quickly increases toward low metallicities. Using the lowest observed metallicity, we adopt an $\alpha_{\text{CO}} \sim 15$ for the metal-poor regime. 
For sources from the literature, we reanalysed the $M_{\mathrm{H}_2}$ estimates using an $\alpha_{\mathrm{CO}}$ value consistent with our methodology.

\section{Properties of H\,\textsc{i}–selected galaxies} \label{sec:results}

The selection of H\,\textsc{i} absorbers through absorption features in a background quasar spectrum allows the detection of diffuse, low-surface brightness neutral gas that often eludes direct emission observations due to the low density of gas \citep{wolfeDAMPEDLYaSYSTEMS2005, prochaskaCOSHalosSurveyMetallicities2017}. 
This method offers redshift-independent sensitivity, {and in the case of DLAs} yields accurate measurements of metallicity that are largely insensitive to temperature or ionisation conditions, unlike emission diagnostics \citep{pettiniMetalAbundances151999, kulkarniHubbleSpaceTelescope2005, lehnerBIMODALMETALLICITYDISTRIBUTION2013}.
For sub-DLAs and systems with $\log N$(H\,\textsc{i}) < 19, {we correct for dust depletion and ionisation effects following the approach of, e.g., \cite{ quiretESOUVESAdvanced2016}, which ensures that the derived metallicities are reliable. A more detailed description of these corrections will be presented in a forthcoming paper (Halley et al., in prep.).}
In this section, we revise the properties of absorption-selected galaxies and relate them to their cold gas content.

\subsection{Molecular gas detection rate}

Despite the high neutral hydrogen column densities of the targeted absorbers in the MUSE-ALMA Haloes survey ($\log$ [$N$(H\,\textsc{i})/cm$^2$] $\gtrsim 18$), CO emission is detected in 12 out of 60 galaxies, corresponding to a detection rate of 20\%.
Although this detection rate is lower than those reported in some previous studies \citep[e.g.,][]{klitschALMACALIIICombined2018}, it represents a substantial number of detections drawn from a broader and less biased sample, {as no metallicity-based preselection was applied}. 
This represents a crucial step toward a more comprehensive understanding of molecular gas in absorption-selected galaxies.

Previous studies have reported high CO detection rates ($>60\%$), particularly in systems pre-selected for high metallicity or H$_2$ absorption.
For instance, \citet{neelemanFIRSTCONNECTIONCOLD2016} reported CO(1$-$0) emission from a $z\sim0.101$ absorber-associated galaxy with a molecular gas mass of $4.2 \times 10^9\,M_{\odot}$ and low star formation rate, indicating a long gas depletion timescale.
Similarly, \citet{mollerALMAVLTObservations2018} detected CO(2$-$1) in a highly metal-rich DLA galaxy at $z=0.716$, finding a large molecular mass ($\sim 2.3 \times 10^{10}\,M_{\odot}$) but suppressed star formation, deviating from canonical SFR–$M_{\rm H_2}$ scaling relations. 
These examples suggest that molecular gas-rich absorption-selected galaxies may undergo inefficient star formation, possibly due to environmental quenching.
\citet{kanekarMassiveAbsorptionselectedGalaxies2018} extended this trend with a sample of high-metallicity H\,\textsc{i} absorbers at $z\sim 0.5$–0.8, detecting CO in five out of seven cases (detection rate $>70\%$). 
These galaxies have molecular gas masses from $0.6$ to $8.2 \times 10^{10}\,M_\odot$, again linked with modest star formation rates, reinforcing the idea of gas-rich but low-efficiency star-forming systems.
\citet{klitschH2MolecularGas2021} found CO emission in five of six H$_2$-bearing absorbers (detection rate $>80\%$), further supporting a strong link between H$_2$ absorption and CO-rich galaxies, particularly in group environments or overdensities. 
They found no clear correlation between CO detection and absorber metallicity and concluded that H$_2$ absorbers trace diffuse molecular gas in the CGM or intragroup medium, rather than the central disk.
Further analysis by \citet{klitschCOExcitationLine2022} of the CO excitation in absorption-selected galaxies showed a broad range of ISM conditions, highlighting the heterogeneous nature of these systems and the complexity of relating CO luminosity to total H$_2$ content.

Collectively, these findings suggest that absorption selected galaxies at $z\lesssim1$ showing CO emission, particularly at high metallicities, may represent a distinct population, characterised by substantial molecular gas reservoirs but often exhibiting suppressed or inefficient star formation activity relative to typical star-forming galaxies at similar epochs.  However, these higher detection rates likely benefit from sample selections that favour CO-rich systems, {such as selecting high-metallicity absorbers \citep{neelemanMolecularEmissionGalaxy2018}}.

The MUSE-ALMA Haloes survey provides a complementary perspective.
By selecting galaxies solely based on their H\,\textsc{i} absorption properties, without bias toward metallicity or known H$_2$ content, the survey probes a broader galaxy population.
Our $\sim${20}\% detection rate thus provides a more representative estimate of molecular gas occurrence in H\,\textsc{i}-rich systems at $z \lesssim 1$, offering a statistically robust view of cold gas in the CGM and galaxy environments traced by quasar absorbers.

To further explore possible factors influencing CO detection, {Figure~\ref{fig:co_detection_properties} shows the distribution of CO detections and non-detections in galaxies as a function of redshift, stellar mass in the left and middle panels. In the third panel, this is plotted as a function of metallicity corresponding to multiple galaxies to evaluate whether any galaxy associated with the absorbers is detected in CO}. 
We find no clear dependence of detection rate on any of these parameters, suggesting that the presence of CO-emitting molecular gas is not simply governed by global galaxy properties, like the {absorber} metallicity alone. 
{This contrasts with studies that do find a correlation between absorber metallicity and CO detection rate, possibly because sub-DLA samples typically probe smaller impact parameters \citep[e.g.,][]{kanekarMassiveAbsorptionselectedGalaxies2018}, where the absorber metallicity more directly traces the ISM metallicity of the associated galaxy. 
While such studies often implicitly assume a one-to-one correspondence between absorbers and galaxies, {our observations reveal that several galaxies may contribute to the absorption seen along a single quasar sightline.}
In our case, the absence of metallicity preselection broadens the probed galaxy–absorber configurations, naturally diluting such correlations.}
{However,} recent results indicate that local conditions and environmental factors, such as gas density, pressure, or group dynamics, may play a more significant role in regulating cold molecular gas content in {H\,\textsc{i}-{selected} systems} \citep{leeALMAACACO2022}. 
{Since the presence of strong H\,\textsc{i} absorption along a quasar sightline reveals substantial column densities of neutral gas in the CGM or ISM of the associated galaxy, it is reasonable to compare our sample with H\,\textsc{i}-rich galaxies studied in emission at low redshift, without excluding that absorption selection may capture a broader range of gas properties.}

\subsection{Absorber-galaxy connection}\label{sec:abs_gal_connection}

\begin{figure*} 
    \centering
    \includegraphics[width=0.45\linewidth]{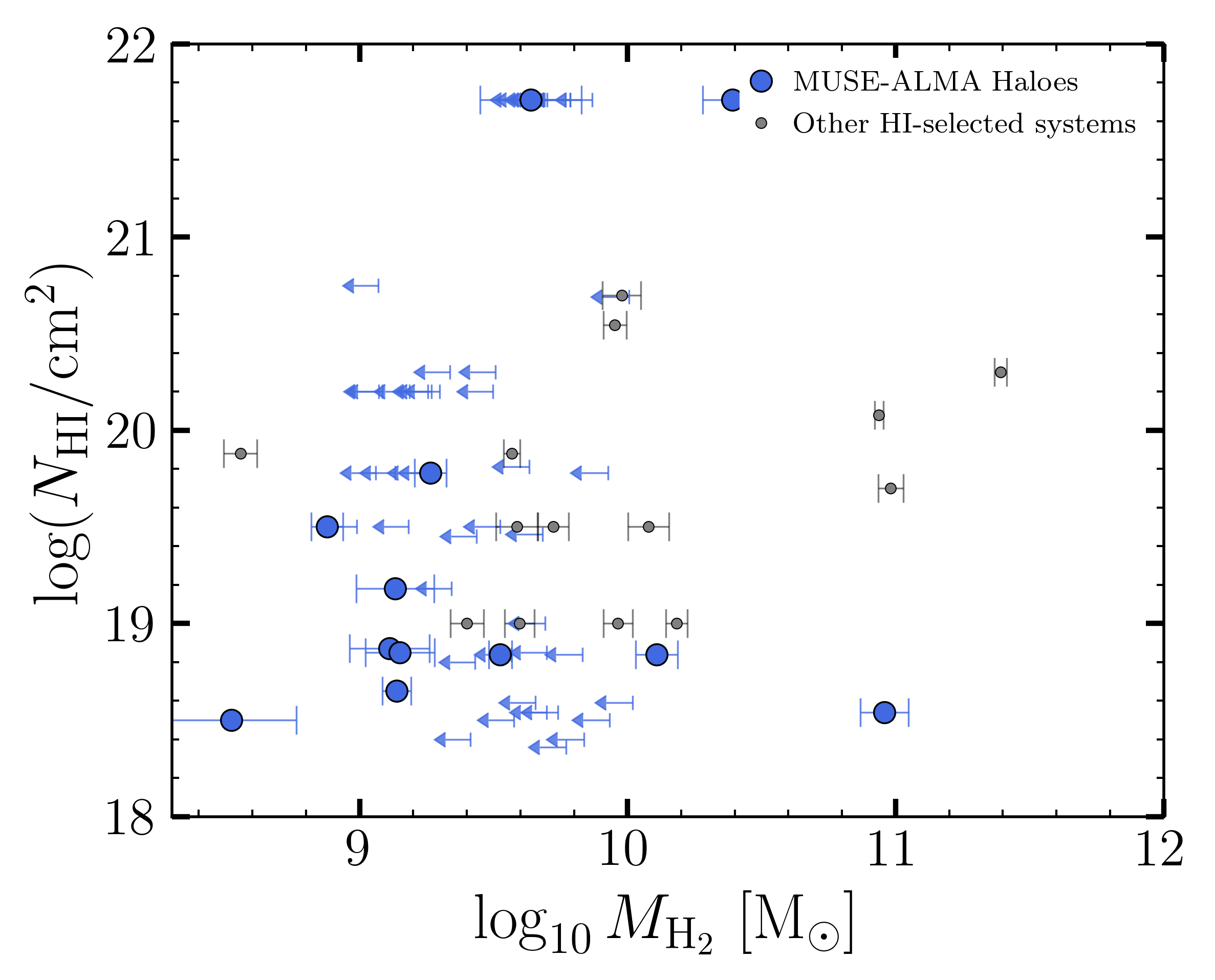}
    \includegraphics[width=0.45\linewidth]{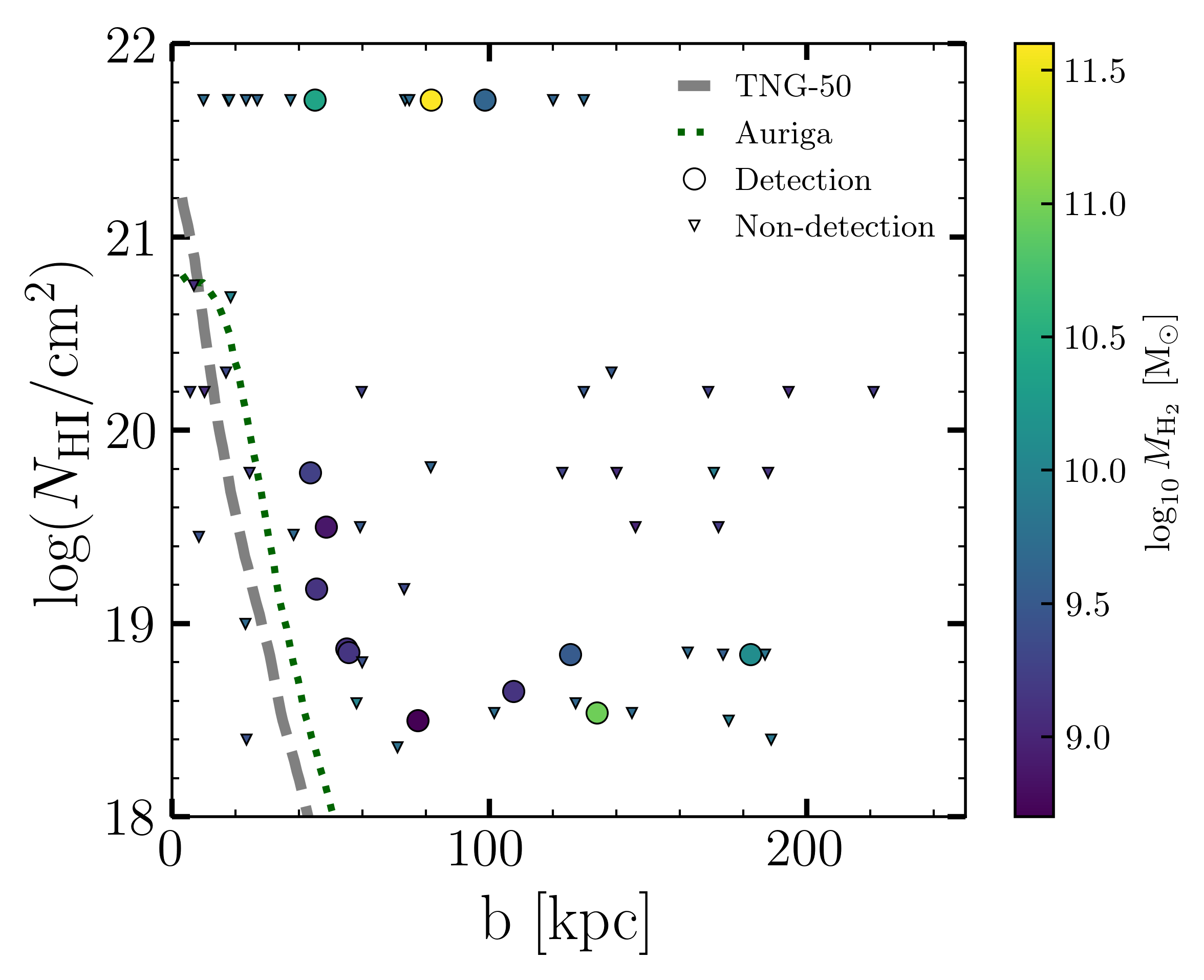}
    \caption{Left: Molecular gas mass ($M_{\text{H}_2}$) as a function of H\,\textsc{i} column density ($N_{\text{HI}}$). Blue circles represent our sample in comparison with previous studies on H\,\textsc{i}-absorption selected systems from \cite{kanekarMassiveAbsorptionselectedGalaxies2018} and \cite{klitschH2MolecularGas2021}, shown in grey.
    Right: $N_{\text{HI}}$ and impact parameter, colour coded by the molecular gas masses.
    In this plot, we include all impact parameters for the sources in our sample, rather than limiting to the smallest $b$ as done in previous studies \citep[e.g.][]{perouxSINFONIIntegralField2016, hamanowiczMUSEALMAHaloesPhysical2020, wengMUSEALMAHaloesVIII2022}. 
    The circles represent CO detections, and the triangle represents non-CO detections.
    We also include the median radial profiles for the neutral hydrogen column density from the simulations by \cite{voortCosmologicalSimulationsCircumgalactic2019} and \cite{nelsonResolvingSmallscaleCold2020} in the dashed and dotted lines, respectively.
    We report no clear correlation between the column density and impact parameter for different molecular gas masses.}
    \label{fig:nhi_properties}
\end{figure*}

The absorbers in the MUSE-ALMA Haloes survey span a range in H\,\textsc{i} column density of $\log$ [$N$(H\,\textsc{i})/cm$^2$] $\sim 18-22$, {covering systems from below the sub-DLA threshold to the DLA regime. We note that including absorbers below the canonical sub-DLA limit ($\log N$(H\,\textsc{i})=19.0) allows us to probe a broader population of gas-rich systems, while comparisons with literature samples restricted to (sub-)DLAs should be treated with this caveat in mind. Within our sample, the detection rate of CO emission appears consistent across both the sub-DLA and lower-column-density regimes, as illustrated in the left panel of Figure \ref{fig:nhi_properties}.}  
Systems with high $N$(H\,\textsc{i}) ($\geq 10^{21}$ cm$^{-2}$) are not necessarily associated with large molecular reservoirs, and conversely, galaxies with substantial $M_{\text{H}_2}$ can show relatively modest $N$(H\,\textsc{i}) values in absorption. 
This suggests that the absorbing H\,\textsc{i} gas, while indicative of the presence of neutral material, does not reliably trace the spatially integrated cold molecular gas content.
This conclusion is consistent with the findings of \cite{klitschH2MolecularGas2021} in a sample of H$_2$-bearing absorbers, where despite the high CO detection rate, they also report no clear correlation between $N$(H\,\textsc{i}), molecular gas fraction, or impact parameter and the host galaxy’s total $M_{\text{H}_2}$.

Our detections span a broad range of impact parameters (from $\sim$40 to 180 kpc), and again, we observe no significant trend between impact parameter and $M_{\text{H}_2}$ or $N$(H\,\textsc{i}). 
Some systems with large molecular reservoirs are found at relatively large projected separations, while others close to the quasar sightline show little molecular gas. 
Moreover, when examining $N$(H\,\textsc{i}) versus impact parameter color-coded by M$_{\text{H}_2}$, shown in the right panel of Figure~\ref{fig:nhi_properties}, we find no coherent trend. 
As a reference, we also include the results from the Auriga zoom-in simulations of a Milky Way-mass galaxy \citep{voortCosmologicalSimulationsCircumgalactic2019}, and from the post-processing of TNG50 around massive halos ($\sim 10^{13.5}M_{\odot}$) at $z\sim0.5$ \citep{nelsonResolvingSmallscaleCold2020}.
Systems with similar impact parameters exhibit large scatter in both $N$(H\,\textsc{i}) and molecular mass, supporting the notion that the location and kinematics of absorbing gas are not strongly tied to the total molecular reservoir of the galaxy.
Note that, unlike typical comparisons of these two quantities \citep[e.g.][]{perouxSINFONIIntegralField2016, hamanowiczMUSEALMAHaloesPhysical2020, wengMUSEALMAHaloesVIII2022}, we do not restrict the plot to only the smallest impact parameter.
In contrast, \citet{klimenkoBaryonicContentGalaxies2023} identified a strong correlation between $N$(H\,\textsc{i}) and impact parameter, found that absorber metallicities matched the emission metallicity gradients derived from integral field spectroscopy (IFS), and reported evidence of co-rotation between absorbing gas and galaxy disks extending out to $\sim$10 effective radii. 
These results suggest that, in some systems, absorbing gas can trace structured, rotating components of the CGM. However, such coherence is not observed across our full sample.
Additionally,
\cite{augustinMUSEALMAHaloesStellar2024} found an anticorrelation between the stellar mass of host galaxies and the $N$(H\,\textsc{i}) of associated absorbers. 
These findings indicate an evolutionary trend in the CGM composition with stellar mass, where lower-mass galaxies tend to host halos rich in cool, dense H\,\textsc{i} gas, while higher-mass systems show a depletion of such gas.

Out of the 12 galaxies with detected CO emission, all but one (ID Q0152m2001\_12) are embedded in environments where other galaxies are present at similar redshifts. 
This {supports} the idea that molecular gas is not only a feature of individual galaxies, but also a product of their interaction with the surrounding cosmic web and galaxy associations, consistent with earlier findings \cite{klitschALMACALAbsorptionselectedGalaxies2019, hamanowiczMUSEALMAHaloesPhysical2020}. 
This environmental connection may facilitate the accumulation or retention of molecular gas through processes such as galaxy-galaxy interactions or group-scale gas accretion.
Gravitational interactions can funnel and compress gas, increasing the molecular gas reservoir, while group environments enable continuous gas inflow and help prevent gas loss due to their deeper potential wells. This can lead to galaxies that harbour large amounts of molecular gas but exhibit relatively low star formation rates, suggesting that environmental factors may promote gas buildup without immediately triggering efficient star formation.
Moreover, gas flow geometry derived from MUSE data \citep{wengMUSEALMAHaloesVIII2022} reveals that some of these CO-rich systems are linked to inflows (e.g., Q0152m2001\_12 and Q2131m1207\_5 at $\sim$50 kpc), while others, like Q1229m021\_8 at $\sim$120 kpc, show outflow-like kinematics, indicating that molecular gas can be present in galaxies that are both accreting and expelling material, depending on the galaxy's location and dynamics within its environment.

\subsection{H\,\textsc{i} absorption selected-galaxies compared to star-forming populations}

\begin{figure*} 
    \centering
    \includegraphics[width=0.47\linewidth]{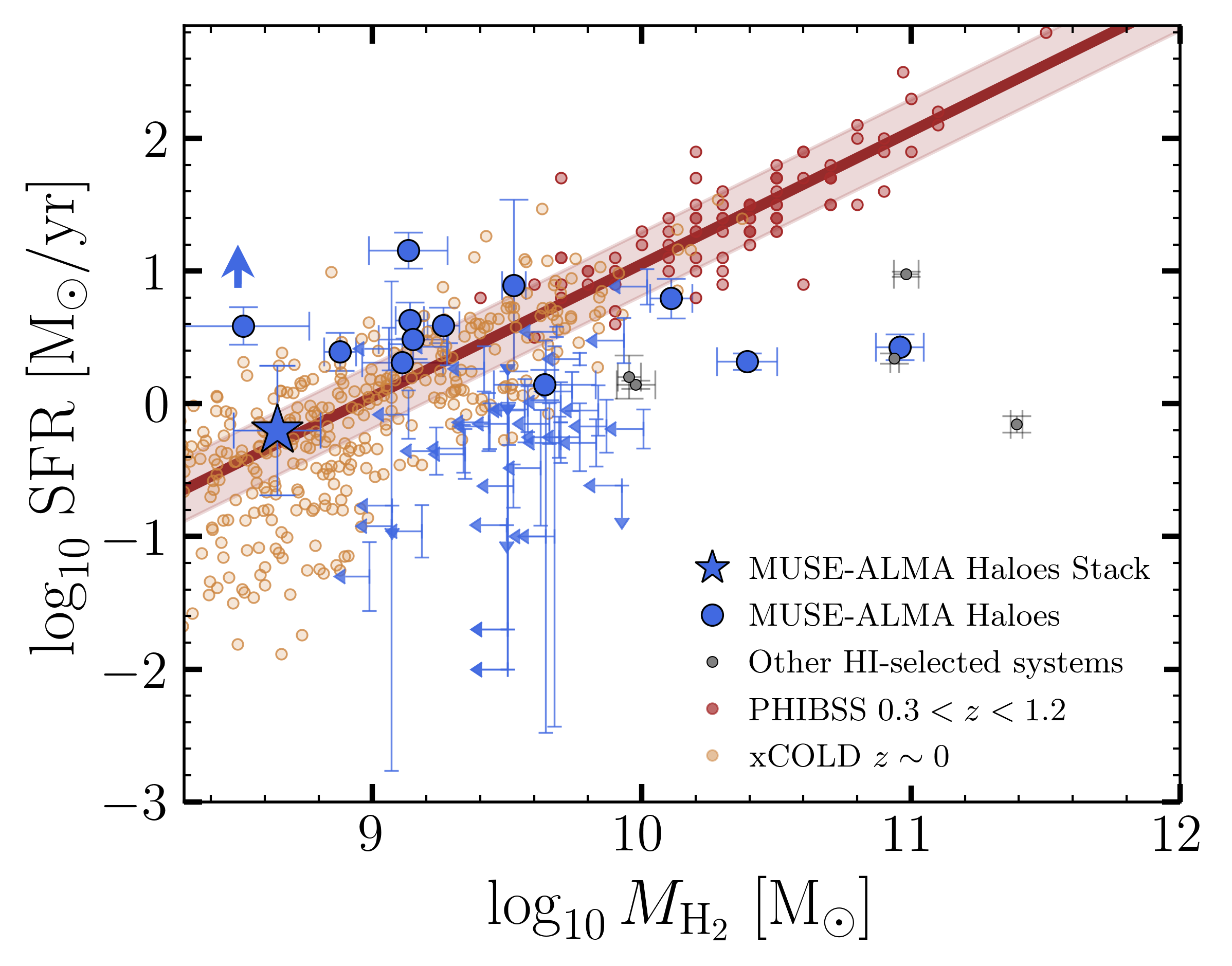}
    \includegraphics[width=0.45\linewidth]{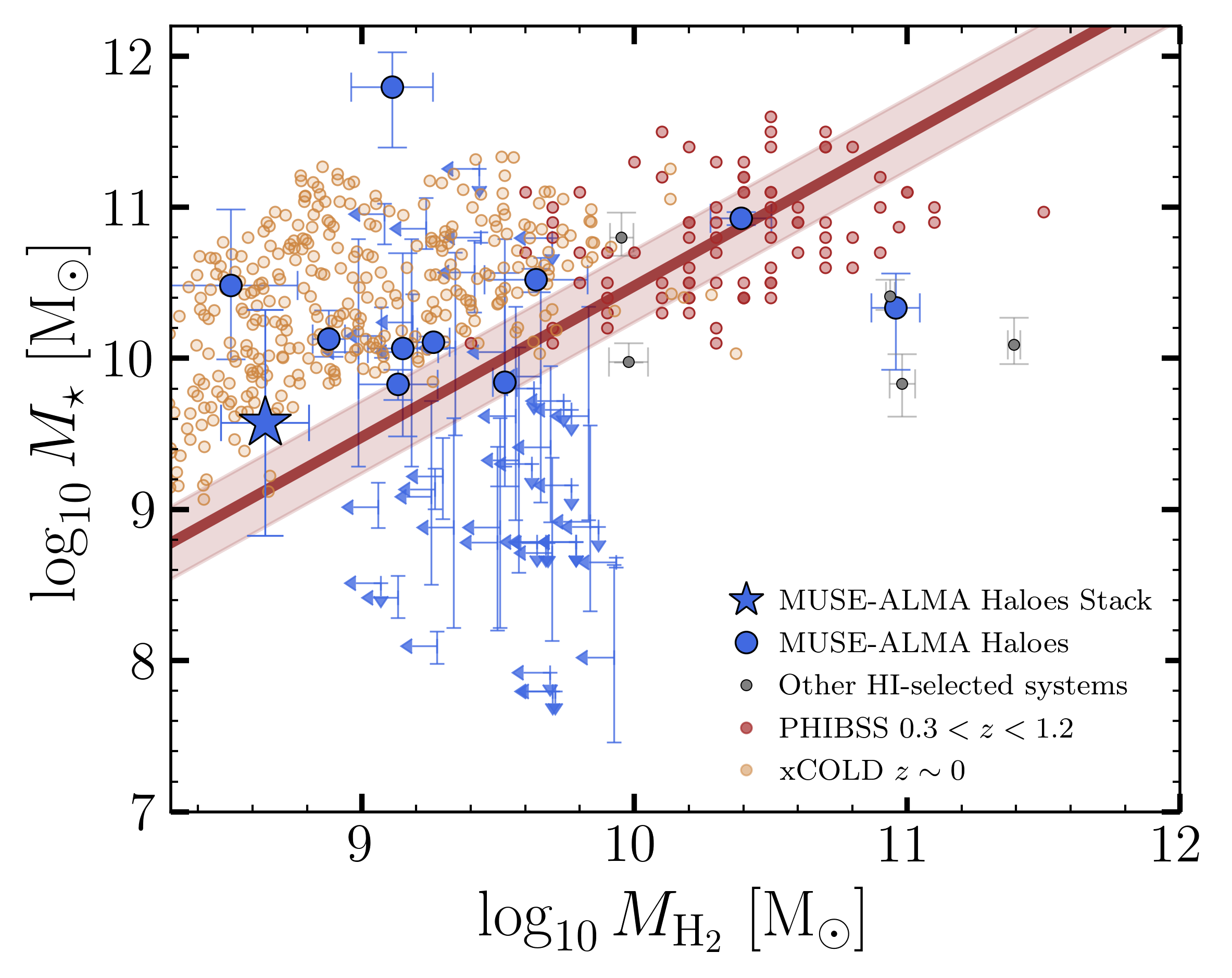}
    \caption{
    Left: Star formation rate (SFR) as a function of molecular gas mass ($M_{\text{H}_2}$). 
    Right: Stellar mass ($M_{\star}$) as a function of molecular gas mass ($M_{\text{H}_2}$).
    Our sample is shown in blue circles, and the blue arrows represent a $3\sigma$ upper limit. 
    {The result from the stacked spectrum described in Sect. \ref{subsec:stack} is shown by the blue star.}
    We also include the mean dust correction for SFR depicted by a large blue arrow on the {corner} of the left panel.
    For comparison, we include data from the xCOLD GASS survey \citep{saintongeXCOLDGASSComplete2017}, which represents low-redshift galaxies, and the PHIBSS survey \citep{tacconiPHIBSSUnifiedScaling2018}, in the redshift range $z\sim 0.3 - 1.2$. The solid line represents the molecular gas main sequence scaling relation from \cite{tacconiPHIBSSUnifiedScaling2018}.
    We also included estimates from \cite{kanekarMassiveAbsorptionselectedGalaxies2018} and \cite{klitschH2MolecularGas2021} in both panels, labelled as other H\,\textsc{i}–selected systems.
    Our sources {with $M_{\rm H_2}/M_{\odot }\lesssim10^{9.8}$ lie $\sim0.3$ dex above the expected SFR$-M_{\text{H}_2}$ relations for `normal star-forming galaxies' at the same redshift range, while the systems with {with $M_{\rm H_2}/M_{\odot }>10^{9.8}$} lie $\sim1.5$ dex below the relation.}
    {This trend suggests that H\,\textsc{i}–selected systems have a dual behaviour. Galaxies with low molecular gas masses form stars efficiently, following both the depletion timescales and the $M_{\star}-M_{\text{H}_2}$ scaling relations of main-sequence galaxies.
    In contrast, galaxies with high molecular gas masses show inefficient star formation and fall below the expected stellar mass growth,} likely because they are still actively accreting gas from the intergalactic medium or interacting within group environments. These systems appear not yet to have reached the equilibrium conditions characteristic of main-sequence galaxies.
    }
   \label{fig:scaling_relations}
\end{figure*}

To contextualise our sample, we compare the properties of the host galaxies associated with the absorbers with the well-established galaxy scaling relations.
We used a subsample of 89 galaxies at $0.3 < z < 1.2$ from the PHIBSS survey \citep{tacconiPHIBSSUnifiedScaling2018} ranging in stellar mass from $10^{9.8}$ to $10^{11.8}$ M$_{\odot}$ and the full sample from xCOLD GASS \citep{saintongeXCOLDGASSComplete2017} at $z\sim0$ ranging in stellar mass from $10^{9}$ to $10^{11.3}$ M$_{\odot}$, which is considered representative of the local galaxy population.
In Figure \ref{fig:scaling_relations}, the left panel shows star formation rate (SFR) as a function of molecular gas mass ($M_{\text{H}_2}$), while the right panel shows the stellar mass against $M_{\text{H}_2}$.
We distinguish the MUSE-ALMA Haloes (blue circles) from other literature H\,\textsc{i}–selected galaxies (grey points) from \cite{kanekarMassiveAbsorptionselectedGalaxies2018} and \cite{klitschALMACALAbsorptionselectedGalaxies2019}, while the red and orange circles represent the PHIBSS and xCOLD GASS samples, respectively.
{The SFR in our sample is derived from MUSE emission lines (H$\alpha$ and [O\,\textsc{iii}] when H$\alpha$ is unavailable, \citealt{wengMUSEALMAHaloesVIII2022}). PHIBSS SFRs combine UV and IR luminosities to account for obscured and unobscured star formation, while xCOLD GASS uses SDSS emission lines, GALEX UV, and WISE IR photometry. Despite minor differences in methodology and dust corrections, the SFRs are broadly consistent across samples, allowing a fair comparison in Figure \ref{fig:scaling_relations}}.

Our H\,\textsc{i}–selected galaxies {span the same general parameter space 
as emission-selected galaxies}.
At fixed $M_{\text{H}_2}$, {we find a clear dependence of SFR offset on molecular gas mass} (Figure \ref{fig:scaling_relations}, left panel). 
{Galaxies with low molecular gas content ($\log M_{\rm H_2} \lesssim 9.8$) lie slightly above ($\sim0.3$ dex) the scaling relation for} normal star-forming galaxies at the same redshift range  \citep{tacconiPHIBSSUnifiedScaling2018}, shown by the solid red line. 
{In contrast, galaxies with high molecular gas content (log $M_{{\rm H}_2} > 9.8$) are typically below the scaling relation by $\sim1.5$ dex. This indicates that while low-$M_{{\rm H}_2}$ galaxies are relatively efficient in forming stars, high-$M_{{\rm H}_2}$ galaxies host abundant molecular gas, which is yet to form stars. 
Even accounting for dust corrections (blue upward arrow), which increases the SFR by $\sim0.24$ dex on average, the offset in high-$M_{{\rm H}_2}$ systems remains significant.}

We investigate the stellar mass–molecular gas mass ($M_{\star}-M_{\text{H}_2}$) relation in the context of the MUSE–ALMA Haloes survey. 
As shown in the right panel of Figure \ref{fig:scaling_relations}, there is a systematic offset in this scaling relation for galaxies selected via H\,\textsc{i} absorption {for the high- and low-molecular gas mass regimes}, compared to the main sequence of star-forming galaxies at similar redshifts. 
{Galaxies with low molecular gas content (log $M_{\rm H_2} \lesssim 9.8$) lie on average $\sim0.9$ dex above the reference relation, indicating they are relatively massive in stars for their gas content. Conversely, galaxies with high molecular gas content (log $M_{\rm H_2} > 9.8$) lie $\sim1.5$ dex below the reference relation, implying that they host abundant molecular gas relative to their comparatively low stellar masses.}

\begin{figure*} [t]
    \centering
    \includegraphics[width=0.52\linewidth]{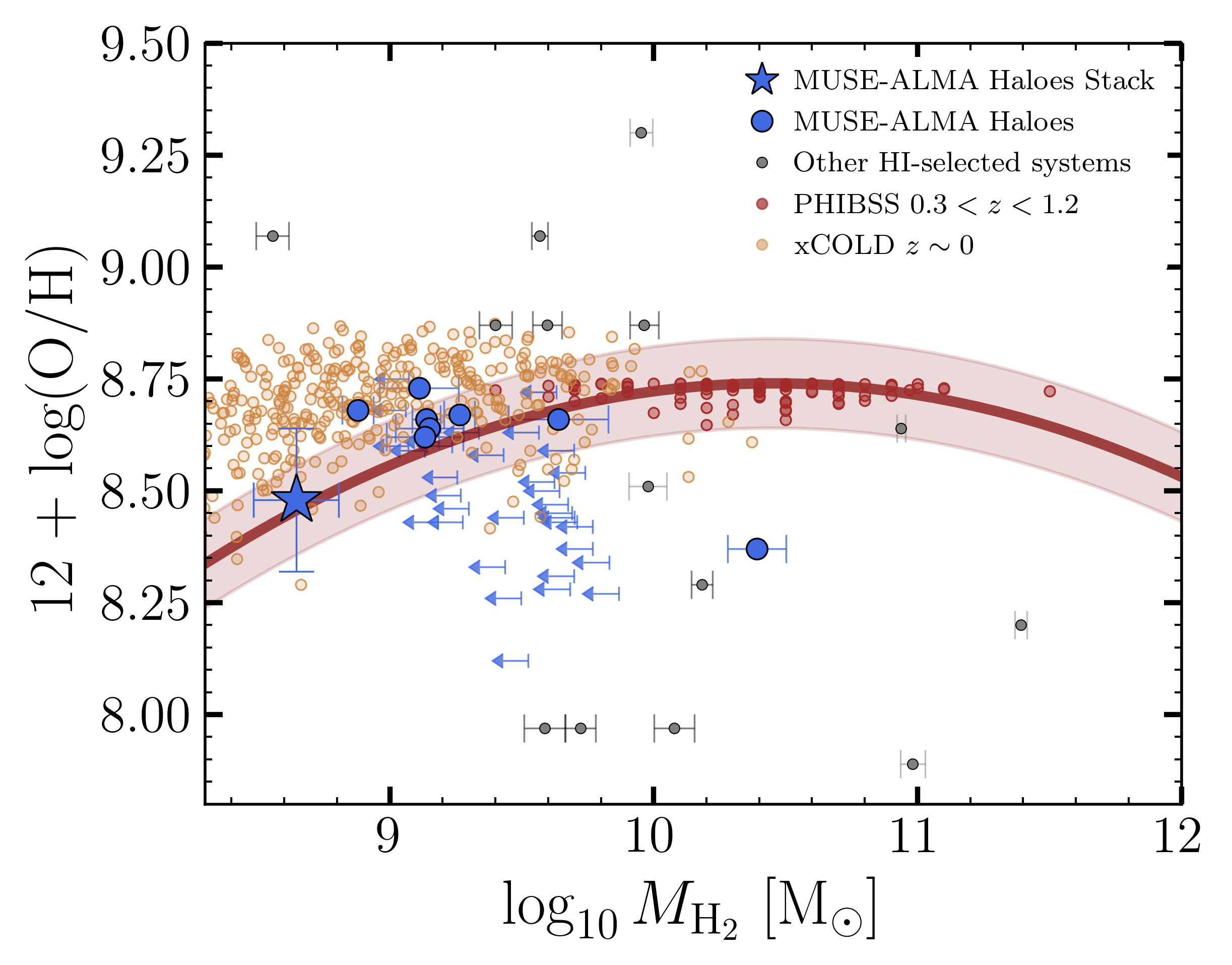}
    \includegraphics[width=0.4
    \linewidth]{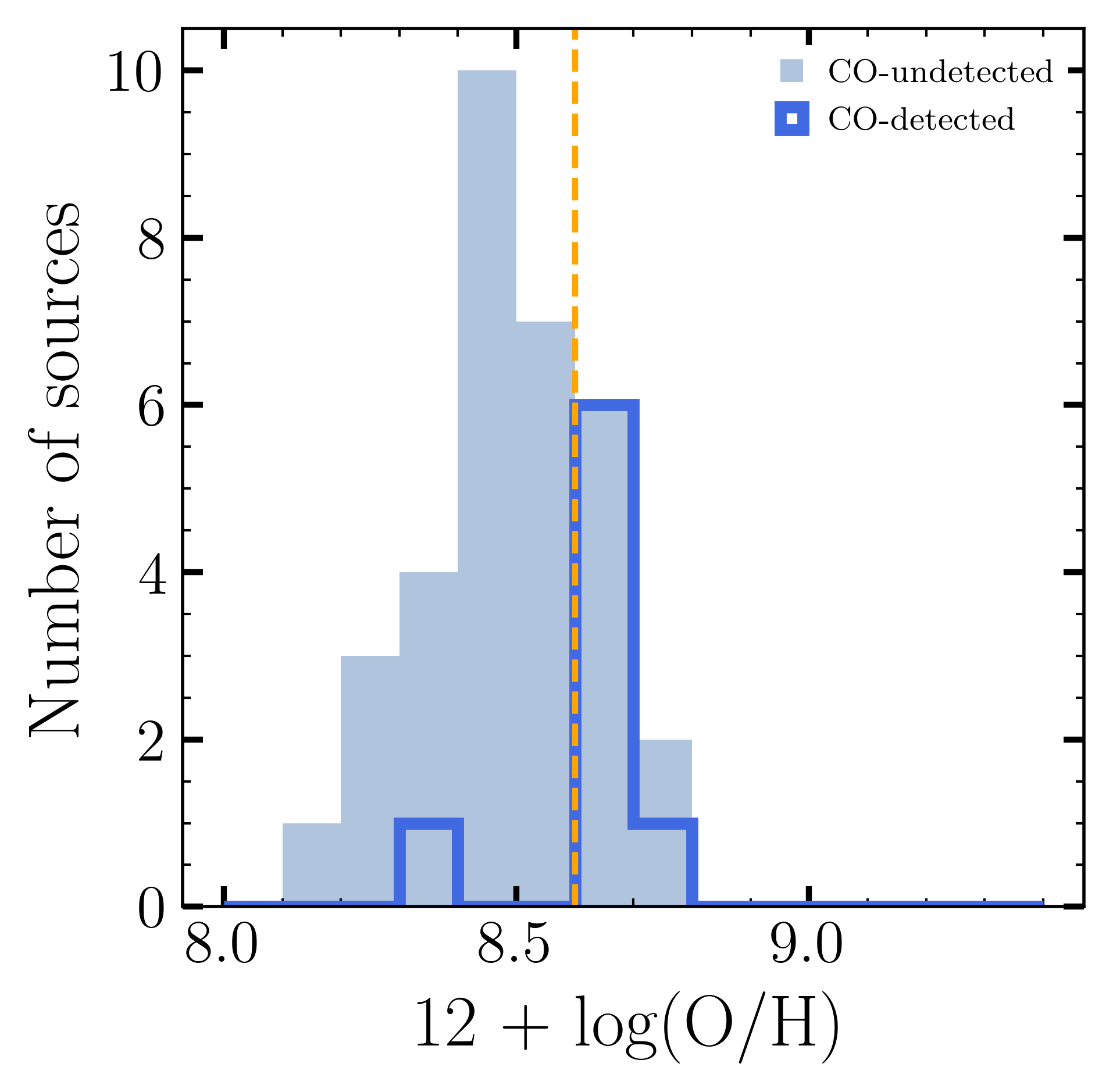}
    \caption{Left: 
    Nebular emission metallicity, $12 + \log$(O/H) {\citep{wengMUSEALMAHaloesVIII2023}}, as a function of molecular gas mass ($M_{\mathrm{H}_2}$) for our H\,\textsc{i}–selected galaxies. Blue circles indicate CO detections, and blue leftward arrows show $3\sigma$ upper limits for CO non-detections.
    {The result from the stacked spectrum described in Sect. \ref{subsec:stack} is shown by the blue star.}
    The solid line indicates the mass-metallicity relation from \cite{genzelCombinedCODust2015} for the PHIBSS sample \citep{tacconiPHIBSSUnifiedScaling2018}.
    A positive correlation emerges between metallicity and CO detectability, with CO-detected galaxies tending to occupy the higher-metallicity regime ($\log Z\gtrsim 8.5$), as found previously in many studies \citep[e.g.,][]{bolattoCOtoH2ConversionFactor2013}.
    Right: Histogram of emission metallicities for all galaxies, with CO detections in blue and non-detections in light blue filled histogram. 
    {The dashed orange line indicates a metallicity $\log Z\sim 8.5$, which sets the higher and lower metallicity regimes discussed in Sect. \ref{sec:discussion}.}
    Although the distributions overlap significantly, a two-sample Kolmogorov-Smirnov test yields a p-value of 0.004, indicating a statistically significant difference in the underlying metallicity distributions. This reinforces the interpretation that metallicity plays a key role in governing the observability of molecular gas in H\,\textsc{i}–selected systems.
    % p-value: 0.011
    }
    \label{fig:metallicity}
\end{figure*}

We examine the molecular gas mass-metallicity relation for our H\,\textsc{i}–selected sample, as shown in the left panel of Figure \ref{fig:metallicity}. We use the derived gas-phase metallicities from nebular lines presented in \cite{wengMUSEALMAHaloesVIII2022}, and compare these with molecular gas masses for the CO-detected systems.
The CO-detected galaxies in our sample generally follow the same locus as normal star-forming galaxies in the molecular gas mass–metallicity plane. 
Most of the non-detections lie at lower metallicities, falling below the expected trend for their gas content. 
This suggests that emission metallicity may influence the detectability of molecular gas via CO lines, since CO is easily photo-dissociated in low metallicity environments, and there may be a lot of molecular gas that is not traced by CO \citep[e.g.][]{maddenTracingTotalMolecular2020}.
To quantify this, we construct an emission metallicity distribution for both CO detections and non-detections as shown in the right panel of Figure \ref{fig:metallicity}. 
A Kolmogorov–Smirnov (KS) test reveals a statistically significant difference between the two populations, with a p-value of 0.004. 
This indicates that the metallicity distributions of the CO-detected and CO non-detected galaxies are unlikely to be drawn from the same parent population.

\section{Discussion}\label{sec:discussion}

\subsection{Implication of CO detections for H\,\textsc{i}–selected samples}

Understanding the possible relation between H\,\textsc{i}-absorption selected galaxies and their CO emission is essential to interpret their physical properties in a wider context.
Several studies have suggested a trend between high-metallicity absorption systems and CO emitters at $z<1$ \citep[e.g.,][]{kanekarMassiveAbsorptionselectedGalaxies2018, mollerALMAVLTObservations2018, klitschH2MolecularGas2021}.
Interestingly, this trend at high redshift ($z\geq2$) has also been explored.
CO emission has been detected in $\sim 7$ out of $\sim20$ DLA fields surveyed to date, with detection predominantly arising in systems with relatively high metallicities ([M/H]$\geq -0.72$) and large molecular gas masses (M$_{\text{mol}} \geq 5 \times 10^{10} M_{\odot}$). For instance, \cite{neelemanMolecularEmissionGalaxy2018} reported the detection of CO(3--2) and far-infrared continuum emission from a galaxy associated with a DLA at $z = 2.192$. 
Located at an impact parameter of 30 kpc from the quasar sightline, the galaxy exhibits a dust-corrected star formation rate (SFR) of $\sim110M_{\odot}$ yr$^{-1}$ and a molecular gas mass of $\sim 1.4\times 10^{11}M_{\odot}$.
\cite{kanekarHighMolecularGas2020} conducted a systematic ALMA survey of 12 high-metallicity DLAs ($z\sim$ 1.7--2.6), detecting CO emission in five fields. The galaxies have molecular gas masses ranging from 1.3 $\times 10^{10}$ to 2.1 $\times 10^{11}$ M$_{\odot}$ and impact parameters between 5.6 and 100 kpc. The study confirmed a strong correlation between DLA metallicity and the likelihood of CO detection, with high-metallicity systems more frequently associated with massive, gas-rich galaxies.
In another example, \citet{kaurMassiveDustyHi2022} used NOEMA to detect CO(3--2) emission from the galaxy DLA0201+365g at $z = 2.4604$. The galaxy, located $\sim$7\,kpc from the quasar sightline, was found to contain a molecular gas mass of $\sim 5 \times 10^{10}$\,M$_\odot$. Despite its gas richness, the galaxy had an SFR upper limit of just 2.3\,M$_\odot$\,yr$^{-1}$, suggesting either significant dust obscuration or a long gas depletion timescale.
Further support for large molecular gas reservoirs in H\,\textsc{i}–selected galaxies at high redshift comes from JVLA detections of CO(1--0) and CO(3--2) in two DLAs at $z \approx 2.193$ and $z \approx 2.356$ \citep{kaurHIabsorptionselectedColdRotating2024, kaurMassiveIabsorptionselectedGalaxy2025}.
These galaxies exhibited CO velocity widths of $\sim$\,500--600\,km\,s$^{-1}$ and molecular gas masses approaching $10^{11}$\,M$_\odot$. Interestingly, while DLA1228--113g showed near-thermal excitation of mid-$J$ CO transitions, DLA1020+2733g showed sub-thermal excitation, indicating a relatively low star formation surface density and possible differences in gas excitation and star formation efficiency.
More recently, \cite{neelemanII158Mm2025} reported a [C\,\textsc{ii}] detection rate of $>70\%$ for galaxies at $z\sim 4.1-4.5$ selected to have an absorption metallicity [M/H]$>-1.5$.

{These studies have primarily targeted H$_2$ absorbers or metal-rich DLAs and sub-DLAs, inherently biasing detections towards galaxies with high molecular gas content ($M_{\rm H_2} \gtrsim 10^{10} M_{\odot}$).
This strategy yields high detection rates, but it has also limited our understanding of the broader population of H\,\textsc{i}–selected systems. 
These systems with high molecular gas masses exhibit star formation activity that is not always balanced with their gas reservoirs, lying systematically below the star-forming main sequence in terms of depletion time. This behaviour suggests that H\,\textsc{i} absorption associated with metal-rich DLAs and sub-DLAs preferentially traces systems at particular evolutionary stages, either in early phases of gas accretion or in temporarily inefficient modes of star formation, not typically represented in flux-limited, emission-selected surveys \citep[e.g.,][]{kulkarniHubbleSpaceTelescope2005}.}

{In contrast, the MUSE–ALMA Haloes survey does not preselect galaxies by metallicity when following up on their molecular gas content. In addition, our observations probe a factor of $\sim1.2$ dex deeper in $M_{\mathrm{H}_2}$ than earlier absorber studies, reaching a regime where such galaxies would have been missed in previous surveys. 
The considerable number of non-detections enables our stacked analysis to yield a $\sim3\sigma$ detection consistent with the scaling relations of normal star-forming galaxies and demonstrates the deep rms levels achieved by our survey.}

The stringent detection limits in our sample probe the intermediate molecular gas mass range ($M_{\mathrm{mol}} \lesssim 10^{9}-10^{10} M_{\odot}$), and notably, several non-detections occur in galaxies with high stellar masses ($>10^{10}M_{\odot}$).
The lack of CO detections in most absorber-selected galaxies does not necessarily imply the absence of molecular gas. 
Instead, it likely reflects the presence of ``CO-dark" molecular gas, defined as regions where H$_2$ exists without detectable CO emission due to {low metallicity, reduced dust shielding, or diffuse, extended gas distributions.}
This interpretation is supported by {both theoretical models and observations of dwarf and metal-poor galaxies \citep{maddenTracingTotalMolecular2020}, which are frequently associated with high H\,\textsc{i} column density absorbers \citep[e.g.,][]{yorkOriginQSOAbsorption1986, kulkarniNICMOSImagingDamped2000}.}
Physical conditions, such as sub-critical gas densities or beam dilution in extended, low-surface-brightness structures, can further suppress the observable CO.
Simulations by \cite{liDarkMolecularGas2018} found that a large fraction of H$_2$ resides in diffuse regions where gas densities are below the critical threshold required to excite CO transitions, where CO emission may be weak or absent.
Observations by \cite{smitEVIDENCEUBIQUITOUSHIGHEQUIVALENTWIDTH2014} also demonstrated that CO-dark gas frequently occurs in filamentary structures, which can be overlooked in high-resolution surveys. 
If molecular gas is spatially extended with low surface brightness, the CO emission may be spread over a large area and diluted below the detection limits of the observing beam.
Given that absorber-selected galaxies are biased toward tracing diffuse, metal-poor, or extended gas phases, the absence of CO detections is best explained not by a true lack of molecular gas but by its predominantly CO-dark nature.
Overall, the environmental conditions and observational limitations strongly support the scenario in which CO-dark molecular gas dominates the molecular content in {H\,\textsc{i}-selected} systems.

{An alternative explanation is that the H\,\textsc{i}-to-H$_2$ transition itself is inefficient or delayed in these galaxies. }
The efficiency of converting atomic hydrogen to molecular hydrogen is a crucial bottleneck in the star formation process. Our observations might be suggesting that the H\,\textsc{i}-to-H$_2$ transition is delayed in these systems, potentially due to sub-solar metallicities that reduce dust abundance and limit the shielding required for H$_2$ formation. This is consistent with theoretical predictions by \citet{krumholzAtomictoMolecularTransitionGalaxies2009} and \citet{sternbergITOH2TRANSITIONSCOLUMN2014}, which show that in low-metallicity environments, the H\,\textsc{i}-to-H$_2$ transition can be suppressed or delayed, with timescales extending to tens or hundreds of Myr.
Additionally, turbulent gas conditions can affect molecule formation. Simulations by \citet{gloverSimulatingFormationMolecular2007} demonstrate that while turbulence can enhance local densities and promote H$_2$ formation, it can also increase mixing and disrupt the formation process in lower-density regions, leading to longer conversion times.
The presence of detectable CO emission in some of these systems suggests that at least partial molecular gas formation has occurred. 
However, the inefficiency in converting H\,\textsc{i} to H$_2$ is likely due to low metallicity, a low dust-to-gas ratio, and disrupted interstellar medium conditions, leading to subdued star formation despite significant gas reservoirs.

\subsection{Physical diversity of H\,\textsc{i} absorbers}

The wide range of galaxy–absorber configurations in our sample emphasises the physical diversity inherent to H\,\textsc{i}–selected systems. 
11 of the 12 CO-detected galaxies are associated with additional galaxies {at the same absorber redshift. }
This implies that absorption selection is particularly sensitive to complex environments, such as group-scale structures or filaments, rather than isolated field galaxies. 
Consistent with this, \cite{klitschH2MolecularGas2021} found that H$_2$-bearing absorbers often trace overdensities or galaxy groups. 
Kinematic classifications from \citet{wengMUSEALMAHaloesVIII2022} further reveal that the absorbing gas in several of these systems likely originates from outflows or inflows, rather than rotationally supported disks. Thus, the cold gas reservoirs probed via H\,\textsc{i} absorption are embedded in a diverse range of physical contexts, and absorption-selected galaxies do not represent a uniform or simple population.

\begin{figure}[h!]
    \centering
    \includegraphics[width=1\linewidth]{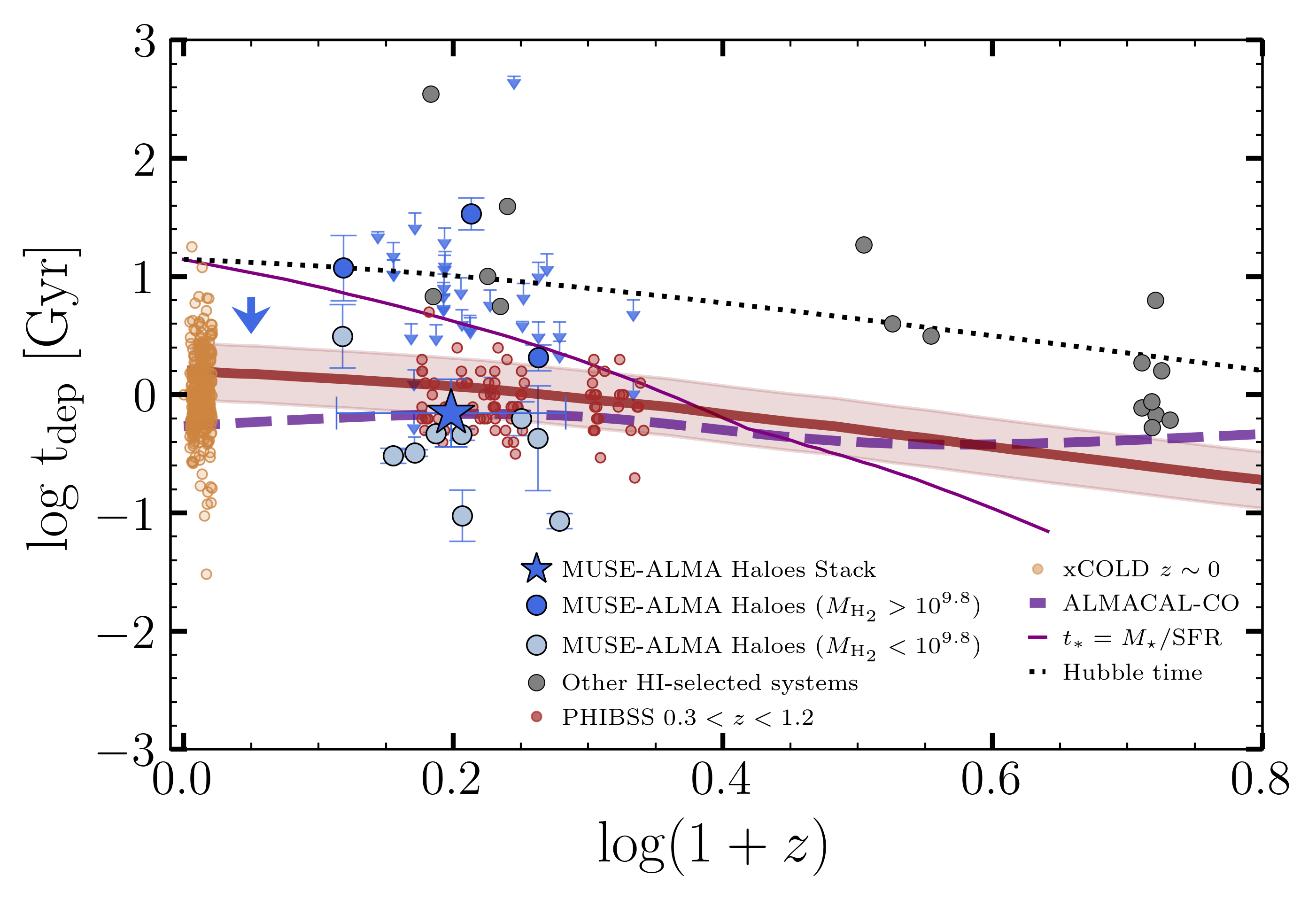}
    \caption{Molecular gas depletion time ($t_{\text{dep}} = M_{\text{H}_2} / \text{SFR}$) as a function of redshift for our H\,\textsc{i}–selected sample (blue/light-blue circles for high/low molecular gas mass systems), compares with literature scaling relations and emission-selected samples \citep{saintongeXCOLDGASSComplete2017, tacconiPHIBSSUnifiedScaling2018}. 
    {The result from the stacked spectrum described in Sect. \ref{subsec:stack} is shown by the blue star.}
    The solid red curve shows the empirically derived $t_{\text{dep}}$ evolution for main-sequence galaxies from \citet{tacconiPHIBSSUnifiedScaling2018}, the dotted line marks the Hubble time at each redshift, the thin solid line represents the stellar mass doubling time ($t_* = M_{\star} / \text{SFR}$) from \cite{tacconiEvolutionStarFormingInterstellar2020}, and the purple dashed line is from the cosmic depletion timescales derived from the ALMACAL survey \citep{bolloALMACALXIIIEvolution2025}. H\,\textsc{i}–selected galaxies, both from our sample and from high-redshift literature sources (gray dots; \citealt{klitschH2MolecularGas2021, neelemanMolecularEmissionGalaxy2018,neelemanColdMassiveRotating2020, neelemanII158Mm2025, fynboALMAObservationsMetalrich2018, kanekarHighMolecularGas2020, kanekarMassiveAbsorptionselectedGalaxies2018, kaurMassiveDustyHi2022, kaurMassiveIabsorptionselectedGalaxy2025}), consistently lie above the main-sequence relation, indicating lower star formation efficiencies. Several galaxies exhibit $t_{\text{dep}}$ approaching or exceeding the Hubble time, suggesting inefficient molecular gas consumption or suppressed star formation.
    }
     \label{fig:depletion_timescale}
\end{figure}

While some of these galaxies, {particularly those with low molecular gas masses (log $M_{\rm H_2} < 9.8$)}, are actively star-forming and {are roughly consistent with the trends for normal star-forming galaxies}, others {with high molecular gas masses show suppressed star formation relative to their gas reservoirs.
The low-$M_{\rm H_2}$ systems exhibit depletion times that are comparable to those of main-sequence galaxies at similar redshifts, as indicated by the light blue circles in Figure \ref{fig:depletion_timescale}. This suggests that they represent more evolved systems that have already accumulated substantial stellar mass and are efficiently forming stars.
By contrast, galaxies with high $M_{\rm H_2}$} show a mismatch between SFR and $M_{\text{H}_2}$, implying that they lie well above the star formation ``main sequence'' in depletion time, as shown in the {blue points} of Figure \ref{fig:depletion_timescale}. 
Therefore, they form stars less efficiently than expected for their molecular gas mass by a factor of {$\sim1.2$} dex, placing them in a slower, more quiescent evolutionary phase.

{The stacking analysis reinforces this emerging picture: the stacked spectrum is consistent with the depletion times of normal star-forming galaxies, confirming that the observed suppression is confined to the most gas-rich individual systems. 
{In the stack, the average molecular gas mass is estimated from the integrated CO flux of the combined spectrum, scaled by the mean redshift of the sample and adopting an $\alpha_{\text{CO}}$ calibrated on the mean metallicity of the stacked systems.
Similarly, the average SFR corresponds to the mean of the individual SFRs prior to stacking, ensuring that the stacked measurement reflects the typical star-formation level of the sample rather than being dominated by the brightest objects. For consistency, the metallicity shown in Fig. \ref{fig:metallicity} is computed as the median of the individual gas-phase metallicities.}
It is worth noting that our deep ALMA observations push the molecular gas mass sensitivity by more than an order of magnitude lower than previous H\,\textsc{i}-selected studies, extending the probed regime toward galaxies with $M_{\rm H_2}$ below $10^{9}$ M$_\odot$. This allows us to capture the diversity of gas properties, from more evolved, star-forming systems to inefficient, gas-rich galaxies.}

As noted by \citet{kanekarMassiveAbsorptionselectedGalaxies2018} and \citet{klitschCOExcitationLine2022}, this apparent disconnect between molecular gas content and star formation activity points to environmental or dynamical suppression of star formation.
Several mechanisms may contribute to the observed inefficiency of star formation in these systems. 
One possibility is that they are in an early phase of gas accretion, where molecular gas has recently condensed, but star formation has not yet fully ignited.
Alternatively, internal feedback or environmental influences, such as turbulence, ram pressure, or low enrichment, may act to suppress or delay the conversion of gas into stars. 
In low-metallicity environments, an overestimated CO-to-H$_2$ conversion factor ($\alpha_{\text{CO}}$) could also contribute to the apparent excess in molecular gas, though this alone is unlikely to explain the consistent suppression across multiple systems.

At the same time, our H\,\textsc{i}–selected galaxies {with high $M_{\text{H}_2}$} have stellar masses generally lower than their molecular gas mass, being more gas-rich than typical emission-selected field galaxies of the same stellar mass.
This offset supports the scenario of H\,\textsc{i}–selected systems being at an early stage of stellar mass buildup, possibly following recent gas accretion from filamentary structures or group environments.
A similar offset was also reported by \citet{kulkarniDampedLyaAbsorbers2022} in the H\,\textsc{i} versus stellar mass relation, which they interpreted as evidence for recent accretion fuelling the gas reservoirs in absorption-selected galaxies.
These trends align with findings from other absorption-selected samples {at similar $M_{\text{H}_2}$} \citep[e.g.,][]{neelemanFIRSTCONNECTIONCOLD2016, neelemanIi158Mm2019, kanekarMassiveAbsorptionselectedGalaxies2018}, reinforcing the notion that H\,\textsc{i} absorbers trace {both normal star forming galaxies and unusually} gas-rich galaxies. 
{Importantly, our deep observations also extend the parameter space by more than an order of magnitude toward lower stellar masses, allowing us to probe gas-rich galaxies below the stellar mass range typically reached by previous absorption- and emission-selected surveys.}

Finally, the strong segregation in emission metallicity between CO-detected and non-detected galaxies adds further nuance to this picture. 
The detectability of CO is clearly influenced by metallicity, which governs the abundance and survival of CO relative to H$_2$ in diffuse, unshielded regions \citep{bolattoCOtoH2ConversionFactor2013, schrubaLOWCOLUMINOSITIES2012}. 
Even when molecular gas is present, low-metallicity environments may have CO emission too faint to detect, particularly in the outskirts or CGM of galaxies.
The fact that stellar mass and redshift distributions are comparable between CO detections and non-detections, while metallicity differs, suggests that local ISM conditions, such as enrichment, shielding, and density, are the primary regulators of molecular gas observability. This is consistent with both theoretical expectations and recent observational evidence emphasising the role of small-scale environmental factors in shaping cold gas content \citep[e.g.,][]{sternUniversalDensityStructure2016, diemerAtomicMolecularGas2019}.

\subsection{Role in cosmic baryon cycle}

Our analysis of H\,\textsc{i}–selected galaxies provides valuable insights into the cosmic baryon cycle, particularly regarding gas accretion, star formation efficiency, and the conversion of atomic to molecular gas. These systems, often overlooked in emission-selected surveys, provide a unique window into the processes governing galaxy evolution.

In the context of the equilibrium gas regulator model \citep{boucheImpactColdGas2010, forbesOriginFundamentalMetallicity2014, lillyGasRegulationGalaxies2013, tacchellaEvolutionDensityProfiles2016}, which describes the balance between gas inflow, star formation, and outflows, our H\,\textsc{i}–selected galaxies {separate into two distinct regimes.
Galaxies with low molecular gas masses (log $M_{\rm H_2} \lesssim 9.8$) show depletion timescales that are shorter than the stellar mass doubling time (thin purple line in Figure \ref{fig:depletion_timescale}), consistent with the regime where gas is consumed efficiently and star formation proceeds at rates similar to those of normal star-forming galaxies (thick solid line). 
In contrast, galaxies with high molecular gas masses (log $M_{\rm H_2} > 9.8$) exhibit significantly longer depletion timescales, well above $t_\star$, indicating inefficient star formation despite large gas reservoirs. These two regimes suggest that while some H\,\textsc{i}–selected galaxies are forming stars in a manner consistent with the gas regulator model, others accumulate molecular gas without efficiently converting it into stars, undergoing early stages of gas buildup or entering a temporarily inefficient star-forming phase.
}

The observed trend of depletion timescale with redshift in our sample {of high molecular gas masses}  appears to persist across cosmic time. Several H\,\textsc{i}–selected galaxies at high redshift show unusually long depletion timescales, as shown in the grey points of Figure \ref{fig:depletion_timescale}.
We included estimates from \citet{kaurMassiveIabsorptionselectedGalaxy2025}, who report gas depletion timescales of $\gtrsim0.6$ Gyr for a galaxy at $z = 2.357$ (DLA 1020+2733g), where the star formation rate (SFR) is only weakly constrained. 
Similarly, other $z > 2$ H\,\textsc{i}–selected galaxies, such as DLA1228-113g at $z \approx 2.193$ with $t_{\text{dep}} \approx 1.2$ Gyr \citep{neelemanMolecularEmissionGalaxy2018, kaurMassiveDustyHi2022}, 
DLA 0918+1636g at $z \approx 2.5832$ with $t_{\text{dep}} \approx 0.5$ Gyr \citep{fynboALMAObservationsMetalrich2018, kaurMassiveDustyHi2022}, and DLA 0817+1351g at $z \approx 4.258$ with $t_{\text{dep}} \approx 0.8$ Gyr \citep{neelemanColdMassiveRotating2020}, also exhibit star formation efficiencies lower than those typical of main-sequence galaxies at similar redshifts, which often have depletion times as short as $\sim$0.2 Gyr.
From the DLAs studies by \cite{neelemanII158Mm2025} at $z>4$ and assuming that the [C\,\textsc{ii}] luminosity is tracing the molecular gas reliably in normal star-forming galaxies \citep{zanellaIIEmissionMolecular2018}, we estimated their depletion timescales to range from $t_{\text{dep}} \sim 0.53$ to $1.8$.

{Although some H\,\textsc{i} selected systems show $\sim 1.5$ dex higher molecular gas masses than normal star-forming galaxies, this does not necessarily indicate unusually gas-rich galaxies.
Since these systems are selected based on strong H\,\textsc{i} absorption, a significant atomic gas component is expected (O’Beirne et al., in prep.).} 
Supporting this, \cite{messiasContentCosmicNoon2024} reports that the H\,\textsc{i}-to-H$_2$ mass ratio remains $\sim 1-3$ over a wide redshift range, implying that the molecular gas we observe likely coexists with comparable H\,\textsc{i} reservoirs.
This suggests that {some} H\,\textsc{i}–selected galaxies occupy the upper range of total gas content, with large supplies of gas that are not yet fully ignited for star formation.

\subsection{Implications for molecular gas simulations}

{The diversity in star formation efficiencies observed in H\,\textsc{i}-selected galaxies, with some systems consistent with normal star-forming galaxies and others showing significantly suppressed efficiencies, deviates from simple expectations of standard scaling relations.
This dual behaviour presents new challenges and opportunities for theoretical modelling.
Current cosmological simulations struggle to capture such diversity, particularly in predicting resolved CO emission in galaxies \citep{poppingArtModellingCO2019}.}
While large-volume simulations \citep[e.g., ][]{keatingReproducingCOtoH2Conversion2020, inoueCOUniverseModelling2020, olsenSigameV3Gas2021, pallottiniSurveyHigh$z$Galaxies2022} can reproduce global CO trends using sub-resolutions and post-processed radiative transfer models, they often lack the resolution and detailed physics required to track the formation and distribution of molecular gas self-consistently.
To overcome these limitations, several recent studies have begun incorporating non-equilibrium chemistry solvers into high-resolution simulations of isolated or zoom-in galaxies \citep[e.g.,][]{richingsEffectsMetallicityUV2016, lupiNaturalEmergenceCorrelation2018, lupiPredictingFIRLines2020, richingsEffectsLocalStellar2022, pallottiniSurveyHigh$z$Galaxies2022}, enabling more realistic predictions of the CO-to-H$_2$ abundances and star formation efficiencies across a range of galactic environments.

For example, \cite{thompsonPredictionsCOEmission2024} uses non-equilibrium chemistry with FIRE-2 simulations of isolated disk galaxies that can reproduce the observed CO-H$_2$ and $X_{\mathrm{CO}}$-metallicity relations in the Milky Way and the xCOLD GASS sample. 
Similarly, \cite{richingsEffectsLocalStellar2022} shows that the H\,\textsc{i}-to-H$_2$ transition is sensitive to local stellar radiation and dust depletions.
\cite{lupiNaturalEmergenceCorrelation2018, lupiPredictingFIRLines2020} provide further insights through GIZMO+KROME simulations of isolated galaxies and the post-processing of cosmological boxes with CLOUDY. They demonstrate how local radiation fields and shielding strongly regulate molecular gas content and its link to star formation. 
This supports the interpretation that the CGM–ISM interface in our absorption-selected galaxies may host molecular gas not yet forming stars efficiently, {possibly because local physical conditions such as turbulence, pressure balance, or radiative feedback inhibit star formation}.

On cosmological scales, \cite{maioAtomicMolecularGas2022} uses high-resolution simulations (ColdSim) with time-dependent non-equilibrium chemistry, capturing the evolution of H$_2$ and depletion times from $z\sim 7$ to 2. 
Their models reproduce the observed H\,\textsc{i} and H$_2$ densities and show that even low-metallicity gas can form significant molecular fractions when self-shielded, {which is consistent with the detection of CO in some of our low-metallicity systems}.
Meanwhile, \cite{inoueCOUniverseModelling2020} models molecular gas cloud populations within IllustrisTNG galaxies and computes CO line emission without relying on a fixed $\alpha_{\mathrm{CO}}$.
Their approach reproduces the CO luminosity functions at $z\sim0$ but finds that dwarf galaxies, often CO-dark, contribute non-negligibly to the molecular mass budget, a feature also observed in the metal-poor galaxies in our survey.
Lastly, \cite{olsenSigameV3Gas2021} presents \textsc{sígame3}, which models FIR emission from cosmological simulations with more detailed radiative transfer. Their models generally agree with observations but still overpredict CO luminosities, highlighting how emission-based diagnostics may be biased at low SFRs or in CGM-dominated regimes like those {we observe in our high-$M_{\rm H_2}$, low-SFR systems}.

In this context, the MUSE-ALMA Haloes survey’s empirical constraints on molecular gas content, star formation efficiency, and CGM structure offer a valuable dataset for informing and calibrating future simulations. {Our results reveal a dual behaviour among} absorption-selected galaxies: {some systems exhibit star formation efficiencies comparable to the standard main-sequence of star-forming galaxies, while others remain inefficient at converting their sizeable molecular reservoirs into stars.
This diversity highlights the need for new theoretical efforts to investigate how feedback, gas accretion, and environmental effects regulate the baryon cycle within and around galaxies, and to understand the physical mechanisms that drive galaxies into these distinct regimes.}

\section{Conclusions}\label{sec:conclusions}

Understanding the molecular gas content of galaxies selected via H\,\textsc{i} absorption offers a critical piece in the broader puzzle of galaxy evolution.
In this paper, we present new ALMA observations of CO(2--1), CO(3--2), or CO(4--3) emission from galaxies at $0.3 < z < 1.2$ observed in the ALMA Large Program MUSE-ALMA Haloes survey (Cycle 10, PI: C.Péroux).
These galaxies were initially selected through H\,\textsc{i} absorption along the line of sight of bright optical quasars \citep{wengMUSEALMAHaloesVIII2022}.
Notably, this study represents the first non-metallicity-biased survey of molecular gas in H\,\textsc{i}–selected galaxies, as the targets were selected based solely on neutral hydrogen absorption.

Our main findings are as follows:

\begin{enumerate}
    \item  We report a CO detection rate of {20}\% for our sample, highlighting the heterogeneous nature of H\,\textsc{i}–selected galaxies. {Thanks to the depth of our observations, we have pushed detections of molecular gas by more than 1 dex below the $M_{\rm H_2}$ limits reached in previous H\,\textsc{i}–selected studies.}
    While some absorbers are associated with metal-rich, actively star-forming galaxies with substantial molecular reservoirs, others correspond to gas-rich but {inefficient star-forming} systems, potentially in early evolutionary stages. This diversity reflects the wide range of physical conditions accessible through absorption selection, from chemically enriched disks to diffuse CGM structures.

    \item CO-detected galaxies exhibit gas-phase metallicities consistent with the mass–metallicity relation. {However,} their star formation rates and stellar masses {span a broad range relative to} emission-selected star-forming galaxies at similar redshifts. {Systems with low molecular gas masses tend to align with the scaling relations of normal star-forming galaxies, whereas high-$M_{\rm H_2}$ systems fall significantly below, suggesting distinct evolutionary pathways.  }

    \item The inferred depletion timescales {show a dual behaviour: low-$M_{\rm H_2}$ galaxies exhibit depletion times consistent with main-sequence star-forming galaxies, while high-$M_{\rm H_2}$ galaxies show longer timescales ($\sim 1.5$ dex above), pointing to suppressed star formation efficiency.
    This diversity indicates that absorption selection captures galaxies in both relatively evolved, star-forming phases and in earlier stages where gas reservoirs are not efficiently converted into stars.  } 
    {For the high-$M_{\rm H_2}$ regime, this result} is consistent with prior studies of absorption-selected galaxies \citep[e.g.,][]{neelemanColdMassiveRotating2020, kaurMassiveDustyHi2022}{, which generally pre-selected metal-rich absorbers and thus preferentially traced gas-rich but inefficient systems.}
    {By contrast, the MUSE–ALMA Haloes survey extends this picture to lower-$M_{\rm H_2}$ systems without metallicity preselection, revealing that some absorber-selected galaxies follow the normal depletion times of main-sequence galaxies.}

    \item The properties of H\,\textsc{i}–selected systems (elevated gas fractions, {a wide range of} star formation efficiencies, and agreement with expected metallicities) stress their importance in the baryon cycle. H\,\textsc{i}–selected galaxies may be actively assembling gas from the intergalactic medium or interacting within group environments \citep[e.g., ][]{perouxMultiphaseCircumgalacticMedium2019, hamanowiczMUSEALMAHaloesPhysical2020}, {sometimes reaching a main-sequence-like behaviour, while in other cases remaining inefficient and offset from equilibrium conditions typical of star-forming galaxies.}
    
\end{enumerate}

H\,\textsc{i}–selected systems, often missed in emission-selected surveys, serve as crucial laboratories for studying the interplay between cold gas accretion, feedback, and star formation regulation. By combining absorption-line selection with molecular gas observations, we can access lower-luminosity, lower-SFR systems and build a more complete picture of the processes shaping galaxy evolution.

Interpreting the stringent CO non-detections afforded by these deep data remains challenging: it is unclear whether they result from genuinely low molecular content, low excitation conditions, or suppressed CO emission due to low metallicity. To disentangle these scenarios, it is essential to explore alternative tracers such as neutral carbon ([C\,\textsc{i}]) and dust continuum, which are less sensitive to metallicity and excitation than CO. Such complementary observations will help refine our understanding of the molecular gas content in diffuse, metal-poor, or quiescent systems.

\begin{acknowledgements}
      This research was supported by the International Space Science Institute (ISSI) in Bern, through ISSI International Team project \#564 (The Cosmic Baryon Cycle from Space).

      This paper makes use of the following ALMA data: ADS/JAO.ALMA\#2023.1.00127.L ALMA is a partnership of ESO (representing its member states), NSF (USA) and NINS (Japan), together with NRC (Canada), MOST and ASIAA (Taiwan), and KASI (Republic of Korea), in cooperation with the Republic of Chile. The Joint ALMA Observatory is operated by ESO, AUI/NRAO and NAOJ. 
     
      N.M.F.S., C.B., J.C. acknowledge funding by the European Union (ERC Advanced Grant GALPHYS, 101055023). Views and opinions expressed are, however, those of the author(s) only and do not necessarily reflect those of the European Union or the European Research Council. Neither the European Union nor the granting authority can be held responsible for them.
      R.A. acknowledges funding from the European Research Council (ERC) under the European Union's Horizon 2020 research and innovation programme (grant agreement 101020943, SPECMAP-CGM).
      
\end{acknowledgements}

\section*{Data Availability}

% WARNING
%-------------------------------------------------------------------
% Please note that we have included the references to the file aa.dem in
% order to compile it, but we ask you to:
%
% - use BibTeX with the regular commands:
  \bibliographystyle{aa} % style aa.bst
  \bibliography{aa5653125} % your references Yourfile.bib

@article{accursoDerivingMultivariateACO2017,
  title = {Deriving a Multivariate {{$\alpha$CO}} Conversion Function Using the [{{C II}}]/{{CO}} (1-0) Ratio and Its Application to Molecular Gas Scaling Relations},
  author = {Accurso, G. and Saintonge, A. and Catinella, B. and Cortese, L. and Dav{\'e}, R. and Dunsheath, S. H. and Genzel, R. and {Gracia-Carpio}, J. and Heckman, T. M. and {Jimmy} and Kramer, C. and Li, Cheng and Lutz, K. and Schiminovich, D. and Schuster, K. and Sternberg, A. and Sturm, E. and Tacconi, L. J. and Tran, K. V. and Wang, J.},
  year = 2017,
  month = oct,
  journal = {MNRAS},
  volume = {470},
  pages = {4750--4766},
  publisher = {OUP},
  issn = {0035-8711},
  doi = {10.1093/mnras/stx1556},
  urldate = {2025-05-02}
}

@article{adamsRadioSurveysNow2019,
  title = {Radio Surveys Now Both Deep and Wide},
  author = {Adams, Elizabeth A. K. and {van Leeuwen}, Joeri},
  year = 2019,
  month = feb,
  journal = {Nat Astron},
  volume = {3},
  number = {2},
  pages = {188--188},
  publisher = {Nature Publishing Group},
  issn = {2397-3366},
  doi = {10.1038/s41550-019-0692-4},
  urldate = {2025-06-20},
  copyright = {2019 Springer Nature Limited},
  langid = {english}
}

@article{allisonFLASHEarlyScience2020,
  title = {{{FLASH Early Science}} -- {{Discovery}} of an Intervening {{HI}} 21-Cm Absorber from an {{ASKAP}} Survey of the {{GAMA}} 23 Field},
  author = {Allison, J. R. and Sadler, E. M. and Bellstedt, S. and Davies, L. J. M. and Driver, S. P. and Ellison, S. L. and Huynh, M. and Kapinska, A. D. and Mahony, E. K. and Moss, V. A. and Robotham, A. S. G. and Whiting, M. T. and Curran, S. J. and Darling, J. and Hotan, A. W. and Hunstead, R. W. and Koribalski, B. S. and Lagos, C. D. P. and Pettini, M. and Pimbblet, K. A. and Voronkov, M. A.},
  year = 2020,
  month = may,
  journal = {MNRAS},
  volume = {494},
  number = {3},
  eprint = {2004.00847},
  primaryclass = {astro-ph},
  pages = {3627--3641},
  issn = {0035-8711, 1365-2966},
  doi = {10.1093/mnras/staa949},
  urldate = {2025-06-20},
  archiveprefix = {arXiv},
  langid = {english}
}

@article{arnoutsMeasuringModellingRedshift1999,
  title = {Measuring and Modelling the Redshift Evolution of Clustering: The {{Hubble Deep Field North}}},
  shorttitle = {Measuring and Modelling the Redshift Evolution of Clustering},
  author = {Arnouts, S. and Cristiani, S. and Moscardini, L. and Matarrese, S. and Lucchin, F. and Fontana, A. and Giallongo, E.},
  year = 1999,
  month = dec,
  journal = {MNRAS},
  volume = {310},
  pages = {540--556},
  publisher = {OUP},
  issn = {0035-8711},
  doi = {10.1046/j.1365-8711.1999.02978.x},
  urldate = {2025-05-08}
}

@article{augustinMUSEALMAHaloesStellar2024,
  title = {{{MUSE-ALMA Haloes X}}: {{The}} Stellar Masses of Gas-Rich Absorbing Galaxies},
  shorttitle = {{{MUSE-ALMA Haloes X}}},
  author = {Augustin, Ramona and P{\'e}roux, C{\'e}line and Karki, Arjun and Kulkarni, Varsha and Weng, Simon and Hamanowicz, A. and Hayes, M. and Howk, J. C. and Kacprzak, G. G. and Klitsch, A. and Zwaan, M. A. and Fox, A. and Biggs, A. and Fresco, A. Y. and Kassin, S. and Kuntschner, H.},
  year = 2024,
  month = feb,
  journal = {MNRAS},
  issn = {0035-8711},
  doi = {10.1093/mnras/stae387},
  urldate = {2024-02-23}
}

@inproceedings{baconMUSESecondgenerationVLT2010,
  title = {The {{MUSE}} Second-Generation {{VLT}} Instrument},
  booktitle = {Ground-Based and {{Airborne Instrumentation}} for {{Astronomy III}}},
  author = {Bacon, R. and Accardo, M. and Adjali, L. and Anwand, H. and Bauer, S. and Biswas, I. and Blaizot, J. and Boudon, D. and {Brau-Nogue}, S. and Brinchmann, J. and Caillier, P. and Capoani, L. and Carollo, C. M. and Contini, T. and Couderc, P. and Daguis{\'e}, E. and Deiries, S. and Delabre, B. and Dreizler, S. and Dubois, J. and Dupieux, M. and Dupuy, C. and Emsellem, E. and Fechner, T. and Fleischmann, A. and Fran{\c c}ois, M. and Gallou, G. and Gharsa, T. and Glindemann, A. and Gojak, D. and Guiderdoni, B. and Hansali, G. and Hahn, T. and Jarno, A. and Kelz, A. and Koehler, C. and Kosmalski, J. and Laurent, F. and Le Floch, M. and Lilly, S. J. and Lizon, J. -L. and Loupias, M. and Manescau, A. and Monstein, C. and Nicklas, H. and Olaya, J. -C. and Pares, L. and Pasquini, L. and {P{\'e}contal-Rousset}, A. and Pell{\'o}, R. and Petit, C. and Popow, E. and Reiss, R. and Remillieux, A. and Renault, E. and Roth, M. and Rupprecht, G. and Serre, D. and Schaye, J. and Soucail, G. and Steinmetz, M. and Streicher, O. and Stuik, R. and Valentin, H. and Vernet, J. and Weilbacher, P. and Wisotzki, L. and Yerle, N.},
  year = 2010,
  month = jul,
  volume = {7735},
  pages = {773508},
  address = {eprint: arXiv:2211.16795},
  doi = {10.1117/12.856027},
  urldate = {2025-06-01}
}

@article{beraAtomicHydrogenStarforming2019,
  title = {Atomic {{Hydrogen}} in {{Star-forming Galaxies}} at {{Intermediate Redshifts}}},
  author = {Bera, Apurba and Kanekar, Nissim and Chengalur, Jayaram N. and Bagla, Jasjeet S.},
  year = 2019,
  month = sep,
  journal = {ApJL},
  volume = {882},
  number = {1},
  pages = {L7},
  publisher = {The American Astronomical Society},
  issn = {2041-8205},
  doi = {10.3847/2041-8213/ab3656},
  urldate = {2025-06-20},
  langid = {english}
}

@article{bigielStarFormationLaw2008,
  title = {The {{Star Formation Law}} in {{Nearby Galaxies}} on {{Sub-Kpc Scales}}},
  author = {Bigiel, F. and Leroy, A. and Walter, F. and Brinks, E. and {de Blok}, W. J. G. and Madore, B. and Thornley, M. D.},
  year = 2008,
  month = dec,
  journal = {AJ},
  volume = {136},
  pages = {2846--2871},
  publisher = {IOP},
  issn = {0004-6256},
  doi = {10.1088/0004-6256/136/6/2846},
  urldate = {2025-06-20}
}

@article{blythLADUMA2016MeerKAT2016,
  title = {{{LADUMA}}: 2016 {{MeerKAT Science}}: {{On}} the {{Pathway}} to the {{SKA}}, {{MeerKAT}} 2016},
  shorttitle = {{{LADUMA}}},
  author = {Blyth, Sarah Louise and Baker, Andrew J. and Holwerda, Benne W. and Bassett, Bruce A. and Bershady, Matthew A. and Bouchard, Antoine and Briggs, Frank H. and Catinella, Barbara and Chemin, Laurent and Crawford, Steven M. and Cress, Catherine M. and Cunnama, Daniel and Darling, Jeremy K. and Dav{\'e}, Romeel and Deane, Roger P. and {de Blok}, W. J.G. and Elson, Ed C. and Faltenbacher, Andreas and February, Sean and Fern{\'a}ndez, Ximena and Frank, Bradley S. and Gawiser, Eric and Henning, Patricia A. and Hess, Kelley M. and Heywood, Ian and Hughes, John P. and Jarvis, Matt J. and Kannappan, Sheila J. and Katz, Neal S. and Kere{\v s}, Du{\v s}an and Kl{\"o}ckner, Hans Rainer and {Kraan-Korteweg}, Ren{\'e}e C. and Lah, Philip and Lehnert, Matthew D. and Leroy, Adam K. and Lochner, Michelle and Maddox, Natasha and Makhathini, Sphesihle and Meurer, Gerhardt R. and Meyer, Martin J. and Moodley, Kavilan and Morganti, Raffaella and Obreschkow, Danail and Oh, Se Heon and Oosterloo, Tom A. and Pisano, D. J. and Popping, Attila and Popping, Gerg{\"o} and Ravindranath, Swara and Schinnerer, Eva and Schr{\"o}der, Anja C. and Sheth, Kartik and Skelton, Rosalind and Smirnov, Oleg M. and Smith, Mathew and Somerville, Rachel S. and Srianand, Raghunathan and {Staveley-Smith}, Lister and Stewart, Ian M. and Vaccari, Mattia and V{\"a}is{\"a}nen, Petri and {van der Heyden}, Kurt J. and {van Driel}, Wim and Verheijen, Marc A.W. and Walter, Fabian and Wilcots, Eric M. and Williams, Theodore B. and Woudt, Patrick A. and Wu, John F. and Zwaan, Martin A. and Zwart, Jonathan T.L. and Rawlings, Steve},
  year = 2016,
  month = jan,
  journal = {Conf.},
  issn = {1824-8039},
  urldate = {2025-06-20}
}

@article{boettcherCosmicUltravioletBaryon2021,
  title = {The {{Cosmic Ultraviolet Baryon Survey}} ({{CUBS}}). {{II}}. {{Discovery}} of an {{H2-bearing DLA}} in the {{Vicinity}} of an {{Early-type Galaxy}} at z = 0.576*},
  author = {Boettcher, Erin and Chen, Hsiao-Wen and Zahedy, Fakhri S. and Cooper, Thomas J. and Johnson, Sean D. and Rudie, Gwen C. and Chen, Mandy C. and Petitjean, Patrick and Cantalupo, Sebastiano and Cooksey, Kathy L. and {Faucher-Gigu{\`e}re}, Claude-Andr{\'e} and Greene, Jenny E. and Lopez, Sebastian and Mulchaey, John S. and Penton, Steven V. and Putman, Mary E. and Rafelski, Marc and Rauch, Michael and Schaye, Joop and Simcoe, Robert A. and Walth, Gregory L.},
  year = 2021,
  month = may,
  journal = {ApJ},
  volume = {913},
  number = {1},
  pages = {18},
  publisher = {The American Astronomical Society},
  issn = {0004-637X},
  doi = {10.3847/1538-4357/abf0a0},
  urldate = {2025-06-01},
  langid = {english}
}

@article{bogdanHOTXRAYCORONAE2013,
  title = {{{HOT X-RAY CORONAE AROUND MASSIVE SPIRAL GALAXIES}}: {{A UNIQUE PROBE OF STRUCTURE FORMATION MODELS}}},
  shorttitle = {{{HOT X-RAY CORONAE AROUND MASSIVE SPIRAL GALAXIES}}},
  author = {Bogd{\'a}n, {\'A}kos and Forman, William R. and Vogelsberger, Mark and Bourdin, Herv{\'e} and Sijacki, Debora and Mazzotta, Pasquale and Kraft, Ralph P. and Jones, Christine and Gilfanov, Marat and Churazov, Eugene and David, Laurence P.},
  year = 2013,
  month = jul,
  journal = {ApJ},
  volume = {772},
  number = {2},
  pages = {97},
  publisher = {The American Astronomical Society},
  issn = {0004-637X},
  doi = {10.1088/0004-637X/772/2/97},
  urldate = {2025-06-01},
  langid = {english}
}

@article{bolattoCOtoH2ConversionFactor2013,
  title = {The {{CO-to-H2 Conversion Factor}}},
  author = {Bolatto, Alberto D. and Wolfire, Mark and Leroy, Adam K.},
  year = 2013,
  journal = {Annu. Rev. Astron. Astrophys.},
  volume = {51},
  number = {1},
  pages = {207--268},
  doi = {10.1146/annurev-astro-082812-140944},
  urldate = {2022-09-28}
}

@article{bolloALMACALXIIData2024,
  title = {{{ALMACAL}} - {{XII}}. {{Data}} Characterisation and Products},
  author = {Bollo, Victoria and Zwaan, Martin and P{\'e}roux, C{\'e}line and Hamanowicz, Aleksandra and Chen, Jianhang and Weng, Simon and Ivison, Rob J. and Biggs, Andrew},
  year = 2024,
  month = oct,
  journal = {A\&A},
  volume = {690},
  pages = {A258},
  publisher = {EDP Sciences},
  issn = {0004-6361, 1432-0746},
  doi = {10.1051/0004-6361/202450336},
  urldate = {2025-12-10},
  copyright = {\copyright{} The Authors 2024},
  langid = {english}
}

@article{bolloALMACALXIIIEvolution2025,
  title = {{{ALMACAL}}: {{XIII}}. {{Evolution}} of the {{CO}} Luminosity Function and the Molecular Gas Mass Density out to {\emph{z}} {$\sim$} 6},
  shorttitle = {{{ALMACAL}}},
  author = {Bollo, Victoria and P{\'e}roux, C{\'e}line and Zwaan, Martin and Hamanowicz, Aleksandra and Chen, Jianhang and Weng, Simon and Del P. Lagos, Claudia and Bravo, Mat{\'i}as and Ivison, Rob J. and Biggs, Andrew},
  year = 2025,
  month = mar,
  journal = {A\&A},
  volume = {695},
  pages = {A163},
  issn = {0004-6361, 1432-0746},
  doi = {10.1051/0004-6361/202453223},
  urldate = {2025-05-30},
  copyright = {https://creativecommons.org/licenses/by/4.0},
  langid = {english}
}

@article{boucheImpactColdGas2010,
  title = {The {{Impact}} of {{Cold Gas Accretion Above}} a {{Mass Floor}} on {{Galaxy Scaling Relations}}},
  author = {Bouch{\'e}, N. and Dekel, A. and Genzel, R. and Genel, S. and Cresci, G. and F{\"o}rster Schreiber, N. M. and Shapiro, K. L. and Davies, R. I. and Tacconi, L.},
  year = 2010,
  month = aug,
  journal = {ApJ},
  volume = {718},
  pages = {1001--1018},
  publisher = {IOP},
  issn = {0004-637X},
  doi = {10.1088/0004-637X/718/2/1001},
  urldate = {2024-09-20}
}

@article{bouchePossibleSignaturesColdflow2016,
  title = {Possible {{Signatures}} of a {{Cold-flow Disk}} from {{MUSE Using}} a z {$\sim$} 1 {{Galaxy-Quasar Pair}} toward {{SDSS J1422-0001}}},
  author = {Bouch{\'e}, N. and Finley, H. and Schroetter, I. and Murphy, M. T. and Richter, P. and Bacon, R. and Contini, T. and Richard, J. and Wendt, M. and Kamann, S. and Epinat, B. and Cantalupo, S. and Straka, L. A. and Schaye, J. and Martin, C. L. and P{\'e}roux, C. and Wisotzki, L. and Soto, K. and Lilly, S. and Carollo, C. M. and Brinchmann, J. and Kollatschny, W.},
  year = 2016,
  month = apr,
  journal = {ApJ},
  volume = {820},
  pages = {121},
  publisher = {IOP},
  issn = {0004-637X},
  doi = {10.3847/0004-637X/820/2/121},
  urldate = {2025-06-24}
}

@article{bruzualStellarPopulationSynthesis2003,
  title = {Stellar Population Synthesis at the Resolution of 2003},
  author = {Bruzual, G. and Charlot, S.},
  year = 2003,
  month = oct,
  journal = {MNRAS},
  volume = {344},
  pages = {1000--1028},
  issn = {0035-8711},
  doi = {10.1046/j.1365-8711.2003.06897.x},
  urldate = {2021-04-21}
}

@article{calzettiDustExtinctionStellar1994,
  title = {Dust {{Extinction}} of the {{Stellar Continua}} in {{Starburst Galaxies}}: {{The Ultraviolet}} and {{Optical Extinction Law}}},
  shorttitle = {Dust {{Extinction}} of the {{Stellar Continua}} in {{Starburst Galaxies}}},
  author = {Calzetti, Daniela and Kinney, Anne L. and {Storchi-Bergmann}, Thaisa},
  year = 1994,
  month = jul,
  journal = {ApJ},
  volume = {429},
  pages = {582},
  publisher = {IOP},
  issn = {0004-637X},
  doi = {10.1086/174346},
  urldate = {2025-05-09}
}

@article{carilliCoolGasHighRedshift2013,
  title = {Cool {{Gas}} in {{High-Redshift Galaxies}}},
  author = {Carilli, C. L. and Walter, F.},
  year = 2013,
  month = aug,
  journal = {Annu. Rev. Astron. Astrophys.},
  volume = {51},
  number = {1},
  pages = {105--161},
  issn = {0066-4146},
  doi = {10.1146/annurev-astro-082812-140953},
  urldate = {2022-10-29},
  langid = {english}
}

@article{catinellaXGASSTotalCold2018,
  title = {{{xGASS}}: Total Cold Gas Scaling Relations and Molecular-to-Atomic Gas Ratios of Galaxies in the Local {{Universe}}},
  shorttitle = {{{xGASS}}},
  author = {Catinella, Barbara and Saintonge, Am{\'e}lie and Janowiecki, Steven and Cortese, Luca and Dav{\'e}, Romeel and Lemonias, Jenna J. and Cooper, Andrew P. and Schiminovich, David and Hummels, Cameron B. and Fabello, Silvia and Ger{\'e}b, Katinka and Kilborn, Virginia and Wang, Jing},
  year = 2018,
  month = may,
  journal = {MNRAS},
  volume = {476},
  pages = {875--895},
  publisher = {OUP},
  issn = {0035-8711},
  doi = {10.1093/mnras/sty089},
  urldate = {2025-01-21}
}

@article{chabrierGalacticStellarSubstellar2003,
  title = {Galactic {{Stellar}} and {{Substellar Initial Mass Function}}},
  author = {Chabrier, Gilles},
  year = 2003,
  month = jul,
  journal = {PASP},
  volume = {115},
  pages = {763--795},
  issn = {0004-6280},
  doi = {10.1086/376392},
  urldate = {2021-04-21}
}

@article{chowdhury21centimetreEmissionEnsemble2020,
  title = {H~i 21-Centimetre Emission from an Ensemble of Galaxies at an Average Redshift of One},
  author = {Chowdhury, Aditya and Kanekar, Nissim and Chengalur, Jayaram N. and Sethi, Shiv and Dwarakanath, K. S.},
  year = 2020,
  month = oct,
  journal = {Nat Astron},
  volume = {586},
  number = {7829},
  pages = {369--372},
  publisher = {Nature Publishing Group},
  issn = {1476-4687},
  doi = {10.1038/s41586-020-2794-7},
  urldate = {2025-06-20},
  copyright = {2020 The Author(s), under exclusive licence to Springer Nature Limited},
  langid = {english}
}

@article{cooperCosmicUltravioletBaryon2021,
  title = {The {{Cosmic Ultraviolet Baryon Survey}} ({{CUBS}}) -- {{IV}}. {{The}} Complex Multiphase Circumgalactic Medium as Revealed by Partial {{Lyman}} Limit Systems},
  author = {Cooper, Thomas J and Rudie, Gwen C and Chen, Hsiao-Wen and Johnson, Sean D and Zahedy, Fakhri S and Chen, Mandy C and Boettcher, Erin and Walth, Gregory L and Cantalupo, Sebastiano and Cooksey, Kathy L and {Faucher-Gigu{\`e}re}, Claude-Andr{\'e} and Greene, Jenny E and Lopez, Sebastian and Mulchaey, John S and Penton, Steven V and Petitjean, Patrick and Putman, Mary E and Rafelski, Marc and Rauch, Michael and Schaye, Joop and Simcoe, Robert A},
  year = 2021,
  month = dec,
  journal = {MNRAS},
  volume = {508},
  number = {3},
  pages = {4359--4384},
  issn = {0035-8711},
  doi = {10.1093/mnras/stab2869},
  urldate = {2025-06-01}
}

@article{curtiMassMetallicityFundamental2020,
  title = {The Mass--Metallicity and the Fundamental Metallicity Relation Revisited on a Fully {{Te-based}} Abundance Scale for Galaxies},
  author = {Curti, Mirko and Mannucci, Filippo and Cresci, Giovanni and Maiolino, Roberto},
  year = 2020,
  month = jan,
  journal = {MNRAS},
  volume = {491},
  number = {1},
  pages = {944--964},
  issn = {0035-8711},
  doi = {10.1093/mnras/stz2910},
  urldate = {2025-03-13}
}

@article{curtiNewFullyEmpirical2017,
  title = {New Fully Empirical Calibrations of Strong-Line Metallicity Indicators in Star-Forming Galaxies},
  author = {Curti, M. and Cresci, G. and Mannucci, F. and Marconi, A. and Maiolino, R. and Esposito, S.},
  year = 2017,
  month = feb,
  journal = {MNRAS},
  volume = {465},
  number = {2},
  pages = {1384--1400},
  issn = {0035-8711},
  doi = {10.1093/mnras/stw2766},
  urldate = {2025-03-13}
}

@misc{dekaMeerKATAbsorptionLine2023,
  title = {The {{MeerKAT Absorption Line Survey}} ({{MALS}}) Data Release {{I}}: {{Stokes I}} Image Catalogs at 1-1.4 {{GHz}}},
  shorttitle = {The {{MeerKAT Absorption Line Survey}} ({{MALS}}) Data Release {{I}}},
  author = {Deka, P. P. and Gupta, N. and Jagannathan, P. and Sekhar, S. and Momjian, E. and Bhatnagar, S. and Wagenveld, J. and Kl{\"o}ckner, H.-R. and Jose, J. and Balashev, S. A. and Combes, F. and Hilton, M. and Borgaonkar, D. and Chatterjee, A. and Emig, K. L. and Gaunekar, A. N. and J{\'o}zsa, G. I. G. and Klutse, D. Y. and Knowles, K. and Krogager, J.- K. and Mohapatra, A. and Moodley, K. and Muller, S{\'e}bastien and Noterdaeme, P. and Petitjean, P. and Salas, P. and Sikhosana, S.},
  year = 2023,
  month = aug,
  number = {arXiv:2308.12347},
  eprint = {2308.12347},
  primaryclass = {astro-ph},
  publisher = {arXiv},
  doi = {10.48550/arXiv.2308.12347},
  urldate = {2025-06-20},
  archiveprefix = {arXiv}
}

@article{dessauges-zavadskyNewComprehensiveSet2007,
  title = {A New Comprehensive Set of Elemental Abundances in {{DLAs}} - {{III}}. {{Star}} Formation Histories},
  author = {{Dessauges-Zavadsky}, M. and Calura, F. and Prochaska, J. X. and D'Odorico, S. and Matteucci, F.},
  year = 2007,
  month = aug,
  journal = {A\&A},
  volume = {470},
  number = {2},
  pages = {431--448},
  publisher = {EDP Sciences},
  issn = {0004-6361, 1432-0746},
  doi = {10.1051/0004-6361:20077050},
  urldate = {2025-06-01},
  copyright = {\copyright{} ESO, 2007},
  langid = {english}
}

@article{diemerAtomicMolecularGas2019,
  title = {Atomic and Molecular Gas in {{IllustrisTNG}} Galaxies at Low Redshift},
  author = {Diemer, Benedikt and Stevens, Adam R. H. and Lagos, Claudia del P. and Calette, A. R. and Tacchella, Sandro and Hernquist, Lars and Marinacci, Federico and Nelson, Dylan and Pillepich, Annalisa and {Rodriguez-Gomez}, Vicente and {Villaescusa-Navarro}, Francisco and Vogelsberger, Mark},
  year = 2019,
  month = aug,
  journal = {MNRAS},
  volume = {487},
  number = {2},
  eprint = {1902.10714},
  primaryclass = {astro-ph},
  pages = {1529--1550},
  issn = {0035-8711, 1365-2966},
  doi = {10.1093/mnras/stz1323},
  urldate = {2024-09-30},
  archiveprefix = {arXiv}
}

@article{duttaMUSEAnalysisGas2020,
  title = {{{MUSE Analysis}} of {{Gas}} around {{Galaxies}} ({{MAGG}}) -- {{II}}: Metal-Enriched Halo Gas around z~{$\sim$} 1 Galaxies},
  shorttitle = {{{MUSE Analysis}} of {{Gas}} around {{Galaxies}} ({{MAGG}}) -- {{II}}},
  author = {Dutta, Rajeshwari and Fumagalli, Michele and Fossati, Matteo and Lofthouse, Emma K and Prochaska, J Xavier and Arrigoni~Battaia, Fabrizio and Bielby, Richard M and Cantalupo, Sebastiano and Cooke, Ryan J and Murphy, Michael T and O'Meara, John M},
  year = 2020,
  month = nov,
  journal = {MNRAS},
  volume = {499},
  number = {4},
  pages = {5022--5046},
  issn = {0035-8711},
  doi = {10.1093/mnras/staa3147},
  urldate = {2025-06-01}
}

@article{duttaMUSEQuBESKinematicsVibearing2025,
  title = {{{MUSEQuBES}}: {{The Kinematics}} of {{O}} vi-Bearing {{Gas}} in and around {{Low-redshift Galaxies}}},
  shorttitle = {{{MUSEQuBES}}},
  author = {Dutta, Sayak and Muzahid, Sowgat and Schaye, Joop and Cantalupo, Sebastiano and Chen, Hsiao-Wen and Johnson, Sean},
  year = 2025,
  month = feb,
  journal = {ApJ},
  volume = {980},
  number = {2},
  pages = {264},
  publisher = {The American Astronomical Society},
  issn = {0004-637X},
  doi = {10.3847/1538-4357/adabbd},
  urldate = {2025-06-01},
  langid = {english}
}

@misc{duttaMUSEQuBESMappingDistribution2024,
  title = {{{MUSEQuBES}}: {{Mapping}} the Distribution of Neutral Hydrogen around Low-Redshift Galaxies},
  shorttitle = {{{MUSEQuBES}}},
  author = {Dutta, Sayak and Muzahid, Sowgat and Schaye, Joop and Mishra, Sapna and Chen, Hsiao-Wen and Johnson, Sean and Wisotzki, Lutz and Cantalupo, Sebastiano},
  year = 2024,
  month = jan,
  number = {arXiv:2303.16933},
  eprint = {2303.16933},
  primaryclass = {astro-ph},
  publisher = {arXiv},
  doi = {10.48550/arXiv.2303.16933},
  urldate = {2025-06-01},
  archiveprefix = {arXiv}
}

@article{erbHaObservationsLarge2006,
  title = {H{$\alpha$} {{Observations}} of a {{Large Sample}} of {{Galaxies}} at z \textasciitilde{} 2: {{Implications}} for {{Star Formation}} in {{High-Redshift Galaxies}}*},
  shorttitle = {H{$\alpha$} {{Observations}} of a {{Large Sample}} of {{Galaxies}} at z \textasciitilde{} 2},
  author = {Erb, Dawn K. and Steidel, Charles C. and Shapley, Alice E. and Pettini, Max and Reddy, Naveen A. and Adelberger, Kurt L.},
  year = 2006,
  month = aug,
  journal = {ApJ},
  volume = {647},
  number = {1},
  pages = {128},
  publisher = {IOP Publishing},
  issn = {0004-637X},
  doi = {10.1086/505341},
  urldate = {2022-05-02},
  langid = {english}
}

@article{faucher-giguereKeyPhysicalProcesses2023,
  title = {Key {{Physical Processes}} in the {{Circumgalactic Medium}}},
  author = {{Faucher-Gigu{\`e}re}, Claude-Andr{\'e} and Oh, S. Peng},
  year = 2023,
  month = aug,
  journal = {Annu. Rev. Astron. Astrophys.},
  volume = {61},
  pages = {131--195},
  issn = {0066-4146},
  doi = {10.1146/annurev-astro-052920-125203},
  urldate = {2024-07-04}
}

@article{fernandezHIGHESTREDSHIFTIMAGE2016,
  title = {{{HIGHEST REDSHIFT IMAGE OF NEUTRAL HYDROGEN IN EMISSION}}: {{A CHILES DETECTION OF A STARBURSTING GALAXY AT}} z = 0.376},
  shorttitle = {{{HIGHEST REDSHIFT IMAGE OF NEUTRAL HYDROGEN IN EMISSION}}},
  author = {Fern{\'a}ndez, Ximena and Gim, Hansung B. and van Gorkom, J. H. and Yun, Min S. and Momjian, Emmanuel and Popping, Attila and Chomiuk, Laura and Hess, Kelley M. and Hunt, Lucas and Kreckel, Kathryn and Lucero, Danielle and Maddox, Natasha and Oosterloo, Tom and Pisano, D. J. and Verheijen, M. A. W. and Hales, Christopher A. and Chung, Aeree and Dodson, Richard and Golap, Kumar and Gross, Julia and Henning, Patricia and Hibbard, John and Jaff{\'e}, Yara L. and Meyer, Jennifer Donovan and Meyer, Martin and {Sanchez-Barrantes}, Monica and Schiminovich, David and Wicenec, Andreas and Wilcots, Eric and Bershady, Matthew and Scoville, Nick and Strader, Jay and Tremou, Evangelia and Salinas, Ricardo and Ch{\'a}vez, Ricardo},
  year = 2016,
  month = jun,
  journal = {ApJL},
  volume = {824},
  number = {1},
  pages = {L1},
  publisher = {The American Astronomical Society},
  issn = {2041-8205},
  doi = {10.3847/2041-8205/824/1/L1},
  urldate = {2025-06-20},
  langid = {english}
}

@article{fixsenCOBEFarInfrared1999,
  title = {{{COBE Far Infrared Absolute Spectrophotometer Observations}} of {{Galactic Lines}}},
  author = {Fixsen, D. J. and Bennett, C. L. and Mather, J. C.},
  year = 1999,
  month = nov,
  journal = {ApJ},
  volume = {526},
  pages = {207--214},
  issn = {0004-637X},
  doi = {10.1086/307962},
  urldate = {2024-04-12}
}

@article{forbesOriginFundamentalMetallicity2014,
  title = {On the Origin of the Fundamental Metallicity Relation and the Scatter in Galaxy Scaling Relations},
  author = {Forbes, John C. and Krumholz, Mark R. and Burkert, Andreas and Dekel, Avishai},
  year = 2014,
  month = sep,
  journal = {MNRAS},
  volume = {443},
  number = {1},
  pages = {168--185},
  issn = {0035-8711},
  doi = {10.1093/mnras/stu1142},
  urldate = {2025-05-30}
}

@article{fossatiMUSEAnalysisGas2021,
  title = {{{MUSE}} Analysis of Gas around Galaxies ({{MAGG}}) -- {{III}}. {{The}} Gas and Galaxy Environment of z = 3--4.5 Quasars},
  author = {Fossati, M and Fumagalli, M and Lofthouse, E K and Dutta, R and Cantalupo, S and Arrigoni~Battaia, F and Fynbo, J P U and Lusso, E and Murphy, M T and Prochaska, J X and Theuns, T and Cooke, R J},
  year = 2021,
  month = may,
  journal = {MNRAS},
  volume = {503},
  number = {2},
  pages = {3044--3064},
  issn = {0035-8711},
  doi = {10.1093/mnras/stab660},
  urldate = {2025-06-01}
}

@article{frankObservableSignaturesLowz2012,
  title = {Observable Signatures of the Low-z Circumgalactic and Intergalactic Media: Ultraviolet Line Emission in Simulations},
  shorttitle = {Observable Signatures of the Low-z Circumgalactic and Intergalactic Media},
  author = {Frank, S. and Rasera, Y. and Vibert, D. and Milliard, B. and Popping, A. and Blaizot, J. and Courty, S. and Deharveng, J.-M. and P{\'e}roux, C. and Teyssier, R. and Martin, C. D.},
  year = 2012,
  month = feb,
  journal = {MNRAS},
  volume = {420},
  number = {2},
  pages = {1731--1753},
  issn = {0035-8711},
  doi = {10.1111/j.1365-2966.2011.20172.x},
  urldate = {2025-06-01}
}

@article{fynboALMAObservationsMetalrich2018,
  title = {{{ALMA}} Observations of a Metal-Rich Damped {{Ly}} {$\alpha$} Absorber at z = 2.5832: Evidence for Strong Galactic Winds in a Galaxy Group},
  shorttitle = {{{ALMA}} Observations of a Metal-Rich Damped {{Ly}} {$\alpha$} Absorber at z = 2.5832},
  author = {Fynbo, J. P. U. and Heintz, K. E. and Neeleman, M. and Christensen, L. and {Dessauges-Zavadsky}, M. and Kanekar, N. and M{\o}ller, P. and Prochaska, J. X. and Rhodin, N. H. P. and Zwaan, M.},
  year = 2018,
  month = sep,
  journal = {MNRAS},
  volume = {479},
  pages = {2126--2132},
  publisher = {OUP},
  issn = {0035-8711},
  doi = {10.1093/mnras/sty1520},
  urldate = {2025-05-30}
}

@article{genzelCombinedCODust2015,
  title = {Combined {{CO}} \& {{Dust Scaling Relations}} of {{Depletion Time}} and {{Molecular Gas Fractions}} with {{Cosmic Time}}, {{Specific Star Formation Rate}} and {{Stellar Mass}}},
  author = {Genzel, R. and Tacconi, L. J. and Lutz, D. and Saintonge, A. and Berta, S. and Magnelli, B. and Combes, F. and {Garc{\'i}a-Burillo}, S. and Neri, R. and Bolatto, A. and Contini, T. and Lilly, S. and Boissier, J. and Boone, F. and Bouch{\'e}, N. and Bournaud, F. and Burkert, A. and Carollo, M. and Colina, L. and Cooper, M. C. and Cox, P. and Feruglio, C. and Schreiber, N. M. F{\"o}rster and Freundlich, J. and {Gracia-Carpio}, J. and Juneau, S. and Kovac, K. and Lippa, M. and Naab, T. and Salome, P. and Renzini, A. and Sternberg, A. and Walter, F. and Weiner, B. and Weiss, A. and Wuyts, S.},
  year = 2015,
  month = feb,
  journal = {ApJ},
  volume = {800},
  number = {1},
  eprint = {1409.1171},
  primaryclass = {astro-ph},
  pages = {20},
  issn = {1538-4357},
  doi = {10.1088/0004-637X/800/1/20},
  urldate = {2025-01-31},
  archiveprefix = {arXiv}
}

@article{genzelMETALLICITYDEPENDENCECO2012,
  title = {{{THE METALLICITY DEPENDENCE OF THE CO}} {$\rightarrow$} {{H2 CONVERSION FACTOR IN}} z ⩾ 1 {{STAR-FORMING GALAXIES}}*},
  author = {Genzel, R. and Tacconi, L. J. and Combes, F. and Bolatto, A. and Neri, R. and Sternberg, A. and Cooper, M. C. and Bouch{\'e}, N. and Bournaud, F. and Burkert, A. and Comerford, J. and Cox, P. and Davis, M. and Schreiber, N. M. F{\"o}rster and {Garcia-Burillo}, S. and {Gracia-Carpio}, J. and Lutz, D. and Naab, T. and Newman, S. and Saintonge, A. and Shapiro, K. and Shapley, A. and Weiner, B.},
  year = 2012,
  month = jan,
  journal = {ApJ},
  volume = {746},
  number = {1},
  pages = {69},
  publisher = {The American Astronomical Society},
  issn = {0004-637X},
  doi = {10.1088/0004-637X/746/1/69},
  urldate = {2025-04-29},
  langid = {english}
}

@article{gloverSimulatingFormationMolecular2007,
  title = {Simulating the {{Formation}} of {{Molecular Clouds}}. {{II}}. {{Rapid Formation}} from {{Turbulent Initial Conditions}}},
  author = {Glover, Simon C. O. and Mac Low, Mordecai-Mark},
  year = 2007,
  month = apr,
  journal = {ApJ},
  volume = {659},
  number = {2},
  pages = {1317},
  issn = {0004-637X},
  doi = {10.1086/512227},
  urldate = {2025-06-01},
  langid = {english}
}

@article{glowackiLookingDistantUniverse2022,
  title = {Looking at the {{Distant Universe}} with the {{MeerKAT Array}}: {{Discovery}} of a {{Luminous OH Megamaser}} at z {$>$} 0.5},
  shorttitle = {Looking at the {{Distant Universe}} with the {{MeerKAT Array}}},
  author = {Glowacki, Marcin and Collier, Jordan D. and {Kazemi-Moridani}, Amir and Frank, Bradley and Roberts, Hayley and Darling, Jeremy and Kl{\"o}ckner, Hans-Rainer and Adams, Nathan and Baker, Andrew J. and Bershady, Matthew and Blecher, Tariq and Blyth, Sarah-Louise and Bowler, Rebecca and Catinella, Barbara and Chemin, Laurent and Crawford, Steven M. and Cress, Catherine and Dav{\'e}, Romeel and Deane, Roger and {de Blok}, Erwin and Delhaize, Jacinta and Duncan, Kenneth and Elson, Ed and February, Sean and Gawiser, Eric and Hatfield, Peter and Healy, Julia and Henning, Patricia and Hess, Kelley M. and Heywood, Ian and Holwerda, Benne W. and Hoosain, Munira and Hughes, John P. and Hutchens, Zackary L. and Jarvis, Matt and Kannappan, Sheila and Katz, Neal and Kere{\v s}, Du{\v s}an and Korsaga, Marie and {Kraan-Korteweg}, Ren{\'e}e C. and Lah, Philip and Lochner, Michelle and Maddox, Natasha and Makhathini, Sphesihle and Meurer, Gerhardt R. and Meyer, Martin and Obreschkow, Danail and Oh, Se-Heon and Oosterloo, Tom and Oppor, Joshua and Pan, Hengxing and Pisano, D. J. and Randriamiarinarivo, Nandrianina and Ravindranath, Swara and Schr{\"o}der, Anja C. and Skelton, Rosalind and Smirnov, Oleg and Smith, Mathew and Somerville, Rachel S. and Srianand, Raghunathan and {Staveley-Smith}, Lister and Tanaka, Masayuki and Vaccari, Mattia and {van Driel}, Wim and Verheijen, Marc and Walter, Fabian and Wu, John F. and Zwaan, Martin A.},
  year = 2022,
  month = may,
  journal = {ApJL},
  volume = {931},
  number = {1},
  pages = {L7},
  publisher = {The American Astronomical Society},
  issn = {2041-8205},
  doi = {10.3847/2041-8213/ac63b0},
  urldate = {2025-06-20},
  langid = {english}
}

@article{hamanowiczMUSEALMAHaloesPhysical2020,
  title = {{{MUSE-ALMA}} Haloes {{V}}: Physical Properties and Environment of z {$\leq$} 1.4 {{H I}} Quasar Absorbers},
  shorttitle = {{{MUSE-ALMA}} Haloes {{V}}},
  author = {Hamanowicz, Aleksandra and P{\'e}roux, C{\'e}line and Zwaan, Martin A. and Rahmani, Hadi and Pettini, Max and York, Donald G. and Klitsch, Anne and Augustin, Ramona and Krogager, Jens-Kristian and Kulkarni, Varsha and Fresco, Alejandra and Biggs, Andrew D. and Milliard, Bruno and Vernet, Jo{\"e}l D. R.},
  year = 2020,
  month = feb,
  journal = {MNRAS},
  volume = {492},
  pages = {2347--2368},
  issn = {0035-8711},
  doi = {10.1093/mnras/stz3590},
  urldate = {2024-02-23}
}

@article{haynesNeutralHydrogenIsolated1984,
  title = {Neutral Hydrogen in Isolated Galaxies. {{IV}}. {{Results}} for the {{Arecibo}} Sample.},
  author = {Haynes, M. P. and Giovanelli, R.},
  year = 1984,
  month = jun,
  journal = {AJ},
  volume = {89},
  pages = {758--800},
  publisher = {IOP},
  issn = {0004-6256},
  doi = {10.1086/113573},
  urldate = {2025-06-20}
}

@article{hotanAustralianSquareKilometre2021,
  title = {Australian Square Kilometre Array Pathfinder: {{I}}. System Description},
  shorttitle = {Australian Square Kilometre Array Pathfinder},
  author = {Hotan, A. W. and Bunton, J. D. and Chippendale, A. P. and Whiting, M. and Tuthill, J. and Moss, V. A. and McConnell, D. and Amy, S. W. and Huynh, M. T. and Allison, J. R. and Anderson, C. S. and Bannister, K. W. and Bastholm, E. and Beresford, R. and Bock, D. C.-J. and Bolton, R. and Chapman, J. M. and Chow, K. and Collier, J. D. and Cooray, F. R. and Cornwell, T. J. and Diamond, P. J. and Edwards, P. G. and Feain, I. J. and Franzen, T. M. O. and George, D. and Gupta, N. and Hampson, G. A. and {Harvey-Smith}, L. and Hayman, D. B. and Heywood, I. and Jacka, C. and Jackson, C. A. and Jackson, S. and Jeganathan, K. and Johnston, S. and Kesteven, M. and Kleiner, D. and Koribalski, B. S. and {Lee-Waddell}, K. and Lenc, E. and Lensson, E. S. and Mackay, S. and Mahony, E. K. and {McClure-Griffiths}, N. M. and McConigley, R. and Mirtschin, P. and Ng, A. K. and Norris, R. P. and Pearce, S. E. and Phillips, C. and Pilawa, M. A. and Raja, W. and Reynolds, J. E. and Roberts, P. and Roxby, D. N. and Sadler, E. M. and Shields, M. and Schinckel, A. E. T. and Serra, P. and Shaw, R. D. and Sweetnam, T. and Troup, E. R. and Tzioumis, A. and Voronkov, M. A. and Westmeier, T.},
  year = 2021,
  month = jan,
  journal = {Publ. Astron. Soc. Aust.},
  volume = {38},
  pages = {e009},
  issn = {1323-3580, 1448-6083},
  doi = {10.1017/pasa.2021.1},
  urldate = {2025-06-02},
  langid = {english}
}

@article{ilbertAccuratePhotometricRedshifts2006,
  title = {Accurate Photometric Redshifts for the {{CFHT}} Legacy Survey Calibrated Using the {{VIMOS VLT}} Deep Survey},
  author = {Ilbert, O. and Arnouts, S. and McCracken, H. J. and Bolzonella, M. and Bertin, E. and Le F{\`e}vre, O. and Mellier, Y. and Zamorani, G. and Pell{\`o}, R. and Iovino, A. and Tresse, L. and Le Brun, V. and Bottini, D. and Garilli, B. and Maccagni, D. and Picat, J. P. and Scaramella, R. and Scodeggio, M. and Vettolani, G. and Zanichelli, A. and Adami, C. and Bardelli, S. and Cappi, A. and Charlot, S. and Ciliegi, P. and Contini, T. and Cucciati, O. and Foucaud, S. and Franzetti, P. and Gavignaud, I. and Guzzo, L. and Marano, B. and Marinoni, C. and Mazure, A. and Meneux, B. and Merighi, R. and Paltani, S. and Pollo, A. and Pozzetti, L. and Radovich, M. and Zucca, E. and Bondi, M. and Bongiorno, A. and Busarello, G. and {de La Torre}, S. and Gregorini, L. and Lamareille, F. and Mathez, G. and Merluzzi, P. and Ripepi, V. and Rizzo, D. and Vergani, D.},
  year = 2006,
  month = oct,
  journal = {A\&A},
  volume = {457},
  pages = {841--856},
  issn = {0004-6361},
  doi = {10.1051/0004-6361:20065138},
  urldate = {2025-05-08}
}

@article{inoueCOUniverseModelling2020,
  title = {The {{CO}} Universe: Modelling {{CO}} Emission and {{H2}} Abundance in Cosmological Galaxy Formation Simulations},
  shorttitle = {The {{CO}} Universe},
  author = {Inoue, Shigeki and Yoshida, Naoki and Yajima, Hidenobu},
  year = 2020,
  month = nov,
  journal = {MNRAS},
  volume = {498},
  pages = {5960--5971},
  publisher = {OUP},
  issn = {0035-8711},
  doi = {10.1093/mnras/staa2744},
  urldate = {2024-09-23}
}

@article{kacprzakMorphologicalProperties052011,
  title = {Morphological Properties of Z{$\sim$} 0.5 Absorption-Selected Galaxies: The Role of Galaxy Inclination},
  shorttitle = {Morphological Properties of Z{$\sim$} 0.5 Absorption-Selected Galaxies},
  author = {Kacprzak, Glenn G. and Churchill, Christopher W. and Evans, Jessica L. and Murphy, Michael T. and Steidel, Charles C.},
  year = 2011,
  month = oct,
  journal = {MNRAS},
  volume = {416},
  number = {4},
  pages = {3118--3137},
  issn = {0035-8711},
  doi = {10.1111/j.1365-2966.2011.19261.x},
  urldate = {2025-05-30}
}

@article{kanekarHighMolecularGas2020,
  title = {High {{Molecular Gas Masses}} in {{Absorption-selected Galaxies}} at z {$\approx$} 2},
  author = {Kanekar, N. and Prochaska, J. X. and Neeleman, M. and Christensen, L. and M{\o}ller, P. and Zwaan, M. A. and Fynbo, J. P. U. and {Dessauges-Zavadsky}, M.},
  year = 2020,
  month = sep,
  journal = {ApJL},
  volume = {901},
  number = {1},
  pages = {L5},
  publisher = {The American Astronomical Society},
  issn = {2041-8205},
  doi = {10.3847/2041-8213/abb4e1},
  urldate = {2025-02-25},
  langid = {english}
}

@article{kanekarMassiveAbsorptionselectedGalaxies2018,
  title = {Massive, {{Absorption-selected Galaxies}} at {{Intermediate Redshifts}}},
  author = {Kanekar, N. and Prochaska, J. X. and Christensen, L. and Rhodin, N. H. P. and Neeleman, M. and Zwaan, M. A. and M{\o}ller, P. and {Dessauges-Zavadsky}, M. and Fynbo, J. P. U. and Zafar, T.},
  year = 2018,
  month = mar,
  journal = {ApJL},
  volume = {856},
  number = {2},
  pages = {L23},
  publisher = {The American Astronomical Society},
  issn = {2041-8205},
  doi = {10.3847/2041-8213/aab6ab},
  urldate = {2025-01-29},
  langid = {english}
}

@article{kanekarSpinTemperatureHighredshift2014,
  title = {The Spin Temperature of High-Redshift Damped {{Lyman}} {$\alpha$} Systems},
  author = {Kanekar, N. and Prochaska, J. X. and Smette, A. and Ellison, S. L. and {Ryan-Weber}, E. V. and Momjian, E. and Briggs, F. H. and Lane, W. M. and Chengalur, J. N. and Delafosse, T. and Grave, J. and Jacobsen, D. and {de Bruyn}, A. G.},
  year = 2014,
  month = mar,
  journal = {MNRAS},
  volume = {438},
  number = {3},
  pages = {2131--2166},
  publisher = {Oxford Academic},
  issn = {0035-8711},
  doi = {10.1093/mnras/stt2338},
  urldate = {2025-03-26},
  langid = {english}
}

@article{karkiMUSEALMAHaloesIX2023,
  title = {{{MUSE-ALMA Haloes}} - {{IX}}. {{Morphologies}} and Stellar Properties of Gas-Rich Galaxies},
  author = {Karki, Arjun and Kulkarni, Varsha P. and Weng, Simon and P{\'e}roux, C{\'e}line and Augustin, Ramona and Hayes, Matthew and Ayromlou, Mohammadreza and Kacprzak, Glenn G. and Howk, J. Christopher and Szakacs, Roland and Klitsch, Anne and Hamanowicz, Aleksandra and Fresco, Alejandra and Zwaan, Martin A. and Biggs, Andrew D. and Fox, Andrew J. and Kassin, Susan and Kuntschner, Harald},
  year = 2023,
  month = oct,
  journal = {MNRAS},
  volume = {524},
  pages = {5524--5547},
  issn = {0035-8711},
  doi = {10.1093/mnras/stad2134},
  urldate = {2024-02-23}
}

@misc{kaurHIabsorptionselectedColdRotating2024,
  title = {An {{HI-absorption-selected}} Cold Rotating Disk Galaxy at \$z\textbackslash approx2.193\$},
  author = {Kaur, B. and Kanekar, N. and Neeleman, M. and Rafelski, M. and Prochaska, J. X. and Dutta, R.},
  year = 2024,
  month = aug,
  number = {arXiv:2408.00850},
  eprint = {2408.00850},
  primaryclass = {astro-ph},
  publisher = {arXiv},
  urldate = {2024-08-20},
  archiveprefix = {arXiv}
}

@article{kaurMassiveDustyHi2022,
  title = {A {{Massive}}, {{Dusty}}, {{Hi Absorption}}--{{Selected Galaxy}} at z {$\approx$} 2.46 {{Identified}} in a {{CO Emission Survey}}},
  author = {Kaur, B. and Kanekar, N. and Revalski, M. and Rafelski, M. and Neeleman, M. and Prochaska, J. X. and Walter, F.},
  year = 2022,
  month = jul,
  journal = {ApJ},
  volume = {934},
  number = {1},
  pages = {87},
  publisher = {The American Astronomical Society},
  issn = {0004-637X},
  doi = {10.3847/1538-4357/ac7b2c},
  urldate = {2025-03-26},
  langid = {english}
}

@article{kaurMassiveIabsorptionselectedGalaxy2025,
  title = {A {{Massive H I-absorption-selected Galaxy}} at z {$\approx$} 2.356},
  author = {Kaur, B. and Kanekar, N. and Neeleman, M. and Zhu, Y. and Prochaska, J. X. and Rafelski, M. and Becker, G.},
  year = 2025,
  month = mar,
  journal = {ApJ},
  volume = {982},
  pages = {L26},
  publisher = {IOP},
  issn = {0004-637X},
  doi = {10.3847/2041-8213/adb83d},
  urldate = {2025-05-15}
}

@article{kaurNatureHiabsorptionselectedGalaxies2021,
  title = {The {{Nature}} of {{Hi-absorption-selected Galaxies}} at z {$\approx$} 4},
  author = {Kaur, B. and Kanekar, N. and Rafelski, M. and Neeleman, M. and Revalski, M. and Prochaska, J. X.},
  year = 2021,
  month = nov,
  journal = {ApJ},
  volume = {921},
  number = {1},
  pages = {68},
  publisher = {The American Astronomical Society},
  issn = {0004-637X},
  doi = {10.3847/1538-4357/ac12d2},
  urldate = {2025-03-26},
  langid = {english}
}

@article{keatingReproducingCOtoH2Conversion2020,
  title = {Reproducing the {{CO-to-H2}} Conversion Factor in Cosmological Simulations of {{Milky-Way-mass}} Galaxies},
  author = {Keating, Laura C and Richings, Alexander J and Murray, Norman and {Faucher-Gigu{\`e}re}, Claude-Andr{\'e} and Hopkins, Philip F and Wetzel, Andrew and Kere{\v s}, Du{\v s}an and Benincasa, Samantha and Feldmann, Robert and Loebman, Sarah and Orr, Matthew E},
  year = 2020,
  month = oct,
  journal = {MNRAS},
  volume = {499},
  number = {1},
  pages = {837--850},
  issn = {0035-8711},
  doi = {10.1093/mnras/staa2839},
  urldate = {2025-07-16}
}

@article{kennicuttStarFormationMilky2012,
  title = {Star {{Formation}} in the {{Milky Way}} and {{Nearby Galaxies}}},
  author = {Kennicutt, Robert C. and Evans, Neal J.},
  year = 2012,
  month = sep,
  journal = {Annu. Rev. Astron. Astrophys.},
  volume = {50},
  pages = {531},
  issn = {0066-4146},
  doi = {10.1146/annurev-astro-081811-125610},
  urldate = {2022-06-09},
  langid = {english}
}

@misc{klimenkoBaryonicContentGalaxies2023,
  title = {The {{Baryonic Content}} of {{Galaxies Mapped}} by {{MaNGA}} and the {{Gas Around Them}}},
  author = {Klimenko, Viacheslav V. and Kulkarni, Varsha and Wake, David A. and Poudel, Suraj and Bershady, Matthew A. and Peroux, Celine and Lundgren, Britt},
  year = 2023,
  month = aug,
  number = {arXiv:2308.10992},
  eprint = {2308.10992},
  primaryclass = {astro-ph},
  publisher = {arXiv},
  doi = {10.48550/arXiv.2308.10992},
  urldate = {2023-08-23},
  archiveprefix = {arXiv}
}

@article{klitschALMACALAbsorptionselectedGalaxies2019,
  title = {{{ALMACAL V}}: Absorption-Selected Galaxies with Evidence for Excited {{ISMs}}},
  shorttitle = {{{ALMACAL V}}},
  author = {Klitsch, A. and Zwaan, M. A. and P{\'e}roux, C. and Smail, I. and Oteo, I. and Popping, G. and Swinbank, A. M. and Ivison, R. J. and Biggs, A. D.},
  year = 2019,
  month = jan,
  journal = {MNRAS},
  volume = {482},
  pages = {L65-L69},
  issn = {0035-8711},
  doi = {10.1093/mnrasl/sly187},
  urldate = {2022-01-24}
}

@article{klitschALMACALIIICombined2018,
  title = {{{ALMACAL}} - {{III}}. {{A}} Combined {{ALMA}} and {{MUSE}} Survey for Neutral, Molecular, and Ionized Gas in an {{H I-absorption-selected}} System},
  author = {Klitsch, A. and P{\'e}roux, C. and Zwaan, M. A. and Smail, I. and Oteo, I. and Biggs, A. D. and Popping, G. and Swinbank, A. M.},
  year = 2018,
  month = mar,
  journal = {MNRAS},
  volume = {475},
  pages = {492--507},
  publisher = {OUP},
  issn = {0035-8711},
  doi = {10.1093/mnras/stx3184},
  urldate = {2025-04-30}
}

@article{klitschCOExcitationLine2022,
  title = {{{CO Excitation}} and {{Line Energy Distributions}} in {{Gas-selected Galaxies}}},
  author = {Klitsch, A. and Christensen, L. and Valentino, F. and Kanekar, N. and M{\o}ller, P. and Zwaan, M. A. and Fynbo, J. P. U. and Neeleman, M. and Prochaska, J. X.},
  year = 2022,
  month = jun,
  journal = {MNRAS},
  volume = {514},
  number = {2},
  eprint = {2204.09698},
  primaryclass = {astro-ph},
  pages = {2346--2355},
  issn = {0035-8711, 1365-2966},
  doi = {10.1093/mnras/stac1190},
  urldate = {2024-04-18},
  archiveprefix = {arXiv}
}

@article{klitschH2MolecularGas2021,
  title = {H2 Molecular Gas Absorption-Selected Systems Trace {{CO}} Molecular Gas-Rich Galaxy Overdensities},
  author = {Klitsch, Anne and P{\'e}roux, C{\'e}line and Zwaan, Martin A. and De Cia, Annalisa and Ledoux, C{\'e}dric and Lopez, Sebastian},
  year = 2021,
  month = sep,
  journal = {MNRAS},
  volume = {506},
  pages = {514--522},
  issn = {0035-8711},
  doi = {10.1093/mnras/stab1668},
  urldate = {2022-01-24}
}

@article{koribalskiASKAPRevealsRadio2024,
  title = {{{ASKAP}} Reveals the Radio Tail Structure of the {{Corkscrew Galaxy}} Shaped by Its Passage through the {{Abell}} 3627 Cluster},
  author = {Koribalski, B{\"a}rbel S and Duchesne, Stefan W and Lenc, Emil and Venturi, Tiziana and Botteon, Andrea and Shabala, Stanislav S and Vernstrom, Tessa and Carretti, Ettore and Norris, Ray P and Anderson, Craig and Hopkins, Andrew M and Riseley, C J and Gupta, Nikhel and Velovi{\'c}, Velibor},
  year = 2024,
  month = sep,
  journal = {MNRAS},
  volume = {533},
  number = {1},
  pages = {608--620},
  issn = {0035-8711},
  doi = {10.1093/mnras/stae1838},
  urldate = {2025-06-20}
}

@article{krumholzAtomictoMolecularTransitionGalaxies2009,
  title = {The {{Atomic-to-Molecular Transition}} in {{Galaxies}}. {{II}}: {{H I}} and {{H2 Column Densities}}},
  shorttitle = {The {{Atomic-to-Molecular Transition}} in {{Galaxies}}. {{II}}},
  author = {Krumholz, Mark R. and McKee, Christopher F. and Tumlinson, Jason},
  year = 2009,
  month = mar,
  journal = {ApJ},
  volume = {693},
  pages = {216--235},
  publisher = {IOP},
  issn = {0004-637X},
  doi = {10.1088/0004-637X/693/1/216},
  urldate = {2025-05-31}
}

@article{kulkarniDampedLyaAbsorbers2022,
  title = {Damped {{Ly$\alpha$ Absorbers}} in {{Star-forming Galaxies}} at z {$<$} 0.15 {{Detected}} with the {{Hubble Space Telescope}} and {{Implications}} for {{Galactic Evolution}}},
  author = {Kulkarni, Varsha P. and Bowen, David V. and Straka, Lorrie A. and York, Donald G. and Gupta, Neeraj and Noterdaeme, Pasquier and Srianand, Raghunathan},
  year = 2022,
  month = apr,
  journal = {ApJ},
  volume = {929},
  pages = {150},
  publisher = {IOP},
  issn = {0004-637X},
  doi = {10.3847/1538-4357/ac5fab},
  urldate = {2025-07-14}
}

@article{kulkarniHubbleSpaceTelescope2005,
  title = {Hubble {{Space Telescope Observations}} of {{Element Abundances}} in {{Low-Redshift Damped Ly$\alpha$ Galaxies}} and {{Implications}} for the {{Global Metallicity-Redshift Relation}}*},
  author = {Kulkarni, Varsha P. and Fall, S. Michael and Lauroesch, James T. and York, Donald G. and Welty, Daniel E. and Khare, Pushpa and Truran, James W.},
  year = 2005,
  month = jan,
  journal = {ApJ},
  volume = {618},
  number = {1},
  pages = {68},
  issn = {0004-637X},
  doi = {10.1086/425956},
  urldate = {2025-06-02},
  langid = {english}
}

@article{kulkarniKeckVLTObservations2015,
  title = {Keck and {{VLT Observations}} of {{Super-Damped Lyman-Alpha Absorbers}} at z 2- 2.5: {{Constraints}} on {{Chemical Compositions}} and {{Physical Conditions}}},
  shorttitle = {Keck and {{VLT Observations}} of {{Super-Damped Lyman-Alpha Absorbers}} at z 2- 2.5},
  author = {Kulkarni, Varsha P. and Som, Debopam and Morrison, Sean and P{\'e}roux, Celine and Quiret, Samuel and York, Donald G.},
  year = 2015,
  month = dec,
  journal = {ApJ},
  volume = {815},
  pages = {24},
  publisher = {IOP},
  issn = {0004-637X},
  doi = {10.1088/0004-637X/815/1/24},
  urldate = {2025-06-11}
}

@article{kulkarniNICMOSImagingDamped2000,
  title = {{{NICMOS Imaging}} of the {{Damped Ly$\alpha$ Absorber}} at z = 1.89 toward {{LBQS}} 1210+1731: {{Constraints}} on {{Size}} and {{Star Formation Rate}}},
  shorttitle = {{{NICMOS Imaging}} of the {{Damped Ly$\alpha$ Absorber}} at z = 1.89 toward {{LBQS}} 1210+1731},
  author = {Kulkarni, Varsha P. and Hill, John M. and Schneider, Glenn and Weymann, Ray J. and {Storrie-Lombardi}, Lisa J. and Rieke, Marcia J. and Thompson, Rodger I. and Jannuzi, Buell T.},
  year = 2000,
  month = jun,
  journal = {ApJ},
  volume = {536},
  number = {1},
  pages = {36},
  publisher = {IOP Publishing},
  issn = {0004-637X},
  doi = {10.1086/308904},
  urldate = {2025-03-26},
  langid = {english}
}

@article{langanMusEGAsFLOw2023,
  title = {{{MusE GAs FLOw}} and {{Wind}} ({{MEGAFLOW}}) {{IX}}. {{The}} Impact of Gas Flows on the Relations between the Mass, Star Formation Rate, and Metallicity of Galaxies},
  author = {Langan, Ivanna and Zabl, Johannes and Bouch{\'e}, Nicolas F and Ginolfi, Michele and Popping, Gerg{\"o} and Schroetter, Ilane and Wendt, Martin and Schaye, Joop and Boogaard, Leindert and Freundlich, Jonathan and Richard, Johan and Matthee, Jorryt and Mercier, Wilfried and Contini, Thierry and Guo, Yucheng and Cherrey, Maxime},
  year = 2023,
  month = may,
  journal = {MNRAS},
  volume = {521},
  number = {1},
  pages = {546--557},
  issn = {0035-8711},
  doi = {10.1093/mnras/stad357},
  urldate = {2024-10-16}
}

@article{leeALMAACACO2022,
  title = {{{ALMA}}/{{ACA CO Survey}} of the {{IC}} 1459 and {{NGC}} 4636 {{Groups}}: {{Environmental Effects}} on the {{Molecular Gas}} of {{Group Galaxies}}},
  shorttitle = {{{ALMA}}/{{ACA CO Survey}} of the {{IC}} 1459 and {{NGC}} 4636 {{Groups}}},
  author = {Lee, Bumhyun and Wang, Jing and Chung, Aeree and Ho, Luis C. and Wang, Ran and Michiyama, Tomonari and Molina, Juan and Kim, Yongjung and Shao, Li and Kilborn, Virginia and Wang, Shun and Lin, Xuchen and Kim, Dawoon E. and Catinella, Barbara and Cortese, Luca and Deg, Nathan and Denes, Helga and Elagali, Ahmed and For, Bi-Qing and Kleiner, Dane and Koribalski, B{\"a}rbel S. and {Lee-Waddell}, Karen and Rhee, Jonghwan and Spekkens, Kristine and Westmeier, Tobias and Wong, O. Ivy and Bigiel, Frank and Bosma, Albert and Holwerda, Benne W. and {van der Hulst}, Jan M. and Roychowdhury, Sambit and {Verdes-Montenegro}, Lourdes and Zwaan, Martin A.},
  year = 2022,
  month = sep,
  journal = {ApJS},
  volume = {262},
  pages = {31},
  publisher = {IOP},
  issn = {0067-0049},
  doi = {10.3847/1538-4365/ac7eba},
  urldate = {2025-06-01}
}

@article{lehnerBIMODALMETALLICITYDISTRIBUTION2013,
  title = {{{THE BIMODAL METALLICITY DISTRIBUTION OF THE COOL CIRCUMGALACTIC MEDIUM AT}} z {$\lessequivlnt$} 1*},
  author = {Lehner, N. and Howk, J. C. and Tripp, T. M. and Tumlinson, J. and Prochaska, J. X. and O'Meara, J. M. and Thom, C. and Werk, J. K. and Fox, A. J. and Ribaudo, J.},
  year = 2013,
  month = jun,
  journal = {ApJ},
  volume = {770},
  number = {2},
  pages = {138},
  publisher = {The American Astronomical Society},
  issn = {0004-637X},
  doi = {10.1088/0004-637X/770/2/138},
  urldate = {2025-06-02},
  langid = {english}
}

@article{leroyStarFormationEfficiency2008,
  title = {The {{Star Formation Efficiency}} in {{Nearby Galaxies}}: {{Measuring Where Gas Forms Stars Effectively}}},
  shorttitle = {The {{Star Formation Efficiency}} in {{Nearby Galaxies}}},
  author = {Leroy, Adam K. and Walter, Fabian and Brinks, Elias and Bigiel, Frank and {de Blok}, W. J. G. and Madore, Barry and Thornley, M. D.},
  year = 2008,
  month = dec,
  journal = {AJ},
  volume = {136},
  pages = {2782--2845},
  publisher = {IOP},
  issn = {0004-6256},
  doi = {10.1088/0004-6256/136/6/2782},
  urldate = {2025-06-20}
}

@article{liDarkMolecularGas2018,
  title = {Dark {{Molecular Gas}} in {{Simulations}} of z {$\sim$} 0 {{Disk Galaxies}}},
  author = {Li, Qi and Narayanan, Desika and Dav{\`e}, Romeel and Krumholz, Mark R.},
  year = 2018,
  month = dec,
  journal = {ApJ},
  volume = {869},
  number = {1},
  pages = {73},
  publisher = {The American Astronomical Society},
  issn = {0004-637X},
  doi = {10.3847/1538-4357/aaec77},
  urldate = {2025-06-01},
  langid = {english}
}

@inproceedings{liEnvironmentalEffectsAtomic2013,
  title = {Environmental Effects on the Atomic Gas Content of Galaxies in the Local Universe},
  booktitle = {Molecular {{Gas}}, {{Dust}}, and {{Star Formation}} in {{Galaxies}}},
  author = {Li, Cheng},
  year = 2013,
  month = mar,
  volume = {292},
  pages = {149--152},
  doi = {10.1017/S1743921313000823},
  urldate = {2025-06-01}
}

@article{lillyGasRegulationGalaxies2013,
  title = {Gas {{Regulation}} of {{Galaxies}}: {{The Evolution}} of the {{Cosmic Specific Star Formation Rate}}, the {{Metallicity-Mass-Star-formation Rate Relation}}, and the {{Stellar Content}} of {{Halos}}},
  shorttitle = {Gas {{Regulation}} of {{Galaxies}}},
  author = {Lilly, Simon J. and Carollo, C. Marcella and Pipino, Antonio and Renzini, Alvio and Peng, Yingjie},
  year = 2013,
  month = aug,
  journal = {ApJ},
  volume = {772},
  pages = {119},
  publisher = {IOP},
  issn = {0004-637X},
  doi = {10.1088/0004-637X/772/2/119},
  urldate = {2024-09-20}
}

@article{linInadequateTurbulentSupport2025,
  title = {Inadequate Turbulent Support in Low-Metallicity Molecular Clouds},
  author = {Lin, Lingrui and Zhang, Zhi-Yu and Wang, Junzhi and Papadopoulos, Padelis P. and Shi, Yong and Gong, Yan and Sun, Yan and Sun, Yichen and Bisbas, Thomas G. and Romano, Donatella and Li, Di and Liu, Hauyu Baobab and Qiu, Keping and Liu, Lijie and Luo, Gan and Tsai, Chao-Wei and Wu, Jingwen and Feng, Siyi and Zhang, Bo},
  year = 2025,
  month = jan,
  journal = {Nat Astron},
  pages = {1--11},
  publisher = {Nature Publishing Group},
  issn = {2397-3366},
  doi = {10.1038/s41550-024-02440-3},
  urldate = {2025-02-07},
  copyright = {2025 The Author(s), under exclusive licence to Springer Nature Limited},
  langid = {english}
}

@article{lofthouseMUSEAnalysisGas2020,
  title = {{{MUSE Analysis}} of {{Gas}} around {{Galaxies}} ({{MAGG}}) - {{I}}: {{Survey}} Design and the Environment of a near Pristine Gas Cloud at z {$\approx$} 3.5},
  shorttitle = {{{MUSE Analysis}} of {{Gas}} around {{Galaxies}} ({{MAGG}}) - {{I}}},
  author = {Lofthouse, Emma K. and Fumagalli, Michele and Fossati, Matteo and O'Meara, John M. and Murphy, Michael T. and Christensen, Lise and Prochaska, J. Xavier and Cantalupo, Sebastiano and Bielby, Richard M. and Cooke, Ryan J. and Lusso, Elisabeta and Morris, Simon L.},
  year = 2020,
  month = jan,
  journal = {MNRAS},
  volume = {491},
  pages = {2057--2074},
  publisher = {OUP},
  issn = {0035-8711},
  doi = {10.1093/mnras/stz3066},
  urldate = {2025-06-01}
}

@article{lupiNaturalEmergenceCorrelation2018,
  title = {The Natural Emergence of the Correlation between {{H2}} and Star Formation Rate Surface Densities in Galaxy Simulations},
  author = {Lupi, Alessandro and Bovino, Stefano and Capelo, Pedro R. and Volonteri, Marta and Silk, Joseph},
  year = 2018,
  month = mar,
  journal = {MNRAS},
  volume = {474},
  pages = {2884--2903},
  publisher = {OUP},
  issn = {0035-8711},
  doi = {10.1093/mnras/stx2874},
  urldate = {2025-07-15}
}

@article{lupiPredictingFIRLines2020,
  title = {Predicting {{FIR}} Lines from Simulated Galaxies},
  author = {Lupi, Alessandro and Pallottini, Andrea and Ferrara, Andrea and Bovino, Stefano and Carniani, Stefano and Vallini, Livia},
  year = 2020,
  month = aug,
  journal = {MNRAS},
  volume = {496},
  pages = {5160--5175},
  publisher = {OUP},
  issn = {0035-8711},
  doi = {10.1093/mnras/staa1842},
  urldate = {2025-07-16}
}

@article{maddenTracingTotalMolecular2020,
  title = {Tracing the Total Molecular Gas in Galaxies: [{{CII}}] and the {{CO-dark}} Gas},
  shorttitle = {Tracing the Total Molecular Gas in Galaxies},
  author = {Madden, S. C. and Cormier, D. and Hony, S. and Lebouteiller, V. and Abel, N. and Galametz, M. and Looze, I. De and Chevance, M. and Polles, F. L. and Lee, M.-Y. and Galliano, F. and {Lambert-Huyghe}, A. and Hu, D. and Ramambason, L.},
  year = 2020,
  month = nov,
  journal = {A\&A},
  volume = {643},
  pages = {A141},
  publisher = {EDP Sciences},
  issn = {0004-6361, 1432-0746},
  doi = {10.1051/0004-6361/202038860},
  urldate = {2023-11-03},
  copyright = {\copyright{} S. C. Madden et al. 2020},
  langid = {english}
}

@article{maddoxMIGHTEEHIEmissionProject2021,
  title = {{{MIGHTEE-HI}}: {{The H I}} Emission Project of the {{MeerKAT MIGHTEE}} Survey},
  shorttitle = {{{MIGHTEE-HI}}},
  author = {Maddox, N. and Frank, B. S. and Ponomareva, A. A. and Jarvis, M. J. and Adams, E. a. K. and Dav{\'e}, R. and Oosterloo, T. A. and Santos, M. G. and Blyth, S. L. and Glowacki, M. and {Kraan-Korteweg}, R. C. and Mulaudzi, W. and Namumba, B. and Prandoni, I. and Rajohnson, S. H. A. and Spekkens, K. and Adams, N. J. and Bowler, R. a. A. and Collier, J. D. and Heywood, I. and Sekhar, S. and Taylor, A. R.},
  year = 2021,
  month = feb,
  journal = {A\&A},
  volume = {646},
  pages = {A35},
  publisher = {EDP Sciences},
  issn = {0004-6361, 1432-0746},
  doi = {10.1051/0004-6361/202039655},
  urldate = {2025-06-20},
  copyright = {\copyright{} ESO 2021},
  langid = {english}
}

@article{maioAtomicMolecularGas2022,
  title = {Atomic and Molecular Gas from the Epoch of Reionisation down to Redshift 2},
  author = {Maio, Umberto and P{\'e}roux, C{\'e}line and Ciardi, Benedetta},
  year = 2022,
  month = jan,
  journal = {A\&A},
  volume = {657},
  pages = {A47},
  publisher = {EDP Sciences},
  issn = {0004-6361, 1432-0746},
  doi = {10.1051/0004-6361/202142264},
  urldate = {2023-05-11},
  copyright = {\copyright{} ESO 2022},
  langid = {english}
}

@article{maiolinoAMAZEEvolutionMassmetallicity2008,
  title = {{{AMAZE}}. {{I}}. {{The}} Evolution of the Mass-Metallicity Relation at z {$>$} 3},
  author = {Maiolino, R. and Nagao, T. and Grazian, A. and Cocchia, F. and Marconi, A. and Mannucci, F. and Cimatti, A. and Pipino, A. and Ballero, S. and Calura, F. and Chiappini, C. and Fontana, A. and Granato, G. L. and Matteucci, F. and Pastorini, G. and Pentericci, L. and Risaliti, G. and Salvati, M. and Silva, L.},
  year = 2008,
  month = sep,
  journal = {A\&A},
  volume = {488},
  pages = {463--479},
  issn = {0004-6361},
  doi = {10.1051/0004-6361:200809678},
  urldate = {2025-05-09}
}

@article{martinMultifilamentGasInflows2019,
  title = {Multi-Filament Gas Inflows Fuelling Young Star-Forming Galaxies},
  author = {Martin, D. Christopher and O'Sullivan, Donal and Matuszewski, Mateusz and Hamden, Erika and Dekel, Avishai and Lapiner, Sharon and Morrissey, Patrick and Neill, James D. and Cantalupo, Sebastiano and Prochaska, Jason Xavier and Steidel, Charles and Trainor, Ryan and Moore, Anna and Ceverino, Daniel and Primack, Joel and Rizzi, Luca},
  year = 2019,
  month = sep,
  journal = {Nat Astron},
  volume = {3},
  number = {9},
  pages = {822--831},
  publisher = {Nature Publishing Group},
  issn = {2397-3366},
  doi = {10.1038/s41550-019-0791-2},
  urldate = {2025-06-01},
  copyright = {2019 The Author(s), under exclusive licence to Springer Nature Limited},
  langid = {english}
}

@article{masuiMEASUREMENT21Cm2013,
  title = {{{MEASUREMENT OF}} 21 Cm {{BRIGHTNESS FLUCTUATIONS AT}} z {$\sim$} 0.8 {{IN CROSS-CORRELATION}}},
  author = {Masui, K. W. and Switzer, E. R. and Banavar, N. and Bandura, K. and Blake, C. and Calin, L.-M. and Chang, T.-C. and Chen, X. and Li, Y.-C. and Liao, Y.-W. and Natarajan, A. and Pen, U.-L. and Peterson, J. B. and Shaw, J. R. and Voytek, T. C.},
  year = 2013,
  month = jan,
  journal = {ApJL},
  volume = {763},
  number = {1},
  pages = {L20},
  publisher = {The American Astronomical Society},
  issn = {2041-8205},
  doi = {10.1088/2041-8205/763/1/L20},
  urldate = {2025-06-20},
  langid = {english}
}

@article{mcmullinCASAArchitectureApplications2007,
  title = {{{CASA Architecture}} and {{Applications}}},
  author = {McMullin, J. P. and Waters, B. and Schiebel, D. and Young, W. and Golap, K.},
  year = 2007,
  month = oct,
  volume = {376},
  pages = {127},
  urldate = {2024-02-14}
}

@misc{merloniEROSITAScienceBook2012,
  title = {{{eROSITA Science Book}}: {{Mapping}} the {{Structure}} of the {{Energetic Universe}}},
  shorttitle = {{{eROSITA Science Book}}},
  author = {Merloni, A. and Predehl, P. and Becker, W. and B{\"o}hringer, H. and Boller, T. and Brunner, H. and Brusa, M. and Dennerl, K. and Freyberg, M. and Friedrich, P. and Georgakakis, A. and Haberl, F. and Hasinger, G. and Meidinger, N. and Mohr, J. and Nandra, K. and Rau, A. and Reiprich, T. H. and Robrade, J. and Salvato, M. and Santangelo, A. and Sasaki, M. and Schwope, A. and Wilms, J. and eROSITA Consortium, the German},
  year = 2012,
  month = sep,
  number = {arXiv:1209.3114},
  eprint = {1209.3114},
  primaryclass = {astro-ph},
  publisher = {arXiv},
  doi = {10.48550/arXiv.1209.3114},
  urldate = {2025-06-01},
  archiveprefix = {arXiv}
}

@article{messiasContentCosmicNoon2024,
  title = {H {{I}} Content at Cosmic Noon - a Millimetre-Wavelength Perspective},
  author = {Messias, Hugo and Guerrero, Andrea and Nagar, Neil and Regueiro, Jack and Impellizzeri, Violette and Orellana, Gustavo and Vioque, Miguel},
  year = 2024,
  month = oct,
  journal = {MNRAS},
  volume = {533},
  pages = {3937--3956},
  publisher = {OUP},
  issn = {0035-8711},
  doi = {10.1093/mnras/stae1807},
  urldate = {2024-11-20}
}

@article{mollerALMAVLTObservations2018,
  title = {{{ALMA}} + {{VLT}} Observations of a Damped {{Lyman-$\alpha$}} Absorbing Galaxy: Massive, Wide {{CO}} Emission, Gas-Rich but with Very Low {{SFR}}},
  shorttitle = {{{ALMA}} + {{VLT}} Observations of a Damped {{Lyman-$\alpha$}} Absorbing Galaxy},
  author = {M{\o}ller, P and Christensen, L and {M A Zwaan} and Kanekar, N and Prochaska, J X and Rhodin, N H P and {Dessauges-Zavadsky}, M and Fynbo, J P U and Neeleman, M and Zafar, T},
  year = 2018,
  month = mar,
  journal = {MNRAS},
  volume = {474},
  number = {3},
  pages = {4039--4055},
  issn = {0035-8711},
  doi = {10.1093/mnras/stx2845},
  urldate = {2025-02-27}
}

@article{muzahidHSTCOSSurvey2015,
  title = {An {{HST}}/{{COS}} Survey of Molecular Hydrogen in {{DLAs}} \& Sub-{{DLAs}} at z~{$<~$}1: Molecular Fraction and Excitation Temperature},
  shorttitle = {An {{HST}}/{{COS}} Survey of Molecular Hydrogen in {{DLAs}} \& Sub-{{DLAs}} at z~{$<~$}1},
  author = {Muzahid, S. and Srianand, R. and Charlton, J.},
  year = 2015,
  month = apr,
  journal = {MNRAS},
  volume = {448},
  number = {3},
  pages = {2840--2853},
  issn = {0035-8711},
  doi = {10.1093/mnras/stv133},
  urldate = {2025-05-31}
}

@article{muzahidMUSEQuBESCalibratingRedshifts2020,
  title = {{{MUSEQuBES}}: Calibrating the Redshifts of {{Ly}}\,{$\alpha$} Emitters Using Stacked Circumgalactic Medium Absorption Profiles},
  shorttitle = {{{MUSEQuBES}}},
  author = {Muzahid, Sowgat and Schaye, Joop and Marino, Raffaella Anna and Cantalupo, Sebastiano and Brinchmann, Jarle and Contini, Thierry and Wendt, Martin and Wisotzki, Lutz and Zabl, Johannes and Bouch{\'e}, Nicolas and Akhlaghi, Mohammad and Chen, Hsiao-Wen and Claeyssens, Ad{\'e}la{\^i}de and Johnson, Sean and Leclercq, Floriane and Maseda, Michael and Matthee, Jorryt and Richard, Johan and Urrutia, Tanya and Verhamme, Anne},
  year = 2020,
  month = aug,
  journal = {MNRAS},
  volume = {496},
  number = {2},
  pages = {1013--1022},
  issn = {0035-8711},
  doi = {10.1093/mnras/staa1347},
  urldate = {2025-06-01}
}

@article{naabTheoreticalChallengesGalaxy2017,
  title = {Theoretical {{Challenges}} in {{Galaxy Formation}}},
  author = {Naab, Thorsten and Ostriker, Jeremiah P.},
  year = 2017,
  month = aug,
  journal = {Annu. Rev. Astron. Astrophys.},
  volume = {55},
  number = {1},
  eprint = {1612.06891},
  primaryclass = {astro-ph},
  pages = {59--109},
  issn = {0066-4146, 1545-4282},
  doi = {10.1146/annurev-astro-081913-040019},
  urldate = {2025-06-01},
  archiveprefix = {arXiv}
}

@article{neelemanColdMassiveRotating2020,
  title = {A Cold, Massive, Rotating Disk Galaxy 1.5 Billion Years after the {{Big Bang}}},
  author = {Neeleman, Marcel and Prochaska, J. Xavier and Kanekar, Nissim and Rafelski, Marc},
  year = 2020,
  month = may,
  journal = {Nat Astron},
  volume = {581},
  pages = {269--272},
  issn = {0028-0836},
  doi = {10.1038/s41586-020-2276-y},
  urldate = {2025-02-27}
}

@article{neelemanFIRSTCONNECTIONCOLD2016,
  title = {{{FIRST CONNECTION BETWEEN COLD GAS IN EMISSION AND ABSORPTION}}: {{CO EMISSION FROM A GALAXY}}--{{QUASAR PAIR}}},
  shorttitle = {{{FIRST CONNECTION BETWEEN COLD GAS IN EMISSION AND ABSORPTION}}},
  author = {Neeleman, Marcel and Prochaska, J. Xavier and Zwaan, Martin A. and Kanekar, Nissim and Christensen, Lise and {Dessauges-Zavadsky}, Miroslava and Fynbo, Johan P. U. and van Kampen, Eelco and M{\o}ller, Palle and Zafar, Tayyaba},
  year = 2016,
  month = mar,
  journal = {ApJL},
  volume = {820},
  number = {2},
  pages = {L39},
  publisher = {The American Astronomical Society},
  issn = {2041-8205},
  doi = {10.3847/2041-8205/820/2/L39},
  urldate = {2025-02-27},
  langid = {english}
}

@article{neelemanIi158Mm2019,
  title = {[{{C}} Ii] 158 {$\mu$}m {{Emission}} from z {$\sim$} 4 {{H}} i {{Absorption-selected Galaxies}}},
  author = {Neeleman, Marcel and Kanekar, Nissim and Xavier Prochaska, J. and Rafelski, Marc A. and Carilli, Chris L.},
  year = 2019,
  month = jan,
  journal = {ApJL},
  volume = {870},
  number = {2},
  pages = {L19},
  publisher = {The American Astronomical Society},
  issn = {2041-8205},
  doi = {10.3847/2041-8213/aaf871},
  urldate = {2025-03-26},
  langid = {english}
}

@article{neelemanII158Mm2025,
  title = {A [{{C II}}] 158 {$\mu$}m {{Survey}} of {{Damped Ly$\alpha$ Absorber Galaxies}} at z {$\sim$} 4: {{Observations}} of {{Dense}} and {{Metal-enriched Neutral Gas}} within the {{Circumgalactic Medium}} of {{Star-forming Galaxies}}},
  shorttitle = {A [{{C II}}] 158 {$\mu$}m {{Survey}} of {{Damped Ly$\alpha$ Absorber Galaxies}} at z {$\sim$} 4},
  author = {Neeleman, Marcel and Kanekar, Nissim and Prochaska, J. Xavier and Rafelski, Marc A. and Kahinga, Lordrick A.},
  year = 2025,
  month = apr,
  journal = {ApJ},
  volume = {983},
  pages = {26},
  publisher = {IOP},
  issn = {0004-637X},
  doi = {10.3847/1538-4357/adbbd1},
  urldate = {2025-05-15}
}

@article{neelemanII158mmEmission2017,
  title = {[{{C II}}] 158-{$\mu$}m Emission from the Host Galaxies of Damped {{Lyman-alpha}} Systems},
  author = {Neeleman, Marcel and Kanekar, Nissim and Prochaska, J. Xavier and Rafelski, Marc and Carilli, Chris L. and Wolfe, Arthur M.},
  year = 2017,
  month = mar,
  journal = {Science},
  volume = {355},
  pages = {1285--1288},
  issn = {0036-8075},
  doi = {10.1126/science.aal1737},
  urldate = {2025-03-26}
}

@article{neelemanMolecularEmissionGalaxy2018,
  title = {Molecular {{Emission}} from a {{Galaxy Associated}} with a z {$\sim$} 2.2 {{Damped Ly$\alpha$ Absorber}}},
  author = {Neeleman, Marcel and Kanekar, Nissim and Prochaska, J. Xavier and Christensen, Lise and {Dessauges-Zavadsky}, Miroslava and Fynbo, Johan P. U. and M{\o}ller, Palle and Zwaan, Martin A.},
  year = 2018,
  month = mar,
  journal = {ApJL},
  volume = {856},
  number = {1},
  pages = {L12},
  publisher = {The American Astronomical Society},
  issn = {2041-8205},
  doi = {10.3847/2041-8213/aab5b1},
  urldate = {2025-02-27},
  langid = {english}
}

@article{nelsonResolvingSmallscaleCold2020,
  title = {Resolving Small-Scale Cold Circumgalactic Gas in {{TNG50}}},
  author = {Nelson, Dylan and Sharma, Prateek and Pillepich, Annalisa and Springel, Volker and Pakmor, R{\"u}diger and Weinberger, Rainer and Vogelsberger, Mark and Marinacci, Federico and Hernquist, Lars},
  year = 2020,
  month = sep,
  journal = {MNRAS},
  volume = {498},
  number = {2},
  pages = {2391--2414},
  issn = {0035-8711},
  doi = {10.1093/mnras/staa2419},
  urldate = {2025-05-25}
}

@article{nielsenCGMCosmicNoon2020,
  title = {The {{CGM}} at {{Cosmic Noon}} with {{KCWI}}: {{Outflows}} from a {{Star-forming Galaxy}} at z = 2.071},
  shorttitle = {The {{CGM}} at {{Cosmic Noon}} with {{KCWI}}},
  author = {Nielsen, Nikole M. and Kacprzak, Glenn G. and Pointon, Stephanie K. and Murphy, Michael T. and Churchill, Christopher W. and Dav{\'e}, Romeel},
  year = 2020,
  month = dec,
  journal = {ApJ},
  volume = {904},
  number = {2},
  pages = {164},
  publisher = {The American Astronomical Society},
  issn = {0004-637X},
  doi = {10.3847/1538-4357/abc561},
  urldate = {2025-06-01},
  langid = {english}
}

@article{olsenSigameV3Gas2021,
  title = {S\'igame v3: {{Gas Fragmentation}} in {{Postprocessing}} of {{Cosmological Simulations}} for {{More Accurate Infrared Line Emission Modeling}}},
  shorttitle = {S\'igame V3},
  author = {Olsen, Karen Pardos and Burkhart, Blakesley and Mac Low, Mordecai-Mark and Tre{\ss}, Robin G. and Greve, Thomas R. and Vizgan, David and Motka, Jay and Borrow, Josh and Popping, Gerg{\"o} and Dav{\'e}, Romeel and Smith, Rowan J. and Narayanan, Desika},
  year = 2021,
  month = nov,
  journal = {ApJ},
  volume = {922},
  number = {1},
  pages = {88},
  publisher = {The American Astronomical Society},
  issn = {0004-637X},
  doi = {10.3847/1538-4357/ac20d4},
  urldate = {2025-07-15},
  langid = {english}
}

@article{oppenheimerMassMetalEnergy2008,
  title = {Mass, Metal, and Energy Feedback in Cosmological Simulations},
  author = {Oppenheimer, Benjamin D. and Dav{\'e}, Romeel},
  year = 2008,
  month = jun,
  journal = {MNRAS},
  volume = {387},
  number = {2},
  pages = {577--600},
  issn = {0035-8711},
  doi = {10.1111/j.1365-2966.2008.13280.x},
  urldate = {2025-06-01}
}

@article{oppenheimerSimulatingGroupsIntraGroup2021,
  title = {Simulating {{Groups}} and the {{IntraGroup Medium}}: {{The Surprisingly Complex}} and {{Rich Middle Ground}} between {{Clusters}} and {{Galaxies}}},
  shorttitle = {Simulating {{Groups}} and the {{IntraGroup Medium}}},
  author = {Oppenheimer, Benjamin D. and Babul, Arif and Bah{\'e}, Yannick and Butsky, Iryna S. and McCarthy, Ian G.},
  year = 2021,
  month = jun,
  journal = {Universe},
  volume = {7},
  pages = {209},
  doi = {10.3390/universe7070209},
  urldate = {2025-06-01}
}

@article{pallottiniSurveyHigh$z$Galaxies2022,
  title = {A Survey of High-\$z\$ Galaxies: {{SERRA}} Simulations},
  shorttitle = {A Survey of High-\$z\$ Galaxies},
  author = {Pallottini, A. and Ferrara, A. and Gallerani, S. and Behrens, C. and Kohandel, M. and Carniani, S. and Vallini, L. and Salvadori, S. and Gelli, V. and Sommovigo, L. and D'Odorico, V. and Mascia, F. Di and Pizzati, E.},
  year = 2022,
  month = may,
  journal = {MNRAS},
  eprint = {2201.02636},
  primaryclass = {astro-ph},
  issn = {0035-8711, 1365-2966},
  doi = {10.1093/mnras/stac1281},
  urldate = {2025-07-17},
  archiveprefix = {arXiv}
}

@article{perouxCosmicBaryonMetal2020,
  title = {The {{Cosmic Baryon}} and {{Metal Cycles}}},
  author = {P{\'e}roux, C{\'e}line and Howk, J. Christopher},
  year = 2020,
  month = aug,
  journal = {Annu. Rev. Astron. Astrophys.},
  volume = {58},
  pages = {363--406},
  issn = {0066-4146},
  doi = {10.1146/annurev-astro-021820-120014},
  urldate = {2021-09-01}
}

@article{perouxEvolutionOHIEpoch2003,
  title = {The Evolution of {{\textohm HI}} and the Epoch of Formation of Damped {{Lyman}} {$\alpha$} Absorbers},
  author = {P{\'e}roux, C. and McMahon, R. G. and {Storrie-Lombardi}, L. J. and Irwin, M. J.},
  year = 2003,
  month = dec,
  journal = {MNRAS},
  volume = {346},
  pages = {1103--1115},
  publisher = {OUP},
  issn = {0035-8711},
  doi = {10.1111/j.1365-2966.2003.07129.x},
  urldate = {2024-09-17}
}

@article{perouxMetalrichDampedSubdamped2006,
  title = {Metal-Rich Damped/Subdamped {{Lyman}} {$\alpha$} Quasar Absorbers at z {$<$} 1},
  author = {P{\'e}roux, C. and Meiring, J. D. and Kulkarni, V. P. and Ferlet, R. and Khare, P. and Lauroesch, J. T. and Vladilo, G. and York, D. G.},
  year = 2006,
  month = oct,
  journal = {MNRAS},
  volume = {372},
  pages = {369--380},
  publisher = {OUP},
  issn = {0035-8711},
  doi = {10.1111/j.1365-2966.2006.10865.x},
  urldate = {2025-06-01}
}

@article{perouxMultiphaseCircumgalacticMedium2019,
  title = {Multiphase Circumgalactic Medium Probed with {{MUSE}} and {{ALMA}}},
  author = {P{\'e}roux, C{\'e}line and Zwaan, Martin A and Klitsch, Anne and Augustin, Ramona and Hamanowicz, Aleksandra and Rahmani, Hadi and Pettini, Max and Kulkarni, Varsha and Straka, Lorrie A and Biggs, Andy D and York, Donald G and Milliard, Bruno},
  year = 2019,
  month = may,
  journal = {MNRAS},
  volume = {485},
  number = {2},
  pages = {1595--1613},
  issn = {0035-8711},
  doi = {10.1093/mnras/stz202},
  urldate = {2023-02-15}
}

@misc{perouxMultiScaleMultiPhaseCircumgalactic2024,
  title = {The {{Multi-Scale Multi-Phase Circumgalactic Medium}}: {{Observed}} and {{Simulated}}},
  shorttitle = {The {{Multi-Scale Multi-Phase Circumgalactic Medium}}},
  author = {Peroux, Celine and Nelson, Dylan},
  year = 2024,
  month = nov,
  publisher = {arXiv},
  doi = {10.48550/arXiv.2411.07988},
  urldate = {2025-06-06}
}

@article{perouxMUSEALMAHaloes2022,
  title = {{{MUSE}}--{{ALMA}} Haloes {{VII}}: Survey Science Goals~\& Design, Data Processing and Final Catalogues},
  shorttitle = {{{MUSE}}--{{ALMA}} Haloes {{VII}}},
  author = {P{\'e}roux, C and Weng, S and Karki, A and Augustin, R and Kulkarni, V P and Szakacs, R and Klitsch, A and Hamanowicz, A and Fresco, A Y and Zwaan, M A and Biggs, A and Fox, A J and Hayes, M and Howk, J C and Kacprzak, G G and Kassin, S and Kuntschner, H and Nelson, D and Pettini, M},
  year = 2022,
  month = nov,
  journal = {MNRAS},
  volume = {516},
  number = {4},
  pages = {5618--5636},
  issn = {0035-8711},
  doi = {10.1093/mnras/stac2546},
  urldate = {2022-10-18}
}

@article{perouxNatureAbsorbingGas2017,
  title = {Nature of the Absorbing Gas Associated with a Galaxy Group at Z{$\sim$}0.4},
  author = {P{\'e}roux, C{\'e}line and Rahmani, Hadi and Quiret, Samuel and Pettini, Max and Kulkarni, Varsha and York, Donald G. and Straka, Lorrie and Husemann, Bernd and Milliard, Bruno},
  year = 2017,
  month = jan,
  journal = {MNRAS},
  volume = {464},
  number = {2},
  pages = {2053--2065},
  issn = {0035-8711},
  doi = {10.1093/mnras/stw2444},
  urldate = {2025-03-13}
}

@article{perouxSINFONIIntegralField2016,
  title = {A {{SINFONI}} Integral Field Spectroscopy Survey for Galaxy Counterparts to Damped {{Lyman}} {$\alpha$} Systems -- {{VI}}. {{Metallicity}} and Geometry as Gas Flow Probes\ding{72}},
  author = {P{\'e}roux, C{\'e}line and Quiret, Samuel and Rahmani, Hadi and Kulkarni, Varsha P. and Epinat, Benoit and Milliard, Bruno and Straka, Lorrie A. and York, Donald G. and Rahmati, Alireza and Contini, Thierry},
  year = 2016,
  month = mar,
  journal = {MNRAS},
  volume = {457},
  number = {1},
  pages = {903--916},
  issn = {0035-8711},
  doi = {10.1093/mnras/stw016},
  urldate = {2025-03-13}
}

@article{pettiniMetalAbundances151999,
  title = {Metal {{Abundances}} at z {$<$} 1.5: {{Fresh Clues}} to the {{Chemical Enrichment History}} of {{Damped Ly$\alpha$ Systems}}*},
  shorttitle = {Metal {{Abundances}} at z {$<$} 1.5},
  author = {Pettini, Max and Ellison, Sara L. and Steidel, Charles C. and Bowen, David V.},
  year = 1999,
  month = jan,
  journal = {ApJ},
  volume = {510},
  number = {2},
  pages = {576},
  issn = {0004-637X},
  doi = {10.1086/306635},
  urldate = {2025-06-02},
  langid = {english}
}

@article{pettiniOIIINIIAbundance2004,
  title = {[{{OIII}}]/[{{NII}}] as an Abundance Indicator at High Redshift},
  author = {Pettini, Max and Pagel, Bernard E. J.},
  year = 2004,
  month = mar,
  journal = {MNRAS},
  volume = {348},
  pages = {L59-L63},
  publisher = {OUP},
  issn = {0035-8711},
  doi = {10.1111/j.1365-2966.2004.07591.x},
  urldate = {2025-05-08}
}

@article{poppingArtModellingCO2019,
  title = {The Art of Modelling {{CO}}, [{{C I}}], and [{{C II}}] in Cosmological Galaxy Formation Models},
  author = {Popping, Gerg{\"o} and Narayanan, Desika and Somerville, Rachel S. and Faisst, Andreas L. and Krumholz, Mark R.},
  year = 2019,
  month = feb,
  journal = {MNRAS},
  volume = {482},
  pages = {4906--4932},
  issn = {0035-8711},
  doi = {10.1093/mnras/sty2969},
  urldate = {2024-02-19}
}

@article{prochaskaCOSHalosSurveyMetallicities2017,
  title = {The {{COS-Halos Survey}}: {{Metallicities}} in the {{Low-redshift Circumgalactic Medium}}},
  shorttitle = {The {{COS-Halos Survey}}},
  author = {Prochaska, J. Xavier and Werk, Jessica K. and Worseck, G{\'a}bor and Tripp, Todd M. and Tumlinson, Jason and Burchett, Joseph N. and Fox, Andrew J. and Fumagalli, Michele and Lehner, Nicolas and Peeples, Molly S. and Tejos, Nicolas},
  year = 2017,
  month = mar,
  journal = {ApJ},
  volume = {837},
  pages = {169},
  publisher = {IOP},
  issn = {0004-637X},
  doi = {10.3847/1538-4357/aa6007},
  urldate = {2025-06-02}
}

@article{prochaskaProtogalacticDiskModels1998,
  title = {Protogalactic {{Disk Models}} of {{Damped Lya Kinematics}}},
  author = {Prochaska, Jason X. and Wolfe, Arthur M.},
  year = 1998,
  month = nov,
  journal = {ApJ},
  volume = {507},
  number = {1},
  eprint = {astro-ph/9805293},
  pages = {113--130},
  issn = {0004-637X, 1538-4357},
  doi = {10.1086/306325},
  urldate = {2025-06-01},
  archiveprefix = {arXiv}
}

@article{quiretESOUVESAdvanced2016,
  title = {The {{ESO UVES}} Advanced Data Products Quasar Sample - {{VI}}. {{Sub-damped Lyman}} {$\alpha$} Metallicity Measurements and the Circumgalactic Medium},
  author = {Quiret, S. and P{\'e}roux, C. and Zafar, T. and Kulkarni, V. P. and Jenkins, E. B. and Milliard, B. and Rahmani, H. and Popping, A. and Rao, S. M. and Turnshek, D. A. and Monier, E. M.},
  year = 2016,
  month = jun,
  journal = {MNRAS},
  volume = {458},
  pages = {4074--4121},
  publisher = {OUP},
  issn = {0035-8711},
  doi = {10.1093/mnras/stw524},
  urldate = {2025-06-11}
}

@article{rafelskiMETALLICITYEVOLUTIONDAMPED2012,
  title = {{{METALLICITY EVOLUTION OF DAMPED Ly$\alpha$ SYSTEMS OUT TO}} z {$\sim$} 5},
  author = {Rafelski, Marc and Wolfe, Arthur M. and Prochaska, J. Xavier and Neeleman, Marcel and Mendez, Alexander J.},
  year = 2012,
  month = jul,
  journal = {ApJ},
  volume = {755},
  number = {2},
  pages = {89},
  publisher = {The American Astronomical Society},
  issn = {0004-637X},
  doi = {10.1088/0004-637X/755/2/89},
  urldate = {2025-05-02},
  langid = {english}
}

@article{rahmaniLymanLimitSystem2018,
  title = {A {{Lyman}} Limit System Associated with Galactic Winds},
  author = {Rahmani, Hadi and P{\'e}roux, C{\'e}line and Schroetter, Ilane and Augustin, Ramona and Bouch{\'e}, Nicolas and Krogager, Jens-Kristian and Kulkarni, Varsha P. and Milliard, Bruno and M{\o}ller, Palle and Pettini, Max and Vernet, Jo{\"e}l and York, Donald G.},
  year = 2018,
  month = nov,
  journal = {MNRAS},
  volume = {480},
  pages = {5046--5059},
  publisher = {OUP},
  issn = {0035-8711},
  doi = {10.1093/mnras/sty2216},
  urldate = {2025-02-07}
}

@article{rahmaniObservationalSignaturesWarped2018,
  title = {Observational Signatures of a Warped Disk Associated with Cold-Flow Accretion},
  author = {Rahmani, Hadi and P{\'e}roux, C{\'e}line and Augustin, Ramona and Husemann, Bernd and Kacprzak, Glenn G. and Kulkarni, Varsha and Milliard, Bruno and M{\o}ller, Palle and Pettini, Max and Straka, Lorrie and Vernet, Jo{\"e}l and York, Donald G.},
  year = 2018,
  month = feb,
  journal = {MNRAS},
  volume = {474},
  pages = {254--270},
  publisher = {OUP},
  issn = {0035-8711},
  doi = {10.1093/mnras/stx2726},
  urldate = {2025-05-09}
}

@article{rheeNeutralHydrogenGas2018,
  title = {Neutral Hydrogen ({{H}}\,i) Gas Content of Galaxies at z {$\approx$} 0.32},
  author = {Rhee, Jonghwan and Lah, Philip and Briggs, Frank H. and Chengalur, Jayaram N. and Colless, Matthew and Willner, Steven P. and Ashby, Matthew L. N. and Le F{\`e}vre, Olivier},
  year = 2018,
  month = jan,
  journal = {MNRAS},
  volume = {473},
  number = {2},
  pages = {1879--1894},
  issn = {0035-8711},
  doi = {10.1093/mnras/stx2461},
  urldate = {2025-06-20}
}

@article{richingsEffectsLocalStellar2022,
  title = {The Effects of Local Stellar Radiation and Dust Depletion on Non-Equilibrium Interstellar Chemistry},
  author = {Richings, Alexander J. and {Faucher-Giguere}, Claude-Andre and Gurvich, Alexander B. and Schaye, Joop and Hayward, Christopher C.},
  year = 2022,
  month = oct,
  journal = {MNRAS},
  volume = {517},
  number = {2},
  eprint = {2208.02288},
  primaryclass = {astro-ph},
  pages = {1557--1583},
  issn = {0035-8711, 1365-2966},
  doi = {10.1093/mnras/stac2338},
  urldate = {2025-07-15},
  archiveprefix = {arXiv}
}

@article{richingsEffectsMetallicityUV2016,
  title = {The Effects of Metallicity, {{UV}} Radiation and Non-Equilibrium Chemistry in High-Resolution Simulations of Galaxies},
  author = {Richings, A. J. and Schaye, Joop},
  year = 2016,
  month = may,
  journal = {MNRAS},
  volume = {458},
  number = {1},
  pages = {270--292},
  issn = {0035-8711},
  doi = {10.1093/mnras/stw327},
  urldate = {2025-07-16}
}

@article{richterHSTCOSLegacy2017,
  title = {An {{HST}}/{{COS}} Legacy Survey of High-Velocity Ultraviolet Absorption in the {{Milky Way}}'s Circumgalactic Medium and the {{Local Group}}},
  author = {Richter, P. and Nuza, S. E. and Fox, A. J. and Wakker, B. P. and Lehner, N. and Bekhti, N. Ben and Fechner, C. and Wendt, M. and Howk, J. C. and Muzahid, S. and Ganguly, R. and Charlton, J. C.},
  year = 2017,
  month = nov,
  journal = {A\&A},
  volume = {607},
  eprint = {1611.07024},
  primaryclass = {astro-ph},
  pages = {A48},
  issn = {0004-6361, 1432-0746},
  doi = {10.1051/0004-6361/201630081},
  urldate = {2025-06-01},
  archiveprefix = {arXiv}
}

@article{saintongeColdInterstellarMedium2022,
  title = {The Cold Interstellar Medium of Galaxies in the {{Local Universe}}},
  author = {Saintonge, Amelie and Catinella, Barbara},
  year = 2022,
  month = aug,
  journal = {Annu. Rev. Astron. Astrophys.},
  volume = {60},
  number = {1},
  eprint = {2202.00690},
  primaryclass = {astro-ph},
  pages = {319--361},
  issn = {0066-4146, 1545-4282},
  doi = {10.1146/annurev-astro-021022-043545},
  urldate = {2023-03-01},
  archiveprefix = {arXiv}
}

@article{saintongeXCOLDGASSComplete2017,
  title = {{{xCOLD GASS}}: {{The Complete IRAM}} 30 m {{Legacy Survey}} of {{Molecular Gas}} for {{Galaxy Evolution Studies}}},
  shorttitle = {{{xCOLD GASS}}},
  author = {Saintonge, Am{\'e}lie and Catinella, Barbara and Tacconi, Linda J. and Kauffmann, Guinevere and Genzel, Reinhard and Cortese, Luca and Dav{\'e}, Romeel and Fletcher, Thomas J. and {Graci{\'a}-Carpio}, Javier and Kramer, Carsten and Heckman, Timothy M. and Janowiecki, Steven and Lutz, Katharina and Rosario, David and Schiminovich, David and Schuster, Karl and Wang, Jing and Wuyts, Stijn and Borthakur, Sanchayeeta and Lamperti, Isabella and {Roberts-Borsani}, Guido W.},
  year = 2017,
  month = dec,
  journal = {ApJS},
  volume = {233},
  pages = {22},
  publisher = {IOP},
  issn = {0067-0049},
  doi = {10.3847/1538-4365/aa97e0},
  urldate = {2024-06-21}
}

@article{schroetterMUSEGASFLOW2016,
  title = {{{MUSE GAS FLOW AND WIND}} ({{MEGAFLOW}}). {{I}}. {{FIRST MUSE RESULTS ON BACKGROUND QUASARS}}*},
  author = {Schroetter, I. and Bouch{\'e}, N. and Wendt, M. and Contini, T. and Finley, H. and Pell{\'o}, R. and Bacon, R. and Cantalupo, S. and Marino, R. A. and Richard, J. and Lilly, S. J. and Schaye, J. and Soto, K. and Steinmetz, M. and Straka, L. A. and Wisotzki, L.},
  year = 2016,
  month = dec,
  journal = {ApJ},
  volume = {833},
  number = {1},
  pages = {39},
  publisher = {The American Astronomical Society},
  issn = {0004-637X},
  doi = {10.3847/1538-4357/833/1/39},
  urldate = {2025-06-01},
  langid = {english}
}

@article{schrubaLOWCOLUMINOSITIES2012,
  title = {{{LOW CO LUMINOSITIES IN DWARF GALAXIES}}},
  author = {Schruba, Andreas and Leroy, Adam K. and Walter, Fabian and Bigiel, Frank and Brinks, Elias and {de Blok}, W. J. G. and Kramer, Carsten and Rosolowsky, Erik and Sandstrom, Karin and Schuster, Karl and Usero, Antonio and Weiss, Axel and Wiesemeyer, Helmut},
  year = 2012,
  month = may,
  journal = {AJ},
  volume = {143},
  number = {6},
  pages = {138},
  publisher = {The American Astronomical Society},
  issn = {1538-3881},
  doi = {10.1088/0004-6256/143/6/138},
  urldate = {2025-06-01},
  langid = {english}
}

@article{serraSOFIAFlexibleSource2015,
  title = {{{SOFIA}}: A Flexible Source Finder for {{3D}} Spectral Line Data},
  shorttitle = {{{SOFIA}}},
  author = {Serra, Paolo and Westmeier, Tobias and Giese, Nadine and Jurek, Russell and Fl{\"o}er, Lars and Popping, Attila and Winkel, Benjamin and {van der Hulst}, Thijs and Meyer, Martin and Koribalski, B{\"a}rbel S. and {Staveley-Smith}, Lister and Courtois, H{\'e}l{\`e}ne},
  year = 2015,
  month = apr,
  journal = {MNRAS},
  volume = {448},
  pages = {1922--1929},
  issn = {0035-8711},
  doi = {10.1093/mnras/stv079},
  urldate = {2022-09-09}
}

@article{sharmaGMRTHighResolution2023,
  title = {The {{GMRT High Resolution Southern Sky Survey}} for {{Pulsars}} and {{Transients}}. {{V}}. {{Localization}} of {{Two Millisecond Pulsars}}},
  author = {Sharma, Shyam S. and Roy, Jayanta and Kudale, Sanjay and Bhattacharyya, Bhaswati and Behera, Arpit K. and Singh, Shubham},
  year = 2023,
  month = apr,
  journal = {ApJ},
  volume = {947},
  number = {2},
  pages = {88},
  publisher = {The American Astronomical Society},
  issn = {0004-637X},
  doi = {10.3847/1538-4357/acc10f},
  urldate = {2025-06-20},
  langid = {english}
}

@article{smitEVIDENCEUBIQUITOUSHIGHEQUIVALENTWIDTH2014,
  title = {{{EVIDENCE FOR UBIQUITOUS HIGH-EQUIVALENT-WIDTH NEBULAR EMISSION IN}} {\emph{z}} {$\sim$} 7 {{GALAXIES}}: {{TOWARD A CLEAN MEASUREMENT OF THE SPECIFIC STAR-FORMATION RATE USING A SAMPLE OF BRIGHT}}, {{MAGNIFIED GALAXIES}}},
  shorttitle = {{{EVIDENCE FOR UBIQUITOUS HIGH-EQUIVALENT-WIDTH NEBULAR EMISSION IN}} {\emph{z}} {$\sim$} 7 {{GALAXIES}}},
  author = {Smit, R. and Bouwens, R. J. and Labb{\'e}, I. and Zheng, W. and Bradley, L. and Donahue, M. and Lemze, D. and Moustakas, J. and Umetsu, K. and Zitrin, A. and Coe, D. and Postman, M. and Gonzalez, V. and Bartelmann, M. and Ben{\'i}tez, N. and Broadhurst, T. and Ford, H. and Grillo, C. and Infante, L. and {Jimenez-Teja}, Y. and Jouvel, S. and Kelson, D. D. and Lahav, O. and Maoz, D. and Medezinski, E. and Melchior, P. and Meneghetti, M. and Merten, J. and Molino, A. and Moustakas, L. A. and Nonino, M. and Rosati, P. and Seitz, S.},
  year = 2014,
  month = mar,
  journal = {ApJ},
  volume = {784},
  number = {1},
  pages = {58},
  issn = {0004-637X, 1538-4357},
  doi = {10.1088/0004-637X/784/1/58},
  urldate = {2021-11-30},
  langid = {english}
}

@article{solomonMolecularInterstellarMedium1997,
  title = {The {{Molecular Interstellar Medium}} in {{Ultraluminous Infrared Galaxies}}},
  author = {Solomon, P. M. and Downes, D. and Radford, S. J. E. and Barrett, J. W.},
  year = 1997,
  month = mar,
  journal = {ApJ},
  volume = {478},
  pages = {144--161},
  issn = {0004-637X},
  doi = {10.1086/303765},
  urldate = {2022-11-23}
}

@article{sternbergITOH2TRANSITIONSCOLUMN2014,
  title = {H I-{{TO-H2 TRANSITIONS AND H}} i {{COLUMN DENSITIES IN GALAXY STAR-FORMING REGIONS}}},
  author = {Sternberg, Amiel and Le Petit, Franck and Roueff, Evelyne and Le Bourlot, Jacques},
  year = 2014,
  month = jun,
  journal = {ApJ},
  volume = {790},
  number = {1},
  pages = {10},
  publisher = {The American Astronomical Society},
  issn = {0004-637X},
  doi = {10.1088/0004-637X/790/1/10},
  urldate = {2025-05-31},
  langid = {english}
}

@article{sternUniversalDensityStructure2016,
  title = {A {{Universal Density Structure}} for {{Circumgalactic Gas}}},
  author = {Stern, Jonathan and Hennawi, Joseph F. and Prochaska, J. Xavier and Werk, Jessica K.},
  year = 2016,
  month = oct,
  journal = {ApJ},
  volume = {830},
  pages = {87},
  issn = {0004-637X},
  doi = {10.3847/0004-637X/830/2/87},
  urldate = {2023-04-24}
}

@article{stricklandHighSpatialResolution2004,
  title = {A {{High Spatial Resolution X-Ray}} and {{H$\alpha$ Study}} of {{Hot Gas}} in the {{Halos}} of {{Star-forming Disk Galaxies}}. {{II}}. {{Quantifying Supernova Feedback}}},
  author = {Strickland, David K. and Heckman, Timothy M. and Colbert, Edward J. M. and Hoopes, Charles G. and Weaver, Kimberly A.},
  year = 2004,
  month = may,
  journal = {ApJ},
  volume = {606},
  number = {2},
  pages = {829},
  issn = {0004-637X},
  doi = {10.1086/383136},
  urldate = {2025-06-01},
  langid = {english}
}

@article{suFASTDiscoveryFast2023,
  title = {{{FAST Discovery}} of a {{Fast Neutral Hydrogen Outflow}}},
  author = {Su, Renzhi and Gu, Minfeng and Curran, S. J. and Mahony, Elizabeth K. and Tang, Ningyu and Allison, James R. and Li, Di and Zhu, Ming and Aditya, J. N. H. S. and Yoon, Hyein and Zheng, Zheng and Wu, Zhongzu},
  year = 2023,
  month = oct,
  journal = {ApJ},
  volume = {956},
  pages = {L28},
  publisher = {IOP},
  issn = {0004-637X},
  doi = {10.3847/2041-8213/acf4fa},
  urldate = {2024-08-20}
}

@article{szakacsMUSEALMAHaloesVI2021,
  title = {{{MUSE-ALMA}} Haloes {{VI}}: Coupling Atomic, Ionized, and Molecular Gas Kinematics of Galaxies},
  shorttitle = {{{MUSE-ALMA}} Haloes {{VI}}},
  author = {Szakacs, Roland and P{\'e}roux, C{\'e}line and Zwaan, Martin and Hamanowicz, Aleksandra and Klitsch, Anne and Fresco, Alejandra Y. and Augustin, Ramona and Biggs, Andrew and Kulkarni, Varsha and Rahmani, Hadi},
  year = 2021,
  month = aug,
  journal = {MNRAS},
  volume = {505},
  pages = {4746--4761},
  issn = {0035-8711},
  doi = {10.1093/mnras/stab1434},
  urldate = {2022-10-28}
}

@article{tacchellaEvolutionDensityProfiles2016,
  title = {Evolution of Density Profiles in High-z Galaxies: Compaction and Quenching inside-Out},
  shorttitle = {Evolution of Density Profiles in High-z Galaxies},
  author = {Tacchella, Sandro and Dekel, Avishai and Carollo, C. Marcella and Ceverino, Daniel and DeGraf, Colin and Lapiner, Sharon and Mandelker, Nir and Primack, Joel R.},
  year = 2016,
  month = may,
  journal = {MNRAS},
  volume = {458},
  number = {1},
  pages = {242--263},
  issn = {0035-8711},
  doi = {10.1093/mnras/stw303},
  urldate = {2021-10-22}
}

@article{tacconiEvolutionStarFormingInterstellar2020,
  title = {The {{Evolution}} of the {{Star-Forming Interstellar Medium Across Cosmic Time}}},
  author = {Tacconi, Linda J. and Genzel, Reinhard and Sternberg, Amiel},
  year = 2020,
  month = aug,
  journal = {Annu. Rev. Astron. Astrophys.},
  volume = {58},
  pages = {157},
  issn = {0066-4146},
  doi = {10.1146/annurev-astro-082812-141034},
  urldate = {2022-10-21},
  langid = {english}
}

@article{tacconiHighMolecularGas2010,
  title = {High Molecular Gas Fractions in Normal Massive Star-Forming Galaxies in the Young {{Universe}}},
  author = {Tacconi, L. J. and Genzel, R. and Neri, R. and Cox, P. and Cooper, M. C. and Shapiro, K. and Bolatto, A. and Bouch{\'e}, N. and Bournaud, F. and Burkert, A. and Combes, F. and Comerford, J. and Davis, M. and Schreiber, N. M. F{\"o}rster and {Garcia-Burillo}, S. and {Gracia-Carpio}, J. and Lutz, D. and Naab, T. and Omont, A. and Shapley, A. and Sternberg, A. and Weiner, B.},
  year = 2010,
  month = feb,
  journal = {Nature},
  volume = {463},
  number = {7282},
  pages = {781--784},
  publisher = {Nature Publishing Group},
  issn = {1476-4687},
  doi = {10.1038/nature08773},
  urldate = {2024-02-13},
  copyright = {2010 Macmillan Publishers Limited. All rights reserved},
  langid = {english}
}

@article{tacconiPHIBSSUnifiedScaling2018,
  title = {{{PHIBSS}}: {{Unified Scaling Relations}} of {{Gas Depletion Time}} and {{Molecular Gas Fractions}}*},
  shorttitle = {{{PHIBSS}}},
  author = {Tacconi, L. J. and Genzel, R. and Saintonge, A. and Combes, F. and {Garc{\'i}a-Burillo}, S. and Neri, R. and Bolatto, A. and Contini, T. and Schreiber, N. M. F{\"o}rster and Lilly, S. and Lutz, D. and Wuyts, S. and Accurso, G. and Boissier, J. and Boone, F. and Bouch{\'e}, N. and Bournaud, F. and Burkert, A. and Carollo, M. and Cooper, M. and Cox, P. and Feruglio, C. and Freundlich, J. and {Herrera-Camus}, R. and Juneau, S. and Lippa, M. and Naab, T. and Renzini, A. and Salome, P. and Sternberg, A. and Tadaki, K. and {\"U}bler, H. and Walter, F. and Weiner, B. and Weiss, A.},
  year = 2018,
  month = feb,
  journal = {ApJ},
  volume = {853},
  number = {2},
  pages = {179},
  publisher = {The American Astronomical Society},
  issn = {0004-637X},
  doi = {10.3847/1538-4357/aaa4b4},
  urldate = {2024-02-13},
  langid = {english}
}

@article{thompsonPredictionsCOEmission2024,
  title = {Predictions for {{CO}} Emission and the {{CO-to-H2}} Conversion Factor in Galaxy Simulations with Non-Equilibrium Chemistry},
  author = {Thompson, Oliver A. and Richings, Alexander J. and Gibson, Brad K. and {Faucher-Gigu{\`e}re}, Claude-Andr{\'e} and Feldmann, Robert and Hayward, Christopher C.},
  year = 2024,
  month = aug,
  journal = {MNRAS},
  volume = {532},
  pages = {1948--1965},
  publisher = {OUP},
  issn = {0035-8711},
  doi = {10.1093/mnras/stae1486},
  urldate = {2025-05-28}
}

@article{tumlinsonCircumgalacticMedium2017,
  title = {The {{Circumgalactic Medium}}},
  author = {Tumlinson, Jason and Peeples, Molly S. and Werk, Jessica K.},
  year = 2017,
  month = aug,
  journal = {Annu. Rev. Astron. Astrophys.},
  volume = {55},
  number = {1},
  eprint = {1709.09180},
  pages = {389--432},
  issn = {0066-4146, 1545-4282},
  doi = {10.1146/annurev-astro-091916-055240},
  urldate = {2021-07-14},
  archiveprefix = {arXiv}
}

@article{voortCosmologicalSimulationsCircumgalactic2019,
  title = {Cosmological Simulations of the Circumgalactic Medium with 1 Kpc Resolution: Enhanced {{HI}} Column Densities},
  shorttitle = {Cosmological Simulations of the Circumgalactic Medium with 1 Kpc Resolution},
  author = {van de Voort, Freeke and Springel, Volker and Mandelker, Nir and van den Bosch, Frank C. and Pakmor, R{\"u}diger},
  year = 2019,
  month = jan,
  journal = {MNRAS},
  volume = {482},
  number = {1},
  eprint = {1808.04369},
  primaryclass = {astro-ph},
  pages = {L85-L89},
  issn = {1745-3925, 1745-3933},
  doi = {10.1093/mnrasl/sly190},
  urldate = {2025-06-23},
  archiveprefix = {arXiv}
}

@article{walterEvolutionBaryonsAssociated2020,
  title = {The {{Evolution}} of the {{Baryons Associated}} with {{Galaxies Averaged}} over {{Cosmic Time}} and {{Space}}},
  author = {Walter, Fabian and Carilli, Chris and Neeleman, Marcel and Decarli, Roberto and Popping, Gergo and Somerville, Rachel S. and Aravena, Manuel and Bertoldi, Frank and Boogaard, Leindert and Cox, Pierre and {da Cunha}, Elisabete and Magnelli, Benjamin and Obreschkow, Danail and Riechers, Dominik and Rix, Hans-Walter and Smail, Ian and Weiss, Axel and Assef, Roberto J. and Bauer, Franz and Bouwens, Rychard and Contini, Thierry and Cortes, Paulo C. and Daddi, Emanuele and {Diaz-Santo}, Tanio and {Gonzalez-Lopez}, Jorge and Hennawi, Joseph and Hodge, Jacqueline A. and Ivison, Rob and Oesch, Pascal and Sargent, Mark and {van der Werf}, Paul and Wagg, Jeff and Yung, L. Y. Aaron},
  year = 2020,
  month = oct,
  journal = {ApJ},
  volume = {902},
  number = {2},
  eprint = {2009.11126},
  primaryclass = {astro-ph},
  pages = {111},
  issn = {1538-4357},
  doi = {10.3847/1538-4357/abb82e},
  urldate = {2024-02-22},
  archiveprefix = {arXiv}
}

@article{walterTHINGSNEARBYGALAXY2008,
  title = {{{THINGS}}: {{THE H}} i {{NEARBY GALAXY SURVEY}}},
  shorttitle = {{{THINGS}}},
  author = {Walter, Fabian and Brinks, Elias and {de Blok}, W. J. G. and Bigiel, Frank and Kennicutt, Robert C. and Thornley, Michele D. and Leroy, Adam},
  year = 2008,
  month = nov,
  journal = {AJ},
  volume = {136},
  number = {6},
  pages = {2563},
  publisher = {The American Astronomical Society},
  issn = {1538-3881},
  doi = {10.1088/0004-6256/136/6/2563},
  urldate = {2025-06-20},
  langid = {english}
}

@article{wangFEASTSIGMCooling2023,
  title = {{{FEASTS}}: {{IGM Cooling Triggered}} by {{Tidal Interactions}} through the {{Diffuse H I Phase}} around {{NGC}} 4631},
  shorttitle = {{{FEASTS}}},
  author = {Wang, Jing and Yang, Dong and Oh, S. -H. and {Staveley-Smith}, Lister and Wang, Jie and Wang, Q. Daniel and Hess, Kelley M. and Ho, Luis C. and Hou, Ligang and Jing, Yingjie and Kamphuis, Peter and Li, Fujia and Lin, Xuchen and Liu, Ziming and Shao, Li and Wang, Shun and Zhu, Ming},
  year = 2023,
  month = feb,
  journal = {ApJ},
  volume = {944},
  pages = {102},
  publisher = {IOP},
  issn = {0004-637X},
  doi = {10.3847/1538-4357/acafe8},
  urldate = {2025-07-01}
}

@article{wangFEASTSRadialDistribution2025,
  title = {{{FEASTS}}: {{Radial Distribution}} of {{H I Surface Densities Down}} to 0.01 {{M}}{$\odot$} Pc-2 of 35 {{Nearby Galaxies}}},
  shorttitle = {{{FEASTS}}},
  author = {Wang, Jing and Yang, Dong and Lin, Xuchen and Huang, Qifeng and Qu, Zhijie and Chen, Hsiao-wen and Guo, Hong and Ho, Luis C. and Jiang, Peng and Liang, Zezhong and P{\'e}roux, C{\'e}line and {Staveley-Smith}, Lister and Weng, Simon},
  year = 2025,
  month = feb,
  journal = {ApJ},
  volume = {980},
  pages = {25},
  publisher = {IOP},
  issn = {0004-637X},
  doi = {10.3847/1538-4357/ada95a},
  urldate = {2025-07-01}
}

@article{weissSpectralEnergyDistribution2005,
  title = {The Spectral Energy Distribution of {{CO}} Lines in {{M}} 82},
  author = {Wei{\ss}, A. and Walter, F. and Scoville, N. Z.},
  year = 2005,
  month = aug,
  journal = {A\&A},
  volume = {438},
  pages = {533--544},
  issn = {0004-6361},
  doi = {10.1051/0004-6361:20052667},
  urldate = {2024-04-12}
}

@misc{wengMUSEALMAHaloesVIII2022,
  title = {{{MUSE-ALMA Haloes VIII}}: {{Statistical Study}} of {{Circumgalactic Medium Gas}}},
  shorttitle = {{{MUSE-ALMA Haloes VIII}}},
  author = {Weng, Simon and P{\'e}roux, C{\'e}line and Karki, Arjun and Augustin, Ramona and Kulkarni, Varsha P. and Szakacs, Roland and Zwaan, Martin A. and Klitsch, Anne and Hamanowicz, Aleksandra and Sadler, Elaine M. and Biggs, Andrew and Fresco, Alejandra Y. and Hayes, Mattjew and Howk, J. Christopher and Kacprzak, Glenn G. and Kuntschner, Harald and Nelson, Dylan and Pettini, Max},
  year = 2022,
  month = dec,
  number = {arXiv:2212.01395},
  eprint = {2212.01395},
  primaryclass = {astro-ph},
  publisher = {arXiv},
  doi = {10.48550/arXiv.2212.01395},
  urldate = {2022-12-06},
  archiveprefix = {arXiv}
}

@article{wengMUSEALMAHaloesVIII2023,
  title = {{{MUSE-ALMA Haloes}} - {{VIII}}. {{Statistical}} Study of Circumgalactic Medium Gas},
  author = {Weng, S. and P{\'e}roux, C. and Karki, A. and Augustin, R. and Kulkarni, V. P. and Szakacs, R. and Zwaan, M. A. and Klitsch, A. and Hamanowicz, A. and Sadler, E. M. and Biggs, A. and Fresco, A. Y. and Hayes, M. and Howk, J. C. and Kacprzak, G. G. and Kuntschner, H. and Nelson, D. and Pettini, M.},
  year = 2023,
  month = feb,
  journal = {MNRAS},
  volume = {519},
  pages = {931--947},
  issn = {0035-8711},
  doi = {10.1093/mnras/stac3497},
  urldate = {2023-10-12}
}

@article{wengMUSEALMAHaloesXI2023,
  title = {{{MUSE-ALMA Haloes XI}}: Gas Flows in the Circumgalactic Medium},
  shorttitle = {{{MUSE-ALMA Haloes XI}}},
  author = {Weng, Simon and P{\'e}roux, C{\'e}line and Karki, Arjun and Augustin, Ramona and Kulkarni, Varsha P. and Hamanowicz, Aleksandra and Zwaan, Martin and Sadler, Elaine M. and Nelson, Dylan and Hayes, Matthew J. and Kacprzak, Glenn G. and Fox, Andrew J. and Bollo, Victoria and Casavecchia, Benedetta and Szakacs, Roland},
  year = 2023,
  month = jul,
  journal = {MNRAS},
  volume = {523},
  pages = {676--700},
  issn = {0035-8711},
  doi = {10.1093/mnras/stad1462},
  urldate = {2023-10-12}
}

@article{wengPhysicalOriginsGas2024,
  title = {The Physical Origins of Gas in the Circumgalactic Medium Using Observationally Motivated {{TNG50}} Mocks},
  author = {Weng, Simon and P{\'e}roux, C{\'e}line and Ramesh, Rahul and Nelson, Dylan and Sadler, Elaine M. and Zwaan, Martin and Bollo, Victoria and Casavecchia, Benedetta},
  year = 2024,
  month = jan,
  journal = {MNRAS},
  volume = {527},
  pages = {3494--3516},
  publisher = {OUP},
  issn = {0035-8711},
  doi = {10.1093/mnras/stad3426},
  urldate = {2025-06-11}
}

@article{werkCOSHALOSSURVEYPHYSICAL2014,
  title = {{{THE COS-HALOS SURVEY}}: {{PHYSICAL CONDITIONS AND BARYONIC MASS IN THE LOW-REDSHIFT CIRCUMGALACTIC MEDIUM}}},
  shorttitle = {{{THE COS-HALOS SURVEY}}},
  author = {Werk, Jessica K. and Prochaska, J. Xavier and Tumlinson, Jason and Peeples, Molly S. and Tripp, Todd M. and Fox, Andrew J. and Lehner, Nicolas and Thom, Christopher and O'Meara, John M. and Ford, Amanda Brady and Bordoloi, Rongmon and Katz, Neal and Tejos, Nicolas and Oppenheimer, Benjamin D. and Dav{\'e}, Romeel and Weinberg, David H.},
  year = 2014,
  month = aug,
  journal = {ApJ},
  volume = {792},
  number = {1},
  pages = {8},
  publisher = {The American Astronomical Society},
  issn = {0004-637X},
  doi = {10.1088/0004-637X/792/1/8},
  urldate = {2025-02-27},
  langid = {english}
}

@article{westmeierSOFIA2Automated2021a,
  title = {{{SOFIA}} 2 - an Automated, Parallel {{H I}} Source Finding Pipeline for the {{WALLABY}} Survey},
  author = {Westmeier, T. and Kitaeff, S. and Pallot, D. and Serra, P. and {van der Hulst}, J. M. and Jurek, R. J. and Elagali, A. and For, B.-Q. and Kleiner, D. and Koribalski, B. S. and {Lee-Waddell}, K. and Mould, J. R. and Reynolds, T. N. and Rhee, J. and {Staveley-Smith}, L.},
  year = 2021,
  month = sep,
  journal = {Monthly Notices of the Royal Astronomical Society},
  volume = {506},
  pages = {3962--3976},
  publisher = {OUP},
  issn = {0035-8711},
  doi = {10.1093/mnras/stab1881},
  urldate = {2025-12-10}
}

@article{wolfeDAMPEDLYaSYSTEMS2005,
  title = {{{DAMPED LY$\alpha$ SYSTEMS}}},
  author = {Wolfe, Arthur M. and Gawiser, Eric and Prochaska, Jason X.},
  year = 2005,
  month = aug,
  journal = {Annu. Rev. Astron. Astrophys.},
  volume = {43},
  number = {Volume 43, 2005},
  pages = {861--918},
  publisher = {Annual Reviews},
  issn = {0066-4146, 1545-4282},
  doi = {10.1146/annurev.astro.42.053102.133950},
  urldate = {2025-01-25},
  langid = {english}
}

@article{wolfireDarkMolecularGas2010,
  title = {The {{Dark Molecular Gas}}},
  author = {Wolfire, Mark G. and Hollenbach, David and McKee, Christopher F.},
  year = 2010,
  month = jun,
  journal = {ApJ},
  volume = {716},
  pages = {1191--1207},
  publisher = {IOP},
  issn = {0004-637X},
  doi = {10.1088/0004-637X/716/2/1191},
  urldate = {2025-05-08}
}

@article{wolfireNeutralAtomicPhases1995,
  title = {The {{Neutral Atomic Phases}} of the {{Interstellar Medium}}},
  author = {Wolfire, M. G. and Hollenbach, D. and McKee, C. F. and Tielens, A. G. G. M. and Bakes, E. L. O.},
  year = 1995,
  month = apr,
  journal = {ApJ},
  volume = {443},
  pages = {152},
  publisher = {IOP},
  issn = {0004-637X},
  doi = {10.1086/175510},
  urldate = {2025-05-09}
}

@article{wuytsCONSISTENTSTUDYMETALLICITY2014,
  title = {A {{CONSISTENT STUDY OF METALLICITY EVOLUTION AT}} 0.8 {$<$} z {$<$} 2.6*},
  author = {Wuyts, Eva and Kurk, Jaron and F{\"o}rster Schreiber, Natascha M. and Genzel, Reinhard and Wisnioski, Emily and Bandara, Kaushala and Wuyts, Stijn and Beifiori, Alessandra and Bender, Ralf and Brammer, Gabriel B. and Burkert, Andreas and Buschkamp, Peter and Carollo, C. Marcella and Chan, Jeffrey and Davies, Ric and Eisenhauer, Frank and Fossati, Matteo and Kulkarni, Sandesh K. and Lang, Philipp and Lilly, Simon J. and Lutz, Dieter and Mancini, Chiara and Mendel, J. Trevor and Momcheva, Ivelina G. and Naab, Thorsten and Nelson, Erica J. and Renzini, Alvio and Rosario, David and Saglia, Roberto P. and Seitz, Stella and Sharples, Ray M. and Sternberg, Amiel and Tacchella, Sandro and Tacconi, Linda J. and {van Dokkum}, Pieter and Wilman, David J.},
  year = 2014,
  month = jun,
  journal = {ApJL},
  volume = {789},
  number = {2},
  pages = {L40},
  publisher = {The American Astronomical Society},
  issn = {2041-8205},
  doi = {10.1088/2041-8205/789/2/L40},
  urldate = {2025-05-13},
  langid = {english}
}

@article{xiMostDistantGalaxies2024,
  title = {The {{Most Distant H}} i {{Galaxies Discovered}} by the 500 m {{Dish FAST}}},
  author = {Xi, Hongwei and Peng, Bo and {Staveley-Smith}, Lister and For, Bi-Qing and Liu, Bin and Chen, Ru-Rong and Yu, Lei and Ding, Dejian and Guo, Wei-Jian and Zou, Hu and Xue, Suijian and Wang, Jing and Brink, Thomas G. and Zheng, WeiKang and Filippenko, Alexei V. and Yang, Yi and Wei, Jianyan and Dai, Y. Sophia and Li, Zi-Jian and He, Zizhao and Jiang, Chengzi and Moiseev, Alexei and Kotov, Sergey},
  year = 2024,
  month = may,
  journal = {ApJL},
  volume = {966},
  number = {2},
  pages = {L36},
  publisher = {The American Astronomical Society},
  issn = {2041-8205},
  doi = {10.3847/2041-8213/ad4357},
  urldate = {2025-06-20},
  langid = {english}
}

@misc{yoonFirstLargeAbsorption2024,
  title = {The {{First Large Absorption Survey}} in {{HI}} ({{FLASH}}): {{II}}. {{Pilot Survey}} Data Release and First Results},
  shorttitle = {The {{First Large Absorption Survey}} in {{HI}} ({{FLASH}})},
  author = {Yoon, Hyein and Sadler, Elaine M. and Mahony, Elizabeth K. and Allison, James R. and Glowacki, Marcin and Kerrison, Emily F. and Moss, Vanessa A. and Su, Renzhi and Weng, Simon and Whiting, Matthew and Wong, O. Ivy and Callingham, Joseph R. and Curran, Stephen J. and Darling, Jeremy and Edge, Alastair C. and Ellison, Sara L. and Emig, Kimberly L. and {Garratt-Smithson}, Lilian and German, Gordon and Grasha, Kathryn and Koribalski, Baerbel S. and Morganti, Raffaella and Oosterloo, Tom and P{\'e}roux, C{\'e}line and Pettini, Max and Pimbblet, Kevin A. and Zheng, Zheng and Zwaan, Martin and Ball, Lewis and Bock, Douglas C. -J. and Brodrick, David and Bunton, John D. and Cooray, F. R. and Edwards, Philip G. and Hayman, Douglas B. and Hotan, Aidan W. and {Lee-Waddell}, K. and {McClure-Griffiths}, N. M. and Ng, A. and Phillips, Chris J. and Raja, Wasim and Voronkov, Maxim A. and Westmeier, Tobias},
  year = 2024,
  month = aug,
  publisher = {arXiv},
  doi = {10.48550/arXiv.2408.06626},
  urldate = {2025-07-01}
}

@article{yorkOriginQSOAbsorption1986,
  title = {On the {{Origin}} of {{Some QSO Absorption Lines}}},
  author = {York, Donald G. and Dopita, Michael and Green, Richard and Bechtold, Jill},
  year = 1986,
  month = dec,
  journal = {ApJ},
  volume = {311},
  pages = {610},
  publisher = {IOP},
  issn = {0004-637X},
  doi = {10.1086/164800},
  urldate = {2025-06-01}
}

@article{yuCentrallyConcentratedDistribution2022,
  title = {Centrally {{Concentrated H}} i {{Distribution Enhances Star Formation}} in {{Galaxies}}},
  author = {Yu, Niankun and Ho, Luis C. and Wang, Jing},
  year = 2022,
  month = may,
  journal = {ApJ},
  volume = {930},
  number = {1},
  pages = {85},
  publisher = {The American Astronomical Society},
  issn = {0004-637X},
  doi = {10.3847/1538-4357/ac5f07},
  urldate = {2025-06-20},
  langid = {english}
}

@article{yuFASTObservationsNeutral2024,
  title = {{{FAST}} Observations of Neutral Hydrogen in the Interacting Galaxies {{NGC}} 3395/3396},
  author = {Yu, Nai-Ping and Zhu, Ming and Xu, Jin-Long and Zhang, Chuan-Peng and Yu, Hai-Yang and Liu, Xiao-Lan and Jiang, Peng and Ai, Mei},
  year = 2024,
  month = aug,
  journal = {MNRAS},
  volume = {532},
  pages = {1744--1751},
  publisher = {OUP},
  issn = {0035-8711},
  doi = {10.1093/mnras/stae1623},
  urldate = {2025-06-01}
}

@article{zahedyCosmicUltravioletBaryon2021,
  title = {The {{Cosmic Ultraviolet Baryon Survey}} ({{CUBS}}) -- {{III}}. {{Physical}} Properties and Elemental Abundances of {{Lyman-limit}} Systems at z {$<$} 1},
  author = {Zahedy, Fakhri S and Chen, Hsiao-Wen and Cooper, Thomas M and Boettcher, Erin and Johnson, Sean D and Rudie, Gwen C and Chen, Mandy C and Cantalupo, Sebastiano and Cooksey, Kathy L and {Faucher-Gigu{\`e}re}, Claude-Andr{\'e} and Greene, Jenny E and Lopez, Sebastian and Mulchaey, John S and Penton, Steven V and Petitjean, Patrick and Putman, Mary E and Rafelski, Marc and Rauch, Michael and Schaye, Joop and Simcoe, Robert A and Walth, Gregory L},
  year = 2021,
  month = sep,
  journal = {MNRAS},
  volume = {506},
  number = {1},
  pages = {877--902},
  issn = {0035-8711},
  doi = {10.1093/mnras/stab1661},
  urldate = {2025-06-01}
}

@article{zahidFMOSCOSMOSSURVEYSTARFORMING2014,
  title = {{{THE FMOS-COSMOS SURVEY OF STAR-FORMING GALAXIES AT}} z {$\sim$} 1.6. {{II}}. {{THE MASS}}--{{METALLICITY RELATION AND THE DEPENDENCE ON STAR FORMATION RATE AND DUST EXTINCTION}}},
  author = {Zahid, H. J. and Kashino, D. and Silverman, J. D. and Kewley, L. J. and Daddi, E. and Renzini, A. and Rodighiero, G. and Nagao, T. and Arimoto, N. and Sanders, D. B. and Kartaltepe, J. and Lilly, S. J. and Maier, C. and Geller, M. J. and Capak, P. and Carollo, C. M. and Chu, J. and Hasinger, G. and Ilbert, O. and Kajisawa, M. and Koekemoer, A. M. and Kovac{\textasciibreve}, K. and Le F{\`e}vre, O. and Masters, D. and McCracken, H. J. and Onodera, M. and Scoville, N. and Strazzullo, V. and Sugiyama, N. and Taniguchi, Y. and Team, The COSMOS},
  year = 2014,
  month = aug,
  journal = {ApJ},
  volume = {792},
  number = {1},
  pages = {75},
  publisher = {The American Astronomical Society},
  issn = {0004-637X},
  doi = {10.1088/0004-637X/792/1/75},
  urldate = {2025-05-13},
  langid = {english}
}

@article{zanellaIIEmissionMolecular2018,
  title = {The [{{C II}}] Emission as a Molecular Gas Mass Tracer in Galaxies at Low and High Redshifts},
  author = {Zanella, A. and Daddi, E. and Magdis, G. and Diaz Santos, T. and Cormier, D. and Liu, D. and Cibinel, A. and Gobat, R. and Dickinson, M. and Sargent, M. and Popping, G. and Madden, S. C. and Bethermin, M. and Hughes, T. M. and Valentino, F. and Rujopakarn, W. and Pannella, M. and Bournaud, F. and Walter, F. and Wang, T. and Elbaz, D. and Coogan, R. T.},
  year = 2018,
  month = dec,
  journal = {MNRAS},
  volume = {481},
  pages = {1976--1999},
  publisher = {OUP},
  issn = {0035-8711},
  doi = {10.1093/mnras/sty2394},
  urldate = {2025-01-17}
}

@misc{zwaanALMACALSurveyingUniverse2022,
  type = {Text},
  title = {{{ALMACAL}}: {{Surveying}} the {{Universe}} with {{ALMA Calibrator Observations}}},
  shorttitle = {{{ALMACAL}}},
  author = {Zwaan, Martin and Ivison, Rob and Peroux, C{\'e}line and Chen, Jianhang and Klitsch, Anne and Hamanowicz, Aleksandra and Szakacs, Roland and Weng, Simon and Biggs, Andrew and Smail, Ian},
  year = 2022,
  publisher = {European Southern Observatory (ESO)},
  issn = {0722-6691},
  urldate = {2022-10-19},
  howpublished = {https://doi.eso.org/10.18727/0722-6691/5256}
}
%
% - join the .bib files when you upload your source files
%-------------------------------------------------------------------

\begin{appendix}

% \section{Detections}
% Appendix: Detections 

\begin{sidewaystable*}
\centering
\footnotesize
\setlength{\tabcolsep}{4pt}
\renewcommand{\arraystretch}{0.85}
    \caption{Detections. Measured properties of the new nine detected sources from the MUSE-ALMA Haloes ALMA Large Program.}
    % \resizebox{\linewidth}{!}{%
\begin{tabular}{lcccccccccccc}
\toprule
       Galaxy &    $z_{\mathrm{gal}}$ &   $\log N$(H\,\textsc{i}) &      $b$  & 12 + log(O/H) & [M/H] &
       CO line &     $F_{\text{peak}}$ &                 $F_{\text{CO}}$ &                FWHM &          $\log$ $L^{\prime}_{\text{CO(1–0)}}$ &           $\log M_{\text{H}_2}$ \\

       &    &  cm$^{-2}$  &  kpc &  &  &  &   mJy &      mJy km s$^{-1}$ &  km s$^{-1}$ &  K km s$^{-1}$pc$^2$ &     $ M_{\odot}$ \\
       \midrule
       \midrule
 Q0152m2001\_12 &  0.78018 &  18.87 &   55.0 &                 $8.73 \pm  0.10$ &   $-1.9\pm0.1$ &  CO(2-1) &  0.61 &    $122 \pm 40$ &     $241 \pm 71$ &     $8.53 \pm 0.17$ &      $9.11 \pm 0.15$   \\
 Q0152p0023\_13 &  0.48206 &  19.78 &  140.0 &                 $8.68 \pm      0.10$ &   $-0.1\pm0.01$ &  CO(2-1) & - &   $<90 $ &  - &  $<8.44 $ &   $<9.06$  \\
 Q0152p0023\_20 &  0.48143 &  19.78 &  122.9 &                 $8.43 \pm       0.06$ &    $-0.1\pm0.01$ &  CO(2-1) &  - &   $<90 $ & - &  $<8.44$ &   $<9.28$  \\
 Q0152p0023\_23 &  0.48068 &  19.78 &  187.8 &                 $8.59 \pm      0.12$ &    $-0.1\pm0.01$ &  CO(2-1) &  - &   $<90 $ &  - &  $<8.43$ &   $<9.13$ \\
 Q0152p0023\_44 &  0.48320 &  19.78 &  170.7 &            - &   $-0.1\pm0.01$ &  CO(2-1) &  - &   $<90 $ &  - &  $<8.44 $ &   $<9.63 $ \\
   Q0454m220\_4 &  0.48377 &  18.65 &  107.6 &                 $8.66  \pm    0.05$ &   $-0.7^{+0.1}_{-0.7}$ &  CO(2-1) &  1.10 &    $256 \pm 31$ &     $198 \pm 37$ &     $8.50 \pm 0.05$ &      $9.14 \pm 0.05$ \\
  Q0454m220\_69 &  0.47450 &  19.45 &    8.4 &                 $8.33 \pm    0.12$ &  $-0.48\pm0.1$ &  CO(2-1) &  - &  $<106$ &  - &  $<8.50 $ &   $<9.44$  \\
  Q0454p039\_15 &  1.15497 &  18.59 &   58.0 &   - &  $-0.19\pm0.22$ &  CO(3-2) &  - &   $<88$ &  - &  $<8.84 $ &  $<10.02$  \\
  Q0454p039\_57 &  0.85877 &  20.69 &   18.3 &   - &   $-1.57\pm0.17$ &  CO(4-3) &  - &  $<271$ &  - &  $<8.83 $ &  $<10.00$ \\
  Q0454p039\_65 &  1.15470 &  18.59 &  127.1 &   - &   $-0.19\pm0.22$ &  CO(3-2) &  - &   $<88 $ &  - &  $<8.84 $ &   $<9.66 $ \\
 Q1110p0048\_15 &  0.56035 &  20.20 &  129.7 &                 $8.26 \pm   0.09$ & - &  CO(2-1) &  - &   $<73 $ &  - &  $<8.48 $ &   $<9.50$  \\
 Q1110p0048\_44 &  0.56118 &  20.20 &    5.7 &                 $8.46 \pm   0.10$ & - &  CO(2-1) &  - &   $<73$ &  - &  $<8.49$ &   $<9.30 $\\
  Q1110p0048\_6 &  0.56006 &  20.20 &   59.6 &                 $8.49   0.05$ & - &  CO(2-1) &  - &   $<73 $ &  - &  $<8.48 $ &   $<9.27 $  \\
 Q1211p1030\_13 &  0.89909 &  18.50 &  175.3 &          - &  $<-1.2$ &  CO(4-3) &  - &  $<159 $ &  - &  $<8.63 $ &   $<9.85$ \\
 Q1211p1030\_16 &  0.89940 &  18.50 &   77.4 &  - &   $<-1.2$ &  CO(4-3) &  0.90 &     $90 \pm 40$ &     $112 \pm 46$ &     $7.87 \pm 0.24$ &      $8.70 \pm 0.24$  \\
 Q1211p1030\_38 &  0.63031 &  20.30 &  138.4 &                 $8.44 \pm    0.10$ &   $-2.82\pm0.89$ &  CO(2-1) &  - &   $<89 $ &  - &  $<8.67 $ &   $<9.51 $ \\
 Q1211p1030\_57 &  0.62957 &  20.30 &   16.9 &                 $8.63 \pm  0.20$ &   $-2.82\pm0.89$ &  CO(2-1) &  - &   $<89$ &  - &  $<8.67$ &   $<9.34 $ \\
 Q1211p1030\_58 &  0.89894 &  18.50 &  268.6 &     -    &   $<-1.2$ &  CO(4-3) &  - &  $<159$ & - &  $<8.63 $ &   $<9.58 $  \\
  Q1229m021\_10 &  0.83190 &  18.84 &  173.6 &                 $8.63  \pm    0.16$ &   $<-2.6$ &  CO(2-1) &  - &   $<86 $ &  - &  $<8.90 $ &   $<9.56 $ \\
  Q1229m021\_41 &  0.83199 &  18.84 &  186.8 &                 $8.34 \pm   0.19$ &   $<-2.6$ &  CO(2-1) &  - &   $<86 $ &  - &  $<8.90$ &   $<9.83 $ \\
   Q1229m021\_6 &  0.83170 &  18.84 &  182.2 &   -   &   $<-2.6$ &  CO(2-1) &  0.39 &    $197 \pm 43$ &    $419 \pm 191$ &     $8.93 \pm 0.08$ &      $9.74 \pm 0.08$  \\
   Q1229m021\_8 &  0.83060 &  18.84 &  125.5 &   - &   $<-2.6$ &  CO(2-1) &  0.36 &    $172 \pm 16$ &    $464 \pm 204$ &     $8.76 \pm 0.04$ &      $9.59 \pm 0.04$  \\
  Q1342m0035\_4 &  0.53773 &  19.78 &   43.6 &                 $8.67 \pm   0.06$ &   $-1.2\pm0.1$ &  CO(2-1) &  1.44 &    $273 \pm 40$ &     $169 \pm 34$ &     $8.63 \pm 0.06$ &      $9.26 \pm 0.06$  \\
  Q1342m0035\_9 &  0.53877 &  19.78 &   24.3 &                 $8.60 \pm  0.18$ &   $-1.2\pm0.1$ &  CO(2-1) &  - &   $<92 $ &  - &  $<8.55 $ &   $<9.24$ \\
 Q1345m0023\_13 &  0.60697 &  18.85 &   55.6 &                 $8.64 \pm   0.07$ &   $-0.9\pm0.2$ &  CO(2-1) &  0.94 &    $201 \pm 43$ &     $224 \pm 75$ &     $8.49 \pm 0.12$ &      $9.15 \pm 0.13$  \\
 Q1345m0023\_40 &  0.60540 &  18.85 &  162.4 &                 $8.31 \pm  0.14$ &   $-0.9\pm0.2$ &  CO(2-1) &  - &  $<112 $ &  - &  $<8.74 $ &   $<9.70 $  \\
 Q1431m0050\_10 &  0.60843 &  19.18 &   45.4 &                 $8.62 \pm  0.05$ &   $-0.8\pm0.1$ &  CO(2-1) &  4.13 &    $116 \pm 47$ &  $77 \pm 13$ &     $8.46 \pm 0.13$ &      $9.13 \pm 0.15$  \\
 Q1431m0050\_26 &  0.68710 &  18.40 &   23.3 &   - &   $-1.2^{+0.03}_{-0.01}$ &  CO(2-1) &  - &  $<100 $ &  - &  $<8.80 $ &   $<9.41 $  \\
 Q1431m0050\_68 &  0.60742 &  19.18 &   73.0 &    - &  $-0.8\pm0.1$ &  CO(2-1) &  - &   $<95$ &  - &  $<8.67 $ &   $<9.34 $  \\
 Q1431m0050\_73 &  0.68687 &  18.40 &  188.7 &  - &  $-1.2^{+0.03}_{-0.01}$  &  CO(2-1) &  - &  $<100 $ &  - &  $<8.80 $ &   $<9.84 $  \\
 Q1515p0410\_13 &  0.55860 &  20.20 &  194.1 &                 $8.75 \pm   0.08$ &   $0.2\pm0.24$ &  CO(2-1) &  - &   $<78 $ &  - &  $<8.51 $ &   $<9.07 $  \\
  Q1515p0410\_4 &  0.55859 &  20.20 &   10.1 &   - &   $0.2\pm0.24$ &  CO(2-1) &  - &   $<78$ &  - &  $<8.51 $ &   $<9.16$  \\
 Q1515p0410\_54 &  0.55892 &  20.20 &  168.8 &                 $8.53 \pm    0.15$ &   $0.2\pm0.24$ &  CO(2-1) &  - &   $<78$ &  - &  $<8.51$ &   $<9.26 $ \\
  Q1515p0410\_9 &  0.55866 &  20.20 &  221.0 &                 $8.61 \pm  0.11$ &   $0.2\pm0.24$ &  CO(2-1) &  - &   $<78$ &  - &  $<8.51 $ &   $<9.19 $ \\
  Q1554m203\_51 &  0.78558 &  19.00 &   23.0 &   - &   $<-1.6$ &  CO(2-1) &  - &   $<79$ &  - &  $<8.82 $ &   $<9.69 $  \\
 Q2131m1207\_26 &  0.42971 &  19.50 &  172.1 &                 $8.43 \pm   0.12$ &    $-1.47\pm0.22$ &  CO(2-1) &  - &   $<91$ &  - &  $<8.34 $ &   $<9.18$\\
 Q2131m1207\_34 &  0.43080 &  19.50 &   59.2 &                 $8.12 \pm   0.11$ &     $-1.47\pm0.22$ &  CO(2-1) &  - &   $<91 $ &  - &  $<8.34 $ &   $<9.52 $ \\
 Q2131m1207\_43 &  0.43003 &  19.50 &  145.9 &   - &     $-1.47\pm0.22$  &  CO(2-1) &  - &   $<91$ &  - &  $<8.34$ &   $<8.99 $  \\
  Q2131m1207\_5 &  0.42999 &  19.50 &   48.5 &          $8.68 \pm  0.05$ &     $-1.47\pm0.22$  &  CO(2-1) &  0.85 &    $210 \pm 26$ &     $225 \pm 41$ &     $8.26 \pm 0.06$ &      $8.88 \pm 0.06$ \\
\bottomrule
\end{tabular}
\tablefoot{$b$ is the impact parameter, and $F_{\mathrm{peak}}$ is the measured peak flux}

\label{table:co_prop} 
% }
\end{sidewaystable*}
% \end{table*}
% \end{landscape}

\end{appendix}

\end{document}